\documentclass[]{jfm}

\usepackage{graphicx}
\usepackage{newtxtext}
\usepackage{newtxmath}
\usepackage{natbib}
\usepackage{hyperref}
\usepackage{subcaption,caption}
\usepackage{placeins}
\usepackage{comment}
\usepackage{bm}
\usepackage{float}

\hypersetup{
    colorlinks = true,
    urlcolor   = blue,
    citecolor  = blue,
}

\newcommand{\RomanNumeralCaps}[1]
\linenumbers

\title{On the choking mechanism in supersonic ejectors:\\ a one-dimensional analysis of Reynolds-Averaged Navier Stokes simulations}

\author{Jan Van den Berghe\aff{1,2}
  \corresp{\email{jan.vandenberghe@vki.ac.be}},
  Miguel A. Mendez\aff{1,3,4}
 \and Yann Bartosiewicz\aff{2}}

\affiliation{
\aff{1} von Karman Institute for Fluid Dynamics, 1640 Sint-Genesius-Rode, Belgium
\aff{2} Institute of Mechanics, Materials, and Civil Engineering (iMMC), Université catholique de Louvain (UCLouvain),
1348 Louvain-la-Neuve, Belgium
\aff{3} Aero-Thermo-Mechanics Laboratory, Université Libre de Bruxelles, Elsene, Brussels, 1050, Belgium
\aff{4} Aerospace Engineering Research Group, Universidad Carlos III de Madrid,Leganés, 28911, Spain
}

\begin{document}
\newcommand\Ma{\mbox{{Ma}}}  % Mach number

\maketitle

\begin{abstract}
Ejectors are passive devices used in refrigeration, propulsion, and process industries to compress a secondary stream without moving parts. The engineering modeling of choking in these devices remains an open question, with two mechanisms—Fabri and compound choking—proposed in the literature. This work develops a unified one-dimensional framework that implements both mechanisms and compares them with axisymmetric Reynolds-Averaged Navier Stokes (RANS) data processed by cross-sectional averaging. The compound formulation incorporates wall and inter-stream friction and a local pressure-equalization procedure that enables stable integration through the sonic point, together with a normal-shock reconstruction. 
The Fabri formulation is assessed by imposing the dividing streamline extracted from RANS, isolating the sonic condition while avoiding additional modeling assumptions. The calibrated compound model predicts on-design secondary mass flow typically within ~2\% with respect to the RANS simulations, rising to ~5\% for a strongly under-expanded primary jet due to the equal-pressure constraint.
The Fabri analysis attains less than 1\% error in on-design entrainment but exhibits high sensitivity to the dividing streamline and closure, which limits predictive use beyond on-design. Overall, the results show that Fabri and compound mechanisms can coexist within the same device and operating map, each capturing distinct aspects of the physics and offering complementary modeling value. Nevertheless, compound choking emerges as the more general mechanism governing flow rate blockage, as evidenced by choked flows with a subsonic secondary stream.

\end{abstract}


% {\bf MSC Codes }  {\it(Optional)} Please enter your MSC Codes here

\section{Introduction}

Ejectors are fluid dynamic devices that utilize a high-speed primary jet to entrain and compress a secondary flow without moving parts. These are widely used in refrigeration, propulsion and chemical processes because of their simplicity and robustness (see for example the review by \cite{aidoun2019}). The maximum achievable entrainment of the secondary stream for a given set of inlet conditions is defined by the choked operating regime, which is thus of particular importance. This regime is characterized by the critical back pressure—a threshold below which the secondary mass flow rate becomes insensitive to further reductions in downstream pressure. Ejectors are usually designed to operate in these choked conditions to maximize entrainment efficiency, prevent flow reversal, and ensure stable and robust performance against fluctuations in back pressure.

The choked behavior is analogous to that of a converging-diverging nozzle, where choking occurs when the flow reaches unitary Mach at the throat. Unlike a nozzle, however, the presence of entrainment and mixing complicates the identification of a geometric location or a single Mach number criterion for the onset of choking. Instead, choking in ejectors arises from the complex interaction between the primary and secondary streams and the evolving flow within the mixing chamber and diffuser. Two main theories have been proposed to explain the choking mechanism by postulating different sonic conditions in the two streams: the Fabri theory and the compound theory. 

The Fabri theory, introduced in the seminal work of \cite{fabri1958}, treats the sonic conditions of the primary and secondary streams as independent. According to this theory, an ejector is considered choked when each stream individually attains Mach 1 at its respective throat, while accounting for the expansion of the primary stream within the mixing section. In contrast, the compound choking theory, originally proposed by \cite{pearson1958theory} and subsequently refined by \cite{hoge1965choked} and \cite{bernstein1967compound}, defines choking as a compound sonic condition in which the two streams are dynamically coupled. Under this framework, choking may occur even if one of the streams remains subsonic throughout the mixing chamber.

To date, neither theory has been definitively proven superior or more general than the other. Computational fluid dynamics (CFD) has made it possible to reproduce the mass-flow limitations observed in ejectors (e.g., the simulations of \cite{BARTOSIEWICZ200556} and \cite{HEMIDI2009_part1}), but these simulations have not provided conclusive evidence in favor of one choking mechanism over the other. In fact, both types of behavior have been reported in the literature: \cite{lamberts2018fabri, LAMBERTS2018_compound} documented cases consistent with Fabri choking as well as cases exhibiting compound choking. The sonic line clearly penetrates the core of the secondary stream in the first case, while it stays within the shear layer between the streams in the latter case. These results suggest that the compound theory can be overruled depending on the operating conditions and that both mechanisms may even be active simultaneously. 

The works of \cite{hoge1965choked} and \cite{debroeyer2025axisym} show that higher ratios of primary to secondary total pressures tend to favor Fabri choking. A possible explanation is the increasingly under-expanded nature of the primary stream, which results in stronger expansion within the mixing chamber and thus a restriction of the secondary stream akin to the throat of a converging–diverging nozzle. On the other hand, lower inlet pressure ratios tend to produce over-expanded primary flows, which contract in the mixing channel and hence leave more space for the secondary stream, which then remains subsonic throughout. Wall friction has also been shown to play an important role: \cite{KRACIK2023109168} numerically showed that smooth walls led to Fabri choking, while rough walls resulted in compound choking under the same operating conditions. However, the mechanism behind this switch remains poorly understood.

Historically, the Fabri condition has been the most widely used in integral 0D models, with a long lineage of developments from the 1940s to recent years (see \cite{keenan1942, fabri1958, huang1999, chen2013}, to name but a few). In contrast, the compound flow theory has only gained attention more recently (see \cite{METSUE2021121856, CROQUER2021120396, zhu2024compound}). Notably, \cite{CROQUER2021120396} reported a reduction in prediction error from 17\% with the Fabri theory to 5\% using the compound theory. While this might suggest a superiority of the compound theory, it is important to note that 0D models rely heavily on the calibration of efficiency coefficients.

A comprehensive analysis of the choking mechanism requires, at minimum, one-dimensional models capable of capturing the underlying wave propagation phenomena. These models formulate the conservation of mass, momentum, and energy for cross-section–averaged flow quantities, with the objective of reproducing their streamwise distribution under prescribed ejector operating conditions. However, as discussed in section~\ref{sec:definition}, such models are not intrinsically closed: either the static pressures or the cross-sectional areas of the two streams must be specified at each station. A traditional approach to render the compound choking theory predictive is to assume equal static pressure in both streams throughout the mixing region. This assumption implies planar pressure waves and equal propagation velocities in the two streams, leading to a simplified relation for the streamwise evolution of the dividing streamline and to an equivalent Mach number governing compound choking. This may explain why nearly all one-dimensional models reported in the literature rely on the compound flow framework.

Building upon this foundation, the 1D models of \cite{clark1995application,papamoschou1996analysis,grazzini2015constructal} incorporate momentum and energy exchange between the streams through wall friction and interfacial shear. The formulation proposed by \cite{BANASIAK20112235} further extends these approaches to account for phase change and introduces a hybrid 0D–1D framework, in which static pressures are equalized across a lumped control volume located between the primary nozzle and the constant-area section. A related 0D–1D model was later developed by \cite{vandenberghe2023unsteady} to investigate ejector transients, representing the device as a junction of three unsteady one-dimensional ducts without distinguishing between primary and secondary streams. In this formulation, the distinction between compound and Fabri choking does not explicitly appear, since the flow state is cross-sectionally averaged and choking naturally arises in the unsteady solution when the Mach number reaches unity within the mixing duct. 

Focusing on two-stream formulations, the existing compound-based models (\cite{clark1995application, papamoschou1996analysis, grazzini2015constructal, BANASIAK20112235}) treat off design operation and thus do not delve into the choking conditions. To the authors knowledge, the only two-stream formulation treating on-design conditions in a 1D compound setting is the work of \cite{vandenberghe2024extensioncompoundflowtheory}. Similarly, on the Fabri choking side, the only two-stream formulation in the literature is the model by \cite{delvalle2012}, which leverages potential flow and Prandtl–Meyer expansion theories, assuming that the flow is isentropic, and focuses on on-design operations.

This work presents four configurations of a one-dimensional model, aimed at assessing the range of validity of both choking mechanisms. The first two configurations are based on the recent extension of the compound choking theory by \cite{vandenberghe2024extensioncompoundflowtheory}, which accounts for wall friction effects at the sonic section and provides a practical framework for integrating the governing equations through the singular sonic point. In the first case, the friction forces are prescribed using calibrated correlations, yielding the only fully predictive model considered here. In the second, the friction forces are directly imposed from axisymmetric RANS simulations, thereby removing uncertainties associated with empirical closures and isolating the influence of the choking mechanism itself. The remaining two configurations implement the Fabri choking condition, which requires an external specification of the dividing streamline. This input is provided either from RANS data or from the compound model. The latter case establishes a direct comparison between the two theories under otherwise identical conditions, allowing us to explore possible overlap or convergence between the Fabri and compound frameworks.

The common base of the model and both choking mechanisms are formally introduced in section~\ref{sec:definition}. Practical considerations for implementing the model are presented in section~\ref{sec:practical}. Section~\ref{sec:solution} explains the solution procedure of the governing equations. The reference data are discussed in section~\ref{sec:data}, which are used for calibration in section~\ref{sec:calibration}. The resulting predictions are compared and analyzed in section~\ref{sec:results}. The paper closes with conclusions and perspectives in section~\ref{sec:conclusion}.

\section{A one-dimensional view on choking of parallel streams}\label{sec:definition}
Section~\ref{sec:definition_general} introduces the general one-dimensional governing equations. These leave one degree of freedom associated with the dividing streamline in the mixing pipe. Closure proceeds along two alternative routes: (i) enforcing equal static pressure across the streams, yielding the compound-choking formulation; or (ii) prescribing the streamwise evolution of the cross-sectional partition, yielding the Fabri-choking formulation. Further closure is required to account for the exchange terms between the two streams. This is addressed either by (a) employing friction coefficients and predictive correlations; or (b) imposing the exchanges extracted from an external source--—specifically, the RANS simulations presented in section~\ref{sec:data}. These choices yield four distinct model configurations, which are introduced in section~\ref{sec:definition_models}.

\subsection{General equations}\label{sec:definition_general}
\begin{figure*}%[htb!]
  \centering
  \begin{subfigure}[t]{\textwidth}
    \center
    \includegraphics[width=0.75\textwidth]{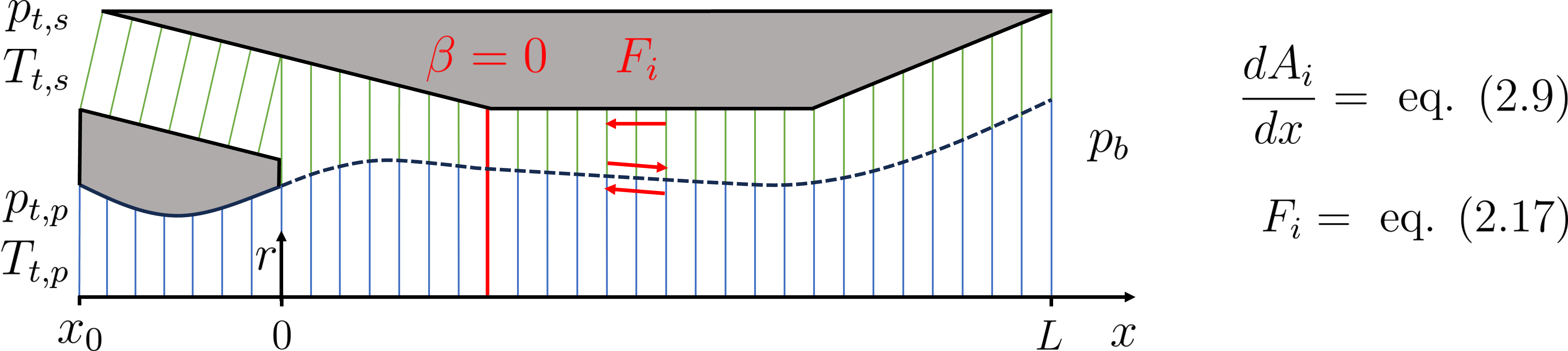}
    \subcaption{Model 1: compound choking with calibrated closure}
    \label{fig:schematics_compound_corr}
  \end{subfigure}
  \begin{subfigure}[t]{\textwidth}
    \center
    \includegraphics[width=0.75\textwidth]{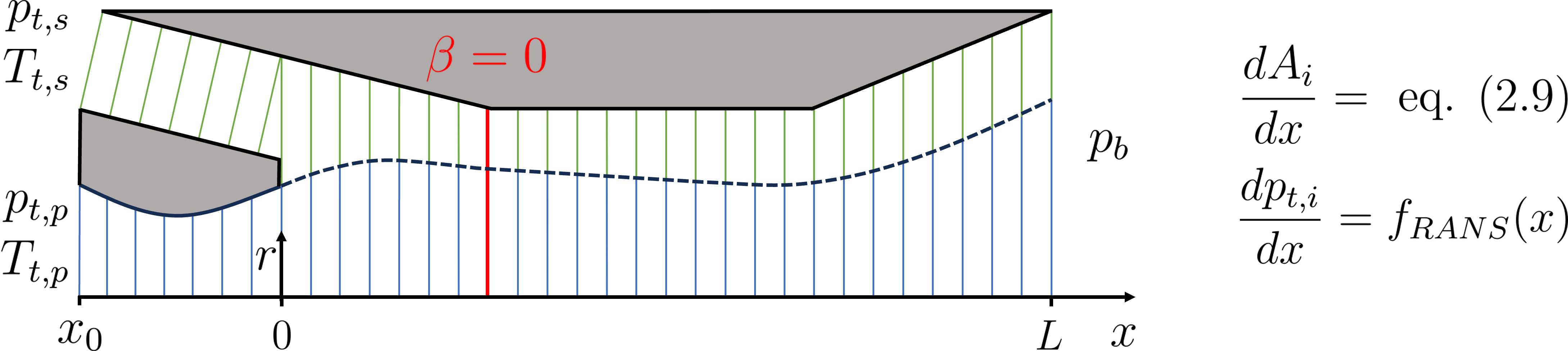}
    \subcaption{Model 2: compound choking with closure from CFD}
    \label{fig:schematics_compound_pt}
  \end{subfigure}
  \begin{subfigure}[t]{\textwidth}
    \center
    \includegraphics[width=0.75\textwidth]{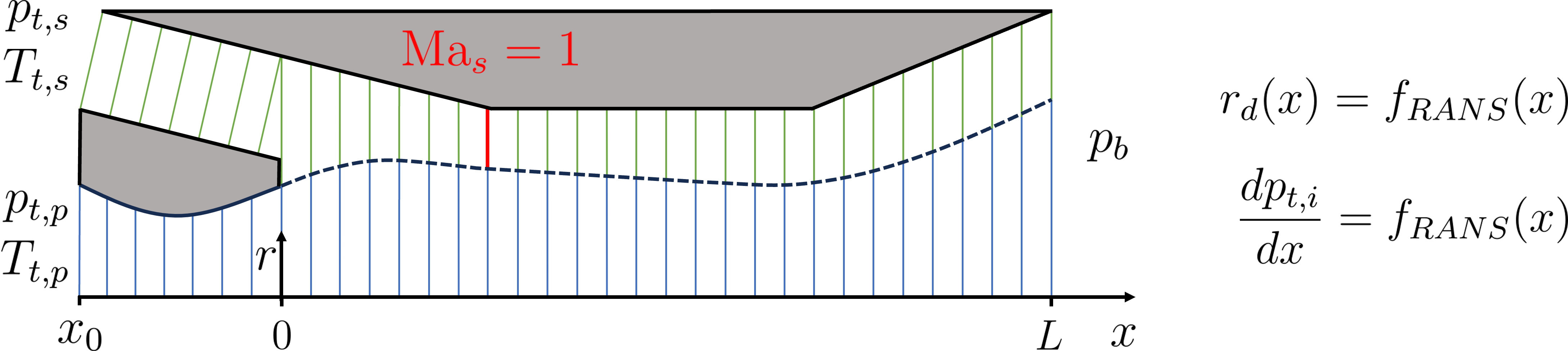}
    \subcaption{Model 3: Fabri choking with closure from CFD}
    \label{fig:schematics_fabri_RANS}
  \end{subfigure}
  \begin{subfigure}[t]{\textwidth}
    \center
    \includegraphics[width=0.75\textwidth]{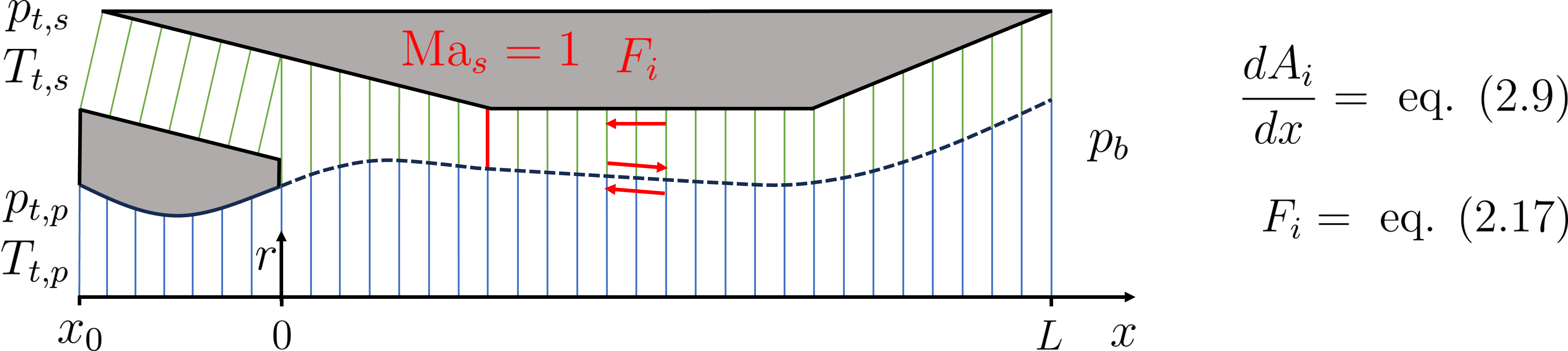}
    \subcaption{Model 4: Fabri choking with closure from compound choking theory}
    \label{fig:schematics_fabri_compound}
  \end{subfigure}
  \captionsetup{width=\textwidth, justification=justified}
  \caption{The ejector is modeled with two 1D domains that are coupled in the mixing pipe. The geometry is assumed to be axisymmetric. The boundary conditions consist of the total pressures $p_{t,i}$ and total temperatures $T_{t,i}$ at the inlets and the static back pressure $p_b$. The dividing streamline, indicated by the dotted line, is computed with the Equation~\eqref{eq:A_i} from the compound choking theory or imposed from the RANS simulations. Momentum exchange is accounted for through friction forces (see \eqref{eq:forces_mix}) or through imposed total pressure gradients from RANS.}
\label{fig:schematics}
\end{figure*}

The 1D modeling approach considered in this work treats the primary and secondary streams as two parallel quasi-1D domains along the axial coordinate $x$, in an axisymmetric geometry with radial coordinate as shown schematically in figure~\ref{fig:schematics}. The domains span the full ejector: single-stream inlets merge at $x=0$ and continue as a coupled, double quasi-1D mixing pipe.

Cross-sections $A_i(x)$, with $i \in [p,s]$, vary smoothly along $x$ and reveal a local discontinuity for the secondary at $x=0$, where inlet inclination and finite primary-nozzle thickness are treated as in section~\ref{sec:practical_jump}. The governing equations are the conservation laws of mass, momentum, and energy in variable-area ducts, with ideal-gas behavior and no inter-stream mass transfer—hence the interface in figure~\ref{fig:schematics} is a dividing streamline. Written in terms of static pressure $p$, the total pressure $p_t$ and the total temperature $T_t$, these read:
\begin{align}
     \frac{1}{p_i} \frac{dp_i}{dx} =  &\left[\frac{\gamma \Ma_i^2}{1-\Ma_i^2}\right]\frac{1}{A_i}\frac{d A_i}{dx} + \left[\frac{1 + \left(\gamma  - 1\right)\Ma_i^2}{1-\Ma_i^2}\right] \frac{F_i}{A_i p_i}- \left[\frac{\left(\gamma-1\right)\Ma_i^2}{1-\Ma_i^2}\right]\frac{Q_i}{p_i A_i u_i}\,, \label{eq:p_i}\\
    \frac{1}{p_{t,i}}\frac{dp_{t,i}}{dx} = &\frac{F_i}{A_i p_i} - \left[\frac{\left(\gamma-1\right)\Ma_i^2}{2 + \left(\gamma-1\right)\Ma_i^2}\right]\frac{Q_i}{p_i A_i u_i}\,,\label{eq:pt_i}\\
    \frac{1}{T_{t,i}}\frac{dT_{t,i}}{dx} = &\left[1 + \frac{\gamma-1}{2}\Ma_i^2\right]^{-1}\frac{Q_i}{p_i A_i u_i}\,,\label{eq:Tt_i}
\end{align}
with $\gamma$ the specific-heat ratio, $F_i$ and $Q_i$ the momentum and energy exchange received by stream $i$ per unit length ($dx$), $u_i$ the axial velocity, and $\Ma_i$ the Mach numbers defined as
\begin{align}
    \Ma_i =  u_i / a_i\,, \quad \mbox{and } \quad a_i = \sqrt{\gamma R T_i} \, ,
\end{align} with $T_i$ the static temperature for each stream and $R$ the specific gas constant. The relations \eqref{eq:p_i}--\eqref{eq:Tt_i} apply both in the separate inlets and in the mixing pipe with the practical difference that the definitions of the momentum and energy exchanges \(F_i\) and \(Q_i\) depend on the domain. For example, wall friction acts where a stream is in contact with a wall, whereas inter-stream shear appears only in the mixing pipe; details are given in section~\ref{sec:practical}. The reader is referred to \cite{vandenberghe2024extensioncompoundflowtheory} and to \cite{shapiro1953dynamics} for more details on the derivation of these equations. 

Energy-exchange terms are retained in \eqref{eq:p_i}--\eqref{eq:Tt_i} for completeness but are set to zero in the present study. The walls are treated as adiabatic and the inlet total temperatures are identical (consistent with the RANS dataset; see section~\ref{sec:data}). Limited inter-stream energy transfer is observed in the reference data; but the corresponding contributions through $Q_i$ in Equations~\eqref{eq:p_i} and \eqref{eq:pt_i} are two orders of magnitude smaller than the terms related to the gradient of the cross-section and the forces $F_i$. This is consistent with the LES of \cite{debroeyer2024analysis}, which reports only a short, localized exchange of total enthalpy in the converging portion of the mixing pipe. The transfer of kinetic energy from the primary to the secondary stream is offset by heat transfer in the opposite direction, owing to the low temperature of the strongly expanded primary stream. Accordingly, \(Q_i=0\) is adopted and no closure terms are introduced. Under this assumption, \eqref{eq:Tt_i} implies uniform total temperature, while \eqref{eq:pt_i} shows that total-pressure variation is governed solely by the force terms \(F_i\). Energy-exchange closures may be required when inlet total temperatures differ significantly but this is left for future work.

The radial position $r_d(x)$ of the dividing streamline determines the local cross-sections of both domains and is defined implicitly from the primary mass flow rate as
\begin{equation}\label{eq:def_r_dividing}
\dot m_p(x)=\int_0^{r_d(x)} \rho(x,r)\,u(x,r)\,2\pi r\,dr,
\end{equation}
so that, at the exit of the primary nozzle \((x=0)\), the streamline originates at the nozzle lip (figure~\ref{fig:schematics}).

The cross-sectional areas \(A_i\) and their axial gradients are prescribed by geometry in the inlet sections but remain unknown in the mixing region, where the position of the dividing streamline must be determined. These unknown areas satisfy the geometric constraint \(A_p(x) + A_s(x) = A(x)\), with \(A(x)\) the total cross-section of the mixing duct, leaving a single degree of freedom. A unique solution can then be obtained by closing the system in one of two ways: by enforcing equal static pressure in both streams---yielding the compound-choking-based models (e.g., \cite{pearson1958theory, bernstein1967compound})---or by prescribing the dividing streamline from an external source, which defines the Fabri-choking-based models.

In this work, both approaches were implemented and evaluated, highlighting their respective advantages and limitations, as well as their potential common ground. Furthermore, two alternatives are considered for the exchange terms in Equations~\eqref{eq:p_i}–\eqref{eq:Tt_i}: they were either defined through correlations or extracted from the axisymmetric RANS simulations presented in section~\ref{sec:data}. The test cases in section~\ref{sec:results} are organized according to these two aspects and are introduced below.

\subsection{Selected models}\label{sec:definition_models}
The four model configurations investigated in this work combine two treatments of momentum exchange with two choking mechanisms, as illustrated in figure~\ref{fig:schematics}. The compound-based models include a fully predictive case using empirical friction correlations (Model~1, section~\ref{sec:definition_compound_predictive}) and a non-predictive one where total pressure gradients are imposed from the RANS simulations (Model~2, section~\ref{sec:definition_compound_analysis}). Comparing these two isolates the role of the closure terms and reveals the intrinsic limitations of the compound formulation.  

The Fabri-based models follow the same logic: the dividing streamline and forces are either imposed from RANS data (Model~3, section~\ref{sec:definition_fabri_analysis}) or taken from the compound model while retaining its closure correlations (Model~4, section~\ref{sec:definition_fabri_rdiv_compound}). These comparisons highlight how the choking mechanism itself influences the predicted flow evolution and help identify conditions where the Fabri and compound frameworks may converge.

\subsubsection{Model 1: Compound choking with calibrated closures}\label{sec:definition_compound_predictive}
Parallel streams choke according to the compound flow theory when a sonic condition involving \emph{all} streams is met. This is commonly modeled through the constraint of uniform static pressure at any streamwise position $x$, as proposed by \cite{pearson1958theory} and \cite{bernstein1967compound} among others. Summing \eqref{eq:p_i} for the primary and secondary streams yields the pressure evolution
\begin{equation}
    \frac{1}{p} \frac{dp}{dx} = \dfrac{1}{\beta}\bigg( \dfrac{dA}{dx} + \sum\limits_{i\in[p,s]}\left[\dfrac{1 + \left(\gamma  - 1\right)\Ma_i^2}{\gamma \Ma_i^2}\right] \dfrac{F_i}{p}- \sum_{i\in[p,s]}\left[\dfrac{\gamma-1}{\gamma}\right]\dfrac{Q_i}{p u_i}\bigg)\,, \label{eq:p_compound}
\end{equation}
where $\beta$ is the compound choking indicator:
\begin{equation} \label{eq:beta}
    \beta = \sum_{i\in[p,s]} A_i\dfrac{1-\Ma_i^2}{\gamma \Ma_i^2}\,.
\end{equation}

This plays a similar role as $1 - \Ma^2$ in Equation~\eqref{eq:p_i}: if $\beta$ equals zero, the system of equations is singular, which corresponds to the sonic point of the parallel streams or the so-called `compound' flow. {Note that the definition \eqref{eq:beta} of $\beta$ is identical to the one in the original isentropic theory by \cite{pearson1958theory} and \cite{bernstein1967compound}. The non-isentropic effects are handled through the exchange terms $F_i$ and $Q_i$. Alternatively, these have been included in the definition of $\beta$ by \cite{debroeyer2025axisym}.} The reader is referred to \cite{vandenberghe2024extensioncompoundflowtheory} for a detailed analysis of the compound flow theory in a 1D formulation. \cite{hedges1974compressible} defined an equivalent Mach number $\Ma_{eq}$ related to $\beta$ as follows:
\begin{equation}
    \Ma_{eq} = \left(\gamma \frac{\beta}{A}+1\right)^{-\frac{1}{2}}\,. \label{eq:Ma_eq}
\end{equation}
The evolution of the dividing streamline is computed explicitly by inverting Equation~\eqref{eq:p_i} to obtain the equation for the individual cross-sections $A_i$:
\begin{equation}
    \frac{dA_i}{dx} = \left[A_i\frac{1-\Ma_i^2}{\gamma \Ma_i^2}\right]\frac{1}{p} \frac{dp}{dx} - \left[\frac{1 + \left(\gamma  - 1\right)\Ma_i^2}{\gamma \Ma_i^2}\right] \frac{F_i}{p} + \left[\dfrac{\gamma-1}{\gamma}\right]\dfrac{Q_i}{p u_i}\,,\label{eq:A_i}
\end{equation}
in which the pressure gradient is computed with Equation~\eqref{eq:p_compound}. The compound choking model therefore provides a direct means to compute the evolution of the dividing streamline under the assumption of equal static pressures between the streams. Moreover, the formulation naturally accommodates the exchange terms, which appear in the governing equations in a generic form. In particular, the pressure constraint remains satisfied whether the friction forces are prescribed from predictive correlations or derived from the total pressure gradients imposed by the RANS simulations.

The constraint of uniform static pressure addresses the eliminates the degree of freedom in \eqref{eq:p_i}-\eqref{eq:Tt_i} introduced by the dividing streamline. This leaves only the exchange terms undefined. The momentum exchange is embodied in this configuration by friction forces between the streams and between the secondary stream the wall as in \cite{vandenberghe2024extensioncompoundflowtheory}. The wall friction is modeled with a classic friction coefficient $f_w$:
\begin{equation}
    F_{w,i} = \frac{1}{2} f_w \gamma p_i \Ma_i^2 l_{w,i}\,, \label{eq:F_i_inlets}
\end{equation}
where $l_{w,i}$ denotes the perimeter of the wall(s) at a given position $x$. Note that the inner and the outer walls are considered for the secondary inlet. Wall friction is the only active force in the inlets ($x < 0$), so $F_i = -F_{w,i}$ in Equations~\eqref{eq:p_i} and \eqref{eq:pt_i}. It only acts on the secondary stream in the mixing pipe, so the local static pressure and Mach number of the secondary stream and the perimeter of the entire cross-section $A$ are used in the equation above. The friction coefficient is computed using the correlation of \cite{vandriest1951turbulent}:
\begin{equation}
    \dfrac{0.242}{\sqrt{f_w}} \sqrt{1 - \lambda^2} \dfrac{\arcsin (\lambda)}{\lambda} = 0.41 + \log_{10} \left(f_w \Rey_x\right) + \log_{10} \left((1 - \lambda^2)\left(1 - \dfrac{\theta \lambda^2}{1 + \theta}\right)\right)\,,\label{eq:f_w_correlation}
\end{equation}
where
\begin{align}
    1-\lambda^2 = \left(1 + \frac{\gamma - 1}{2} \Ma_s^2\right)^{-1}\,,
    \quad
    \Rey_x = \dfrac{\rho_s u_s x}{\mu_s}\,,
    \quad \text{and} \quad
    \theta = \dfrac{S}{T_s}\,, 
    \label{eq:f_w_correlation_2}
\end{align}
where $\rho$ denotes the density, $S = 110.4$ K and the law of \cite{sutherland1893lii} is used for the dynamic viscosity $\mu$, with a reference viscosity $\mu_{ref}$ of $1.716 \ 10^{-5}$ Pa s at $T_{ref} = 273.15$ K. The friction between the streams is defined following \cite{papamoschou1993model}:
\begin{equation} \label{eq:stream_friction}
    F_{ps} = \frac{1}{2} f_{ps} \dfrac{\rho_p+\rho_s}{2} \left(u_p - u_s\right) \left|u_p - u_s\right| l_{ps}\,,
\end{equation}
where $f_{ps}$ denotes a friction coefficient and $l_{ps}$ denotes the perimeter of the primary stream:
\begin{equation}
    l_{ps} = 2 \sqrt{\pi A_p}\,. \label{eq:l_ps}
\end{equation}
The inter-stream friction coefficient $f_{ps}$ is computed as in \cite{papamoschou1993model}:
\begin{equation}
    f_{ps} = 0.013 \dfrac{(1 + \zeta) (1 + \eta)}{1 + \zeta \eta} \left(0.25 + 0.75 \exp(-3 \Ma_c^2)\right)\,, \label{eq:f_ps_correlation}
\end{equation}
where $\zeta=u_s/u_p$, $\eta=\sqrt{\rho_s/\rho_p}$ and the local convective Mach number $\Ma_c$ is defined as follows:
\begin{equation}
    \Ma_c = \dfrac{u_p - u_s}{a_p + a_s}\,.
\end{equation}
The inter-stream friction is defined to be positive if $u_p > u_s$, in which case it counteracts the primary stream. The forces $F_p$ and $F_s$ are thus defined as follows in the mixing pipe:
\begin{equation}\label{eq:forces_mix}
    F_p = -F_{ps} \quad \text{and} \quad F_s = F_{ps} - F_{w,s}\,.
\end{equation}
The effect of these forces on the sonic point is discussed in detail by \cite{vandenberghe2024extensioncompoundflowtheory}.

\subsubsection{Model 2: Compound choking with closure from CFD}\label{sec:definition_compound_analysis}

In this configuration, the compound model from above is used without the friction forces defined in Equation~\eqref{eq:forces_mix}. Instead, the forces are prescribed from the reference RANS simulations (see section~\ref{sec:data}). The model therefore serves as a diagnostic rather than a predictive tool. Rather than computing the forces $F_i$ directly from local shear stresses within the RANS data, the total pressure gradients were extracted and imposed in the model, effectively replacing Equation~\eqref{eq:pt_i} and being incorporated into Equation~\eqref{eq:p_i} through the force term. These gradients were obtained by filtering and differentiating the axial distributions of the cross-stream averaged total pressures. The pressure distributions themselves were obtained using the post-processing tools described in section~\ref{sec:data} (see also \cite{vandenberghe2024extensioncompoundflowtheory}). Practical details of the filtering and numerical differentiation procedures are provided in section~\ref{sec:practical_filtering}.

\subsubsection{Model 3: Fabri choking with closures from CFD}\label{sec:definition_fabri_analysis}
In the Fabri choking framework, the two streams are not constrained to share the same static pressure, and the position of the dividing streamline must be prescribed externally. Consequently, Equation~\eqref{eq:p_i} is integrated independently in each stream, \(i \in [p,s]\). Choking occurs when the secondary stream reaches unit Mach number---the singular point of Equation~\eqref{eq:p_i}---while the primary stream, having expanded through its converging--diverging nozzle, is already supersonic. Beyond this point, no information can propagate upstream in either stream, marking the onset of the Fabri choking regime.

In this third model, the dividing streamline is taken from the axisymmetric RANS simulations described in section~\ref{sec:data} and imposed on the one-dimensional formulation. This configuration is therefore non-predictive and serves instead to analyze the choking mechanism and assess the impact of the simplifying assumptions inherent to one-dimensional models. In particular, it isolates the effect of cross-stream averaging while avoiding any assumptions on the pressure distribution, allowing a direct comparison with the compound-choking model that enforces equal pressures. The implications for choking behavior are discussed in section~\ref{sec:results}.

On the other hand, the Fabri-based models proved highly sensitive to the pressure evolution. Small deviations in the friction forces from the RANS data disrupt the balance in Equation~\eqref{eq:p_i}, especially near the sonic point, where the equation becomes stiff. Consequently, even minor inaccuracies in the closure terms can produce large discrepancies in the pressure distribution or cause numerical failure. This sensitivity stems from the coupled dependence of Equation~\eqref{eq:p_i} on the dividing streamline and the force term, through \(dA_i/dx\) and \(F_i\). As the secondary flow approaches Mach~1, the denominator of Equation~\eqref{eq:p_i} tends to zero, and maintaining a finite pressure gradient requires the combined effect of the area gradient and the force term to vanish simultaneously. Hence, a consistent relation between the force closure and the dividing streamline is essential for numerical stability.

This limitation does not occur in compound-based models, where the dividing streamline evolves under the equal-pressure constraint. Since Equation~\eqref{eq:p_compound} already includes the closure terms, the streamline naturally adapts to them, and the area and force gradients act coherently. In contrast, the Fabri formulation retains an additional degree of freedom, which can lead to instability when the two effects are misaligned near the sonic point. Practically, this means that the imposed forces must be consistent with those determining the dividing streamline, as obtained from the RANS simulations in this model configuration.

\subsubsection{Model 4: Fabri choking with closure from compound choking theory }\label{sec:definition_fabri_rdiv_compound}

In this model, the predictive compound model from section~\ref{sec:definition_compound_predictive} serves as the external source for the dividing streamline, while keeping the Fabri-sonic condition as the criterion for choking. The sensitivity discussed in the previous section remains, so the correlations for the friction forces are used in this configuration to be consistent with the dividing streamline model. This approach implies that the governing equations are identical to those of the predictive compound model. The only remaining difference lies in the sonic condition, which determines the solution for the choked flow (see section~\ref{sec:solution}). Consequently, the corresponding analysis in section~\ref{sec:results_fabri_rdiv_compound} is a direct comparison of the choking mechanisms on a common basis, albeit on an artificial dividing streamline to comply with the constraint of uniform pressure.
\section{Practical implementation}\label{sec:practical}

\subsection{Treatment of the inlets}\label{sec:practical_inlets}

For all the models discussed in the previous section, the friction coefficients in the inlets are taken using the empirical correlations in \eqref{eq:F_i_inlets}-\eqref{eq:f_w_correlation_2}.

The boundary conditions specify the total pressure \(p_{t,i}\) and total temperature \(T_{t,i}\) at each inlet. The governing Equations~\eqref{eq:p_i}--\eqref{eq:Tt_i} form an explicit set of ODEs that are integrated axially from the inlets, where the total quantities are known. A shooting method is used to adjust the inlet static pressure until the target sonic condition is met (see section~\ref{sec:solution}). 

The same formulation applies to both streams. However, since the primary jet is choked in its converging–diverging nozzle, it can be solved independently. The secondary flow, on the other hand, depends on the back pressure \(p_b\) at the outlet of the diffuser and must therefore be solved in conjunction with the mixing section, as discussed below.

\subsection{Jump relations at the inlet of the mixing pipe}\label{sec:practical_jump}
\begin{figure}%[!htb]
	\centering
	\includegraphics[width=0.45\linewidth]{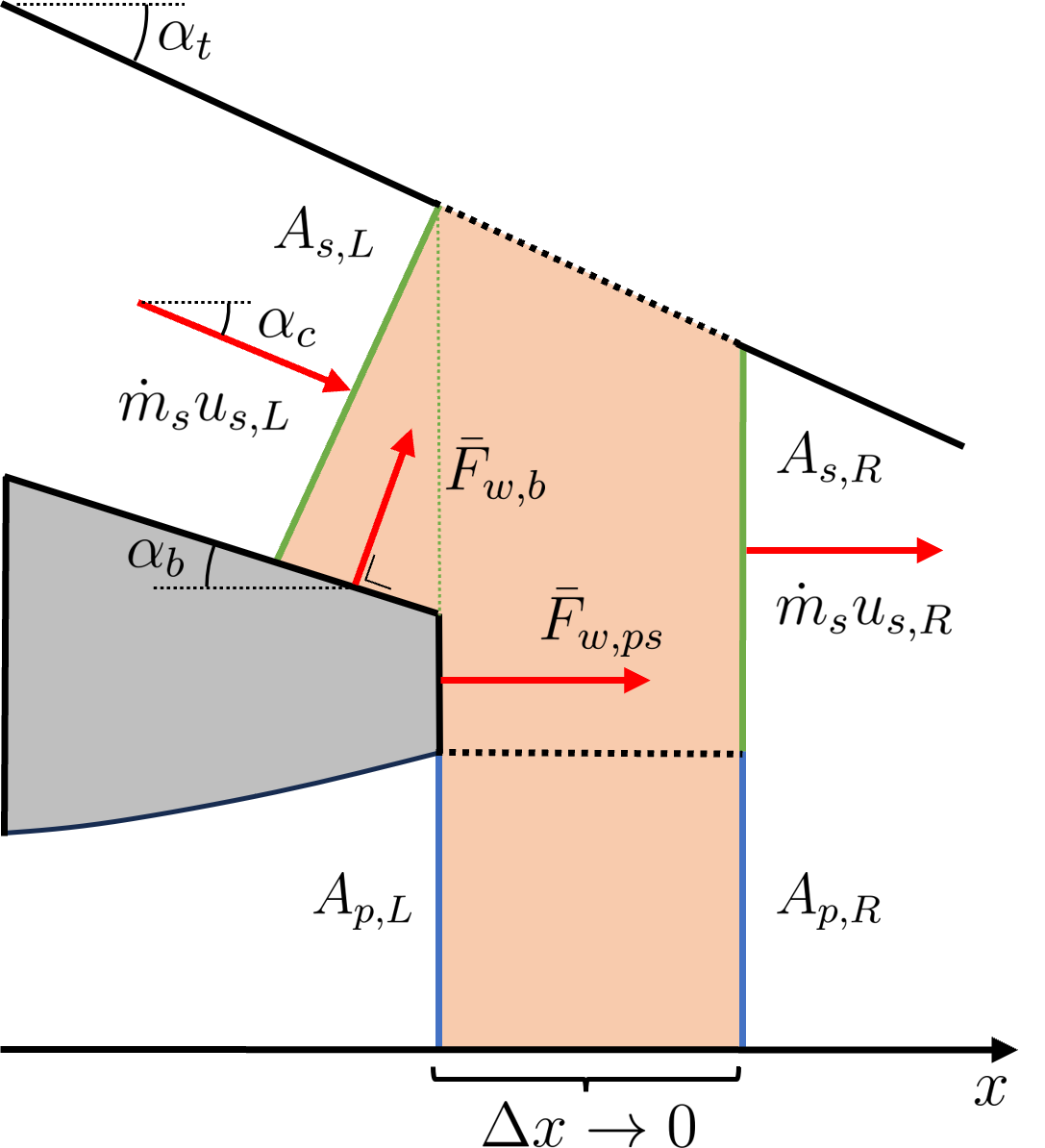}
    \captionsetup{width=\textwidth, justification=justified}
	\caption{The jump conditions from the inlets to the mixing pipe. The primary cross-sections and hence the flow states are identical on both sides. The secondary stream requires a force balance, where the momentum fluxes and pressure forces are projected on the x-axis with their respective angles. The result is a double flow state at the inlet of the mixing pipe $x=0$, which generally has different static pressures.}
	\label{fig:inlets_mixing_pipe}
\end{figure}

The full flow states are known at the outlets of the separate domains ($x=0$) after integrating the respective ODEs in the inlets. The dividing streamline originates at the bottom of the primary nozzle lip, so the primary cross-sections and variables match between the inlet, denoted with the subscript $L$, and in the mixing pipe, with subscript $R$ (see figure~\ref{fig:inlets_mixing_pipe}). This is not the case for the secondary stream due to its inclined inlet and a potential finite thickness $A_w$ of the nozzle. The secondary flow state on the right is found through conservation of mass, momentum and energy. The control volume is infinitesimally thin ($\Delta x \rightarrow 0$ in figure~\ref{fig:inlets_mixing_pipe}). Since $A_{s,R}+A_{p,R}=A(x=0)$, three unknowns remain (e.g., density, velocity and temperature). Two forces intervene in the axial momentum balance: the force on the bottom wall $\bar{F}_{w,b}$ and the force on the nozzle lip $\bar{F}_{w,ps}$. It is assumed that friction acting on these surfaces is negligible compared to the pressure force, which has been confirmed during the post-processing of the RANS simulations in this work. These forces are therefore orthogonal to the surfaces on which they act. The inclinations of the bottom and top wall are generally different and are denoted here with $\alpha_b>0$ and $\alpha_t>0$. The boundary of the secondary inlet is defined as the line originating from the top wall at the exit of the primary nozzle ($x=0$), with a central angle $\alpha_c = 0.5(\alpha_b + \alpha_t)$. Projecting the momentum fluxes and the forces on the x-axis with their respective angles leads to the following balance:
\begin{equation}\label{eq:momentum_jump}
    \left(\dot{m}_s u_{s,L} + p_{s,L} A_{s,L}\right) \cos{\alpha_c} + F_{w,b} \sin{\alpha_b} + F_{w,ps} = \dot{m}_s u_{s,R} + p_{s,R} A_{s,R}\,,
\end{equation}
where $u$ denotes the average velocity as described in the 1D equations. The magnitude of the forces $F_{w,b}$ and $F_{w,ps}$ need to be estimated from the averaged quantities of the left and right states. The expansion of the secondary stream from the inclined section on the left to $x=0$ is generally small, hence the force on the bottom wall is approximated as $F_{w,b} = p_{s,L} A_{w,b}$, where the area can be computed from the geometry. Similarly, the force on the nozzle lip is defined as $F_{w,ps} = p_{s,R} A_{w,ps}$, where the (unknown) static pressure on the right side is applied since it is in direct contact with this wall.

Conservation of energy implies that the total temperature is conserved, since there is no source term and the walls are assumed to be adiabatic. The remaining unknowns in Equation~\eqref{eq:momentum_jump} are the velocity $u_{s,R}$ and the static pressure $p_{s,R}$. These are linked through the known mass flow rate $\dot{m}_s$, the known total temperature $T_{t,s}$ and the ideal gas law:
\begin{equation}
    \dot{m}_s = \frac{p_{s,R}}{R T_{s,R}} A_{s,R} u_{s,R} \,,
    \quad \text{and} \quad
    T_{s,R} = T_{t,s,R} - \frac{u_{s,R}^2}{2c_P}\,.
\end{equation}

The two equations above allow to eliminate $p_{s,R}$ from the momentum balance \eqref{eq:momentum_jump}, leading to a non-linear equation in $u_{s,R}$, which can be solved iteratively. The velocity $u_{s,L}$ from the secondary inlet is a good initial guess. The secondary flow state can then be computed from $\dot{m}_s$, $T_{t,s,R}$ and $u_{s,R}$.

The resulting right primary and secondary states generally have a different static pressure because the primary stream is over- or under-expanded. A uniform static pressure $p(x)=p_p(x)=p_s(x)$ is required for the compound model (see section~\ref{sec:definition_compound_predictive}), so first a pressure equalization mechanism is defined below. This is not required when using the Fabri-based model, where the static pressures are left free in each stream.

\subsection{Pressure equalization mechanism for the compound theory}\label{sec:practical_pressure}

\begin{figure}%[!htb]
  \centering
  \begin{subfigure}[t]{0.39\linewidth}
    \center
    \includegraphics[width=\linewidth]{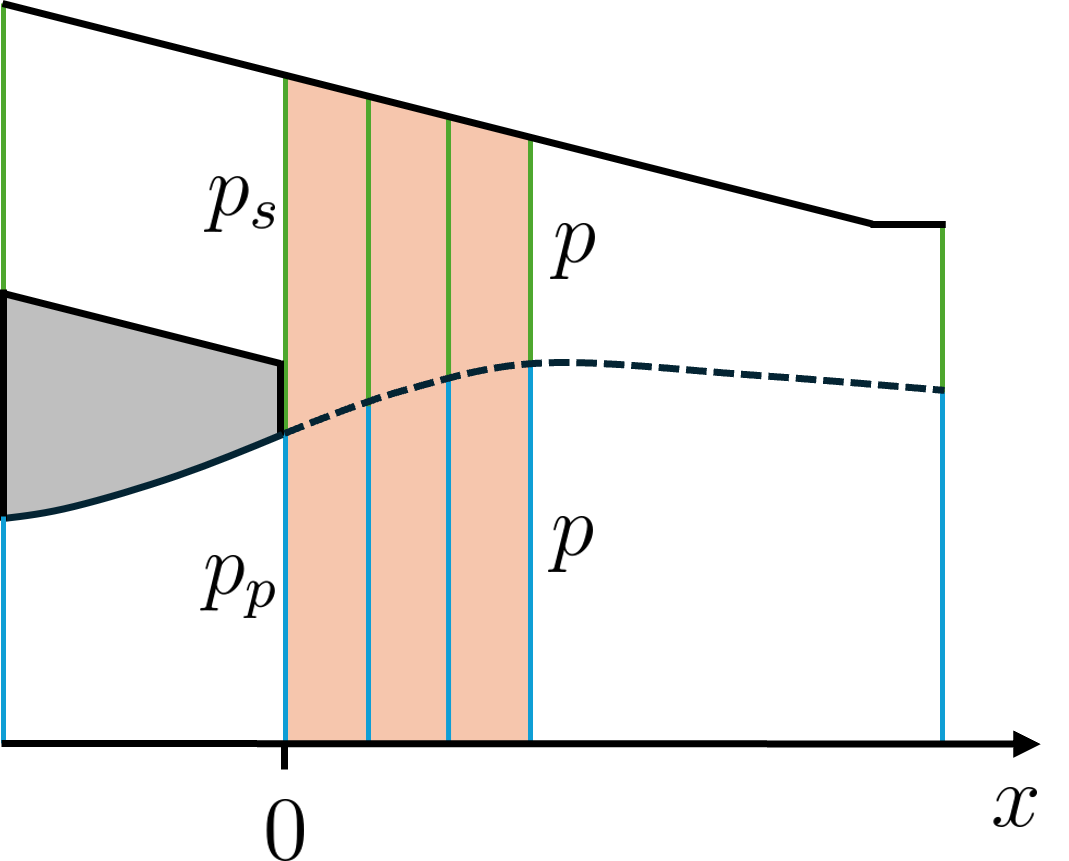}
    \subcaption{Overview.}
    \label{fig:pressure_equalization_goal}
  \end{subfigure}
  \begin{subfigure}[t]{0.29\linewidth}
    \includegraphics[width=\linewidth]{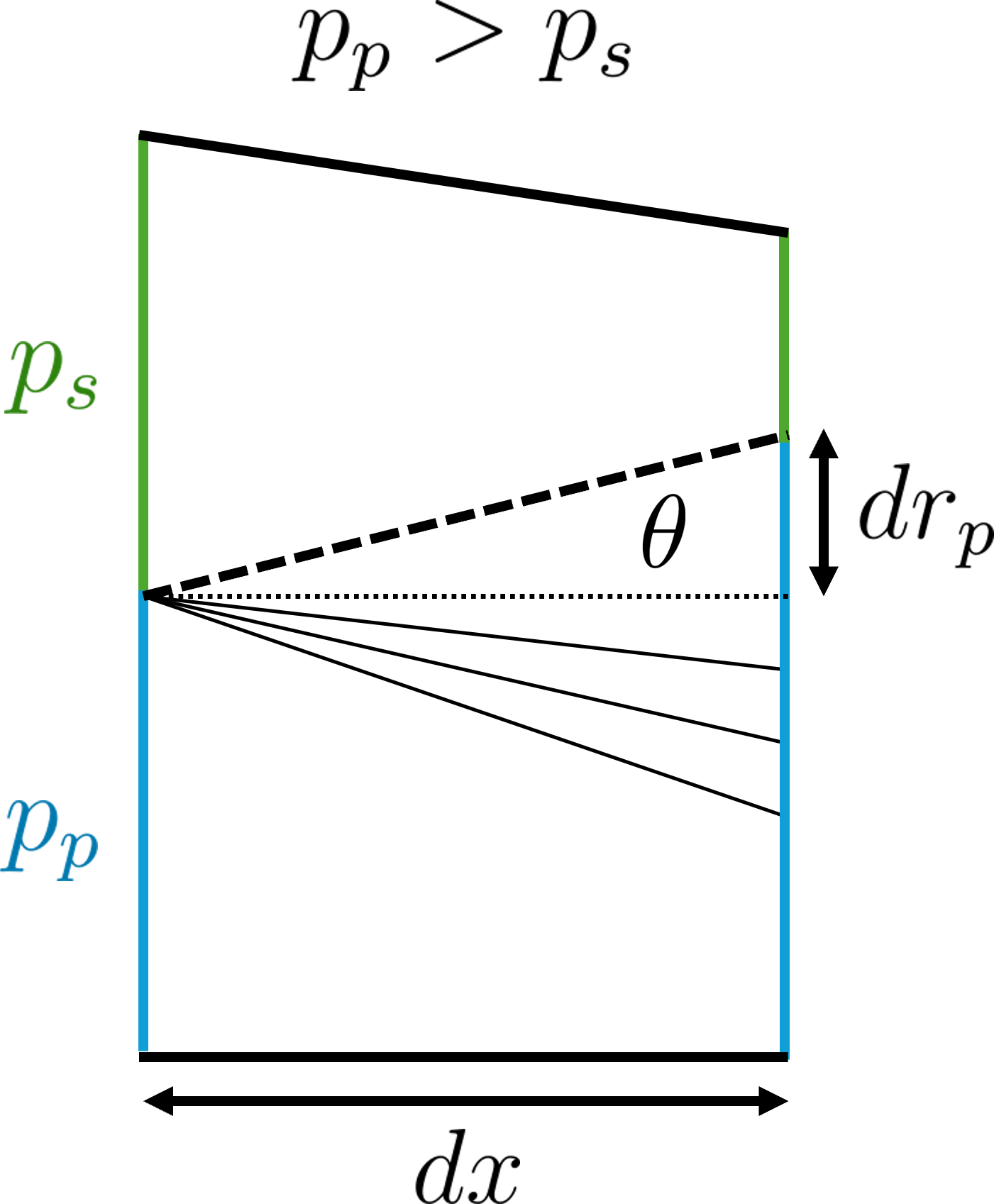}
    \subcaption{Expansion}
    \label{fig:expansion}
  \end{subfigure}
  \begin{subfigure}[t]{0.29\linewidth}
    \includegraphics[width=\linewidth]{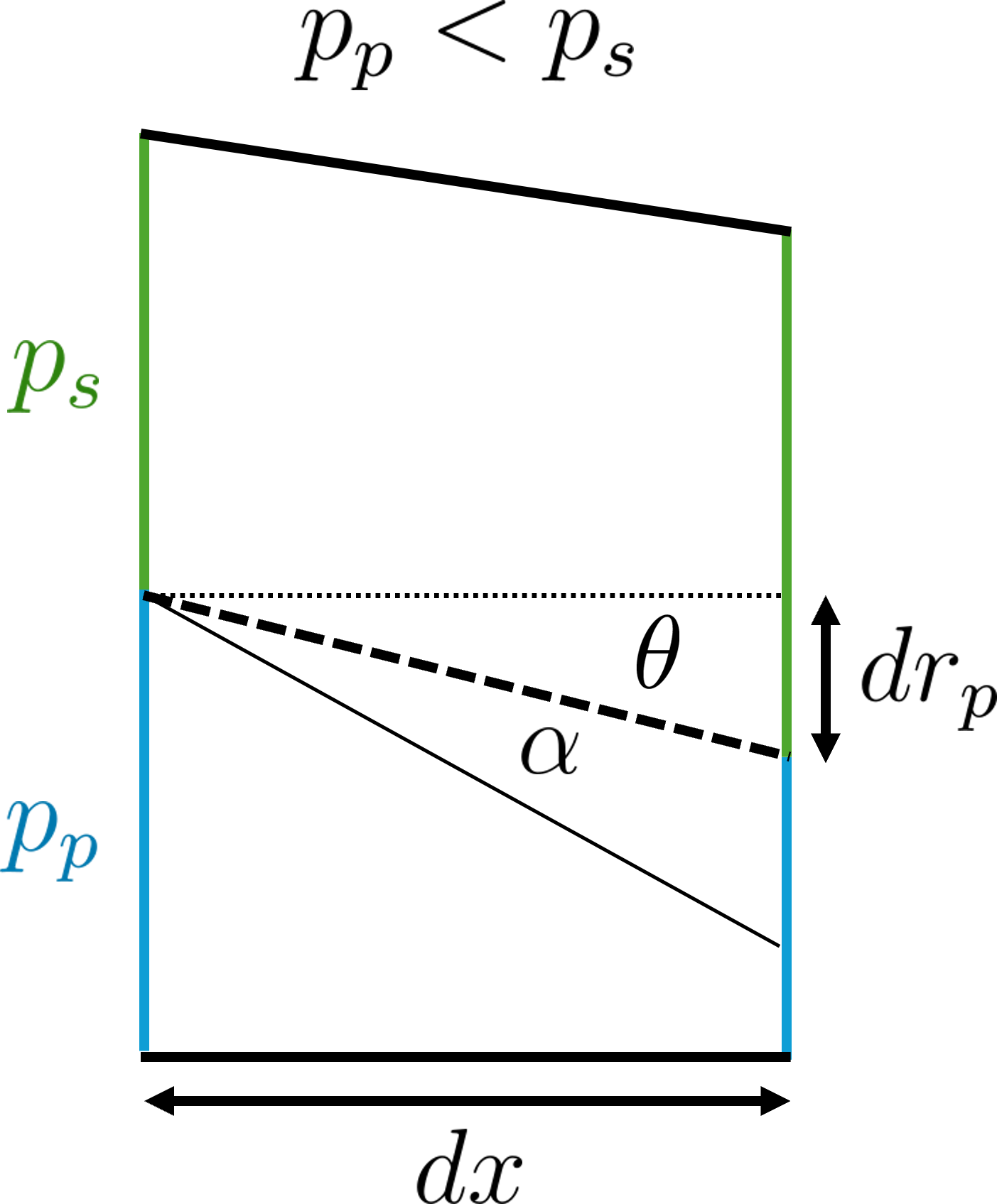}
    \subcaption{Oblique shock}
    \label{fig:oblique_shock}
  \end{subfigure}
  \captionsetup{width=\textwidth, justification=justified}
  \caption{An under- or over-expanded primary stream is brought to a uniform static pressure over a finite distance using Prandtl-Meyer expansion theory or oblique shocks respectively. The procedure results in the angle $\theta$ of the dividing streamline, which suffices to complement the conservation Equations~\eqref{eq:p_i}-\eqref{eq:Tt_i}.}
\label{fig:pressure_equalization_mechanism}
\end{figure}

At the inlet of the mixing pipe, the primary stream is typically either under- or over-expanded relative to the surrounding secondary stream, as evidenced by the shock-train structures observed in numerical studies~\cite{BARTOSIEWICZ200556, HEMIDI2009_part1}. This behavior directly conflicts with the core assumption of equal static pressures in the compound flow theory and thus with model 1 (section~\ref{sec:definition_compound_predictive}) and model 2 (section~\ref{sec:definition_compound_analysis}). To address this limitation, the present work introduces a \emph{pressure equalization mechanism} that brings the two streams to a uniform static pressure over a finite distance downstream of the primary nozzle outlet, as illustrated schematically in figure~\ref{fig:pressure_equalization_goal}.

The proposed mechanism is inspired by the first cell of the shock train and reproduces the local pressure-adjustment process up to the point where the average static pressures of the two streams first become equal. The analysis is local in nature and provides the orientation of the dividing streamline as a function of the local static-pressure ratio, as shown in figure~\ref{fig:pressure_equalization_mechanism}. Separate formulations are adopted for under-expanded and over-expanded primary streams, though the underlying principle remains the same.

For an under-expanded primary stream, further expansion is modeled through a diverging cross-section computed using Prandtl--Meyer expansion theory, following the approach of~\cite{delvalle2012}. The process is assumed isentropic, ensuring conservation of total pressure from the current state $(p_p, \Ma_p)$ to a virtual state $(p_s, \widetilde{\Ma}_p)$ characterized by the same static pressure as the secondary stream:
\begin{equation}
    p_{t,p} = p_s \left(1 + \frac{\gamma - 1}{2}\,\widetilde{\Ma}_p^{\,2} \right)^{\frac{\gamma}{\gamma - 1}},
\end{equation}
from which the expanded Mach number $\widetilde{\Ma}_p$ is determined. The deviation angle $\theta$ is then defined as the difference between the Prandtl--Meyer angles $\nu$ and $\tilde{\nu}$ corresponding to the local and virtual primary Mach numbers $\Ma_p$ and $\widetilde{\Ma}_p$, respectively. For completeness, the Prandtl--Meyer angle is given by~\cite{shapiro1953dynamics}
\begin{equation}
    \nu(\Ma) \;=\; \sqrt{\frac{\gamma + 1}{\gamma - 1}}\,
    \arctan\!\left(\sqrt{\frac{\gamma - 1}{\gamma + 1}\,\big(\Ma^{2}-1\big)}\right)
    \;-\;
    \arctan\!\left(\sqrt{\Ma^{2}-1}\right).
\end{equation}

If the primary and secondary static pressures were equal, the dividing streamline would be tangent to the wall at the outlet of the primary nozzle (without sudden change of angle). The computed deviation angle $\tilde{\nu}-\nu$ should thus be added to the angle $\theta_p$ of the primary nozzle at its outlet to obtain the angle of the dividing streamline:
\begin{equation}
    \theta = \theta_p + \left(\tilde{\nu}-\nu\right)\,.
\end{equation}
An oblique shock is computed for the pressure ratio $p_s/p_p$ and a primary Mach number $\Ma_p$ in case of an over-expanded primary stream. The shock angle $\alpha$ can be computed from the following equation (\cite{shapiro1953dynamics}):
\begin{equation}
    \frac{p_s}{p_p} = 1 + \frac{2\gamma}{\gamma+1} \left( \Ma_p^2 \sin^2(\alpha) - 1 \right)\,.
\end{equation}
The deviation angle $\theta$ is directly related to the shock angle $\alpha$ and the Mach number $\Ma_p$:
\begin{equation}\label{eq:p-eq_theta_shock}
    \tan(\theta) = -\frac{2 \cot(\alpha) \left( \Ma_p^2 \sin^2(\alpha) - 1 \right)}{\Ma_p^2 \left( \gamma + \cos(2\alpha) \right) + 2}\,.
\end{equation}

Note the minus sign for the inward deviation of the primary flow. In the axisymmetric geometry, the axial derivative of the primary cross-section can be decomposed in terms of the deviation angle $\theta$:
\begin{equation}
    \frac{dA_p}{dx} = 2 \pi r_p \frac{dr_p}{dx} = 2 \sqrt{\pi A_p} \tan(\theta)\,.
\end{equation}

The axial derivative of the secondary cross-section follows from the geometry $A(x)$ of the mixing pipe. These local gradients of the cross-sections allow to integrate the governing Equations~\eqref{eq:p_i}-\eqref{eq:Tt_i} (which remain valid in the mixing pipe) to the next grid point, using the closure relations \eqref{eq:forces_mix}. The overall process is not isentropic due to the friction forces; the isentropic pressure equalization mechanism is used only to determine the local angle of the dividing streamline. The procedure above is repeated until the static pressure difference drops below an arbitrary threshold (10 Pa in this work).

This approach reproduces only the first cell of the shock train and cannot capture downstream behavior, as reflections of Mach lines on the centerline and dividing streamline are inherently two-dimensional and thus excluded from the present one-dimensional model. Nevertheless, the proposed pressure equalization mechanism offers two main advantages: (1) its local formulation makes it directly compatible with standard ODE integrators, and (2) the pressure equalization occurs over a short distance, enabling the compound flow theory to be applied as far upstream as possible to accurately locate the sonic point.

\subsection{Normal compound shocks}\label{sec:practical_shock}
An oblique shock train typically forms in the diffuser under on-design conditions, decelerating the flow to subsonic speed and reestablishing the back pressure imposed by the boundary conditions~\citep{BARTOSIEWICZ200556, HEMIDI2009_part1}. In simplified models, this system is often represented as a normal shock in 0D formulations~\citep{huang1999, chen2013, METSUE2021121856}, and it also emerges naturally in the unsteady single-stream model of~\cite{vandenberghe2023unsteady}.

In the compound-flow framework of model 1 and model 2, however, a normal-shock representation cannot be directly applied because the compound supersonic region may include a subsonic component, rendering standard shock relations invalid. Furthermore, shocks of unequal strength would develop in the two streams due to their differing velocities, producing a downstream pressure imbalance and violating the assumption of uniform static pressure. The inherently two-dimensional nature of the oblique shock train and its interaction with boundary layers further preclude an accurate one-dimensional description under the constraint of pressure uniformity.

Since the 1D formulation cannot resolve the flow structure downstream of the shock, the mixing between the two streams also becomes undefined. To apply normal-shock relations in this context, it is therefore necessary to define an \emph{equivalent single-stream state} that represents the combined effect of the primary and secondary streams at the shock location. This equivalent, or “fully mixed,” state is a notional construct used solely for the computation of the normal shock and is defined to have the same total mass flow rate, momentum, and energy as the actual double-stream configuration. The normal-shock relations are then applied between this equivalent upstream state, denoted by the subscript $L$, and the downstream state, denoted by $R$. The properties of the equivalent state upstream of the shock are obtained from the conservation of mass and energy as
\begin{equation}
\dot{m}_L = \dot{m}_p + \dot{m}s,
\quad \text{and} \quad
T_{t,L} = \frac{\dot{m}_p T_{t,p} + \dot{m}_s T_{t,s}}{\dot{m}_p + \dot{m}_s}.
\label{eq:shock_Tt}
\end{equation}

The mass flow rate and the momentum flux can be defined in terms of the static pressure $p$, the total temperature $T_t$, the cross-section $A$ and the Mach number $\Ma$:
\begin{equation}
    \dot{m} = p A \Ma \sqrt{\frac{\gamma}{R T_t}}\sqrt{1+\frac{\gamma-1}{2}\Ma^2}\,,
    \quad \text{and} \quad
    \dot{m} u + p A = pA \left(1 + \gamma \Ma^2\right)\,.
\end{equation}
The static pressure and the cross-section can be eliminated by taking the ratio of the momentum flux and the mass flow rate, which becomes a function of the Mach number and the total temperature:
\begin{equation}
    \frac{\dot{m} u + p A}{\dot{m}} = \frac{1 + \gamma \Ma^2}{\Ma \sqrt{1+\frac{\gamma-1}{2}\Ma^2}} \sqrt{\frac{R T_t}{\gamma}}\,.
\end{equation}
Hence, conservation of momentum between the double state and the equivalent state $L$ implies:
\begin{equation}
    \frac{\sum_{i \in [p,s]}\left(\dot{m}_i u_i + p_i A_i\right)}{\sum_{i \in [p,s]}\dot{m}_i} = \left[\frac{1 + \gamma \Ma_L^2}{\Ma_L \sqrt{1+\frac{\gamma-1}{2}\Ma_L^2}}\right] \sqrt{\frac{R T_{t,L}}{\gamma}}\,, \label{eq:shock_momentum}
\end{equation}
where the total temperature $T_{t,L}$ is known from the Equation~\eqref{eq:shock_Tt}. The term between the brackets is nonlinear in $\Ma_L$ and admits a subsonic and a supersonic solution for a given ratio of momentum to mass flow rate. The minimum corresponds to a sonic flow, indicating a minimal momentum for a given mass flow rate (or a maximal mass flow rate for a given amount of momentum). The supersonic solution is retained, followed by the normal shock relations to compute the right state $R$ (see for example \cite{shapiro1953dynamics}): 
\begin{equation}\label{eq:normshock}
    M_R^2 = \frac{( \gamma - 1 ) M_L^2 + 2}{2\gamma M_L^2 - ( \gamma - 1 )}\,,
    \quad
    p_R = p_L \left(1 + \frac{2 \gamma}{\gamma + 1} \left( M_L^2 - 1 \right)\right)\,,
    \quad \text{and} \quad
    T_{t,R} = T_{t,L}\,.
\end{equation}

The flow downstream is computed as an isentropic single stream, which has an analytical solution considering that the total pressure and total temperature are conserved and that the mass flow rate is known. The Mach number $\Ma_m$ of the single stream respects the following (non-linear) equation in any point downstream of the shock:
\begin{equation}
    \dot{m}_L = p_{t,R} A(x) \sqrt{\frac{\gamma}{R T_{t,R}}} \Ma_m(x) \left[1 + \frac{\gamma - 1}{2} \Ma_m^2(x)\right]^{-\frac{1}{2}\frac{\gamma+1}{\gamma-1}}\,.
\end{equation}

\subsection{Filtering, differentiation and interpolation}\label{sec:practical_filtering}

Imposing cross-section or total-pressure gradients from the RANS simulations requires computing and interpolating these quantities so they can be evaluated by the solver at any axial position \(x\). The primary and secondary cross-sections and total pressures are obtained directly from the geometry, the dividing streamline, and the flow field (see section~\ref{sec:data}). However, their derivatives in Equation~\eqref{eq:p_i}, obtained by numerical differentiation, are sensitive to noise due to interpolation and cross-stream averaging.

The distributions of the total pressures and the primary cross-section are therefore first filtered before applying finite differences with Numpy's \texttt{gradient} function. The cross-section of the secondary stream and its derivatives follow from the known geometry and the primary stream. The filter used in this work is based on the filter of \cite{chambolle2004algorithm} available in the \texttt{denoise\_tv\_chambolle} function from Scikit-image (\cite{van2014scikit}). The total variation denoising formulation avoids spurious overshoot near sharp transitions, but it is subject to boundary effects, which are problematic in the current work because the dividing streamline originates from the lip of the primary nozzle. Therefore, the signal used in the model is a combination of the original and the filtered signal :
\begin{equation}
    \tilde{u}[j] = w[j] u[j] + \hat{w}[j] \hat{u}[j]\,,
\end{equation}
where $\tilde{u}[j]$ denotes the $j^\text{th}$ entry of the final discrete signal, composed of the original and the filtered signals $u$ and $\hat{u}$. These are blended via a partition of unity that balances the unfiltered and the filtered signal such that the sum of the weights equals one at each point:
\begin{equation}
    w[j] = 
    \begin{cases}
        0.5 + 0.5 \cos\left(\pi \frac{i}{60}\right) &\text{ if } j \leq 60\,,\\
        0.5 - 0.5 \cos\left(\pi \frac{i - (N-61)}{60}\right) &\text{ if } j > N-60\,,\\
        0 &\text{ otherwise.}
    \end{cases}
\end{equation}
and
\begin{equation}
    \hat{w}[j] = 
    \begin{cases}
        0.5 - 0.5 \cos\left(\pi \frac{i}{60}\right) &\text{ if } j \leq 60\,,\\
        0.5 + 0.5 \cos\left(\pi \frac{i - (N-61)}{60}\right) &\text{ if } j > N-60\,,\\
        1 &\text{ otherwise.}
    \end{cases}
\end{equation}
The original signal is thus kept near the boundaries, transitioning over 60 entries to the filtered signal which is kept elsewhere. The width of 60 entries has been chosen arbitrarily as a compromise between filter intrusiveness near the boundary and noise tolerance.

Finally, since the full solver requires the evaluation of the governing Equations~\eqref{eq:p_i}-\eqref{eq:Tt_i} at any axial position $x$, an interpolation within the filtered grid was applied. This is a cubic interpolation using Scipy's `interp1d' function (\cite{2020SciPy-NMeth}).

\section{Numerical solution} \label{sec:solution}

The governing equations presented in the section above yields a system of ODEs in space, integrated with a shooting method to match the boundary condition at the outlet. The integration was carried out with the explicit Runge-Kutta 4 scheme using the \texttt{SciPy.integrate} library. The choked flow in the primary nozzle is computed first, since this is independent from the flow in the rest of the ejector. The flows in the secondary inlet and in the mixing pipe are coupled and solved afterwards.

\subsubsection*{Primary inlet}
\begin{figure}%[htb!]
	\centering
	\includegraphics[width=0.6\linewidth]{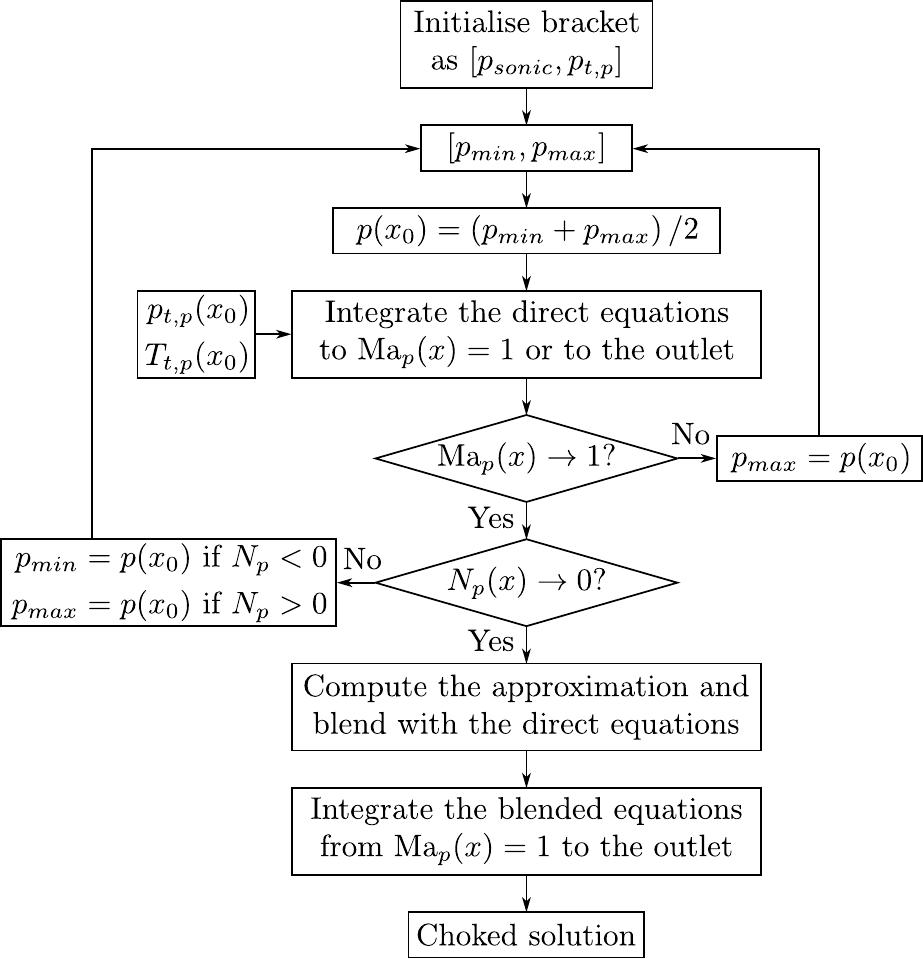}
    \captionsetup{width=\textwidth, justification=justified}
	\caption{Flow chart for computing the choked flow in the primary nozzle. The static pressure at the inlet is found iteratively to satisfy the sonic condition $\Ma_p \rightarrow 1, N_p \rightarrow 0$ in Equation~\eqref{eq:p_i_denom}. The approximations in Appendix~\ref{sec:appendix_single_stream} allow to integrate the governing equations through the sonic point without numerical issues related to division by zero (see also \cite{restrepo2022viscous, vandenberghe2024extensioncompoundflowtheory}).}
	\label{fig:flowchart_primary}
\end{figure}

Given the inlet totals \((p_{t,p},T_{t,p})\), the primary mass flow is obtained by a \emph{bracketed shooting} on the inlet static pressure \(p_p(x_0)\) to enforce \(M_p=1\) (and \(N_p=0\) in Equation~\eqref{eq:N_i}). By definition the upper bound is \(p_p(x_0)\le p_{t,p}\) while the lower bound is the isentropic sonic pressure
\begin{equation}
    p_{\text{sonic}} = p_{t,p}\left(\frac{\gamma+1}{2}\right)^{-\gamma/(\gamma-1)},
\end{equation}
since the flow is subsonic in the converging section. The search therefore brackets \(p_p(x_0)\in[p_{\text{sonic}},\,p_{t,p}]\) and proceeds by bisection: at each iteration a trial \(p_p^{(k)}\) (the bracket midpoint) is used to integrate \eqref{eq:p_i}--\eqref{eq:Tt_i} from \(x_0\) to either the point where \(M_p=1\) or the nozzle exit. Reaching the exit while still subsonic indicates \(p_p^{(k)}\) is too high and hence the upper bound is set to \(p_p^{(k)}\). Otherwise the bracket is tightened around the \(M_p=1\) solution. This monotone update exploits the fact that decreasing \(p_p(x_0)\) increases the mass flow toward choking (see the flowchart in figure~\ref{fig:flowchart_primary}).

As the sonic condition is approached, the denominator in ~\eqref{eq:p_i} tends to zero. To obtain a finite ratio and allow the flow to continuously expand to supersonic conditions, also the numerator must tend to zero (see \cite{shapiro1953dynamics}). Isolating the common denominator and defining the numerator $N_p$:
\begin{equation}
    \frac{1}{p_p} \frac{dp_p}{dx} = \frac{N_p}{1-\Ma_p^2}\,,\label{eq:p_i_denom}
\end{equation}
where
\begin{equation}
    N_p = \gamma \Ma_p^2\frac{1}{A_p}\frac{d A_p}{dx} + \left(1 + \left(\gamma  - 1\right)\Ma_p^2\right) \frac{F_p}{A_p p_p}\,,\label{eq:N_i}
\end{equation} a finite pressure gradient is obtained if:

\begin{equation}\label{eq:hopital_main}
\lim_{(p,\Ma_p)\to (p^\ast,1)} \frac{1}{p_p} \frac{dp_p}{dx}= \frac{dN_p/dx}{-d\Ma_p^2/dx}
\end{equation} following de l'Hôpital's rule. The derivative of the Mach number can be analytically evaluated from the conservation equation (see Equation~\eqref{eq:dMa2_dx}), which introduces the derivative of the static pressure in the denominator. This leads to the quadratic Equation~\eqref{eq:quadratic_dpdx} for the static pressure gradient in the sonic point, which is derived in Appendix~\ref{sec:appendix_single_stream} (see also \cite{shapiro1953dynamics, vandenberghe2024extensioncompoundflowtheory}). The static pressure gradient can thus be calculated in all admissible conditions.

If the procedure in figure~\ref{fig:flowchart_primary} runs into the sonic point, the numerator $N_p$ is computed with the Equation~\eqref{eq:N_i}. If its absolute value drops below an arbitrary threshold ($10^{-6}$ in this work), the correct sonic condition is found and the calculation proceeds. Otherwise, the bracket is adjusted. A negative value of the numerator indicates that the flow would expand if it is subsonic ($dp/dx < 0$ in \eqref{eq:p_i_denom}). This occurs for example in the convergent section for an isentropic flow (see \eqref{eq:N_i} with $F=0$). The fact that the flow already reached $\Ma_p=1$ indicates that the flow rate is too high: the Mach number should be lower, allowing the flow to expand further without running into the singularity of Equation~\eqref{eq:p_i_denom}. In the isentropic example, the sonic point should be reached further downstream at the throat. The guessed static pressure is thus too low if $N_p(x) < 0$ and $M_p(x)=1$. Therefore, the used average pressure $p(x_0)$ becomes the new lower bound $p_{min}$. The inverse is true if $N_p>0$. The friction force $F_p<0$ interferes in the numerator $N_p$, but it does not alter the mechanism discussed above. Note that $dA/dx$ must be positive to compensate the negative term containing the force $F_p$ in Equation~\eqref{eq:N_i}: the wall friction force pushes the sonic section downstream in a divergent section of the nozzle. The algorithm iterates until the tolerance on the numerator is reached or until the bracket reaches the following relative tolerance:
\begin{equation}
    \frac{p_{max}-p_{min}}{0.5 (p_{min} + p_{max})} \leq 10^{-6}\,.
\end{equation}

The procedure above provides the solution up to the sonic point. At this point, the static pressure gradient is undetermined as \eqref{eq:p_i} approaches the limit $0/0$, so a second order approximation of the pressure gradient near the sonic point is computed using Taylor expansions (see Equation~\eqref{eq:single_second_order} in Appendix~\ref{sec:appendix_single_stream}) if $M_p \leq 1.01$. Once the primary stream expands beyond $\Ma_p = 1.01$, the pressure is computed from Equation~\eqref{eq:p_i_denom}. The threshold of $1.01$ was found to provide a good compromise between accuracy and smoothness of the transition from the approximation \eqref{eq:single_second_order} to the governing Equation~\eqref{eq:p_i_denom}. In practice, the approximation \eqref{eq:single_second_order} is blended with Equation~\eqref{eq:p_i_denom} in a single function that automatically switches between both depending on the primary Mach number $\Ma_p$, as indicated in figure~\ref{fig:flowchart_primary}. The interested reader is referred to \cite{vandenberghe2024extensioncompoundflowtheory} for a more detailed description of this approach.

\subsubsection*{On-design operation}
\begin{figure}%[htb!]
    \centering
    \includegraphics[width=0.7\linewidth]{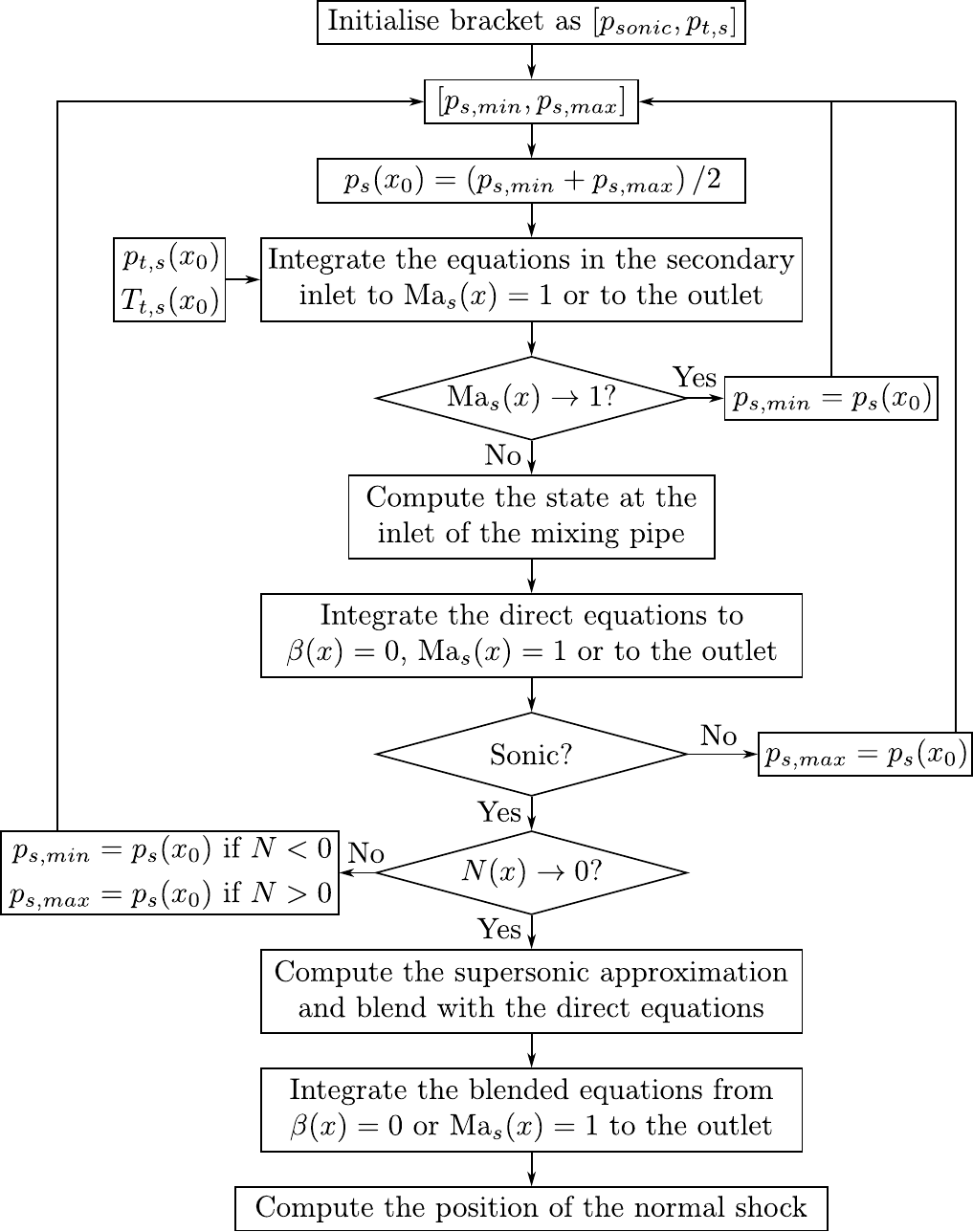}
    \captionsetup{width=\textwidth, justification=justified}
    \caption{The solution procedure iterates on the static pressure at the secondary inlet to find the correct compound- or Fabri-sonic point. A sonic flow in the secondary inlet is not accepted as a valid solution since it is atypical for ejectors. The procedure differs in the compound and Fabri case through the sonic condition and the corresponding approximations of the pressure gradient.}
    \label{fig:flowchart_critical}
\end{figure}

The procedure for the rest of the ejector is similar to the one of the primary nozzle, as shown in figure~\ref{fig:flowchart_critical}: the static pressure is guessed at the secondary inlet, the governing equations are integrated to the compound- or Fabri-sonic point ($\beta=0$ in \eqref{eq:p_compound} or $\Ma_s=1$ in \eqref{eq:p_i}) and the solution is accepted if the numerator of Equation~\eqref{eq:p_compound} or \eqref{eq:p_i} equals zero.

An approximation of the static pressure gradient is then computed near the sonic point similarly as in the primary nozzle. The required derivatives for the compound case can be found in the work of \cite{vandenberghe2024extensioncompoundflowtheory}, and those for the Fabri case in the Appendix~\ref{sec:appendix_single_stream}. The relevant equations are identical to those of the single stream in the primary nozzle, with the difference that the force term also includes the inter-stream friction. The blended equations are integrated to the outlet, switching back to the direct equations once $\Ma_{eq} \geq 1.01$ or $\Ma_s \geq 1.01$.

The flow might become sonic in the secondary inlet, in which case the guessed static pressure is increased to reduce the mass flow rate and reach a sonic condition in the mixing pipe instead. For the geometry analyzed in this work, the sonic point is always located in the mixing pipe, which is the most typical scenario. The analysis of the sign of the numerator in the computed sonic point is identical to the one for the primary nozzle above due to the similarity of Equations~\eqref{eq:p_i} and \eqref{eq:p_compound}. 

The resulting solution is fully supersonic in the diffuser, so the normal shock from section~\ref{sec:practical_shock} is required to match the back pressure. Its axial position is found iteratively to respect the outlet condition.

\section{Calibration of Model 1 }\label{sec:calibration}
The closure equations for model 1 in section~\ref{sec:definition_compound_predictive} include two friction forces, each represented by a friction coefficient. These have to be provided by empirical correlations or by a model calibration.
For the wall friction, reliable correlations exist and this work used the correlation of \cite{vandriest1951turbulent}. For inter-stream friction, however, the only available correlation in the literature is that of \cite{papamoschou1993model}. This correlation is rooted in experimental studies of compressible shear layers and expresses the friction in terms of free-stream quantities. However, such free stream quantities are not readily available from the present 1D model (as opposed to quasi-2D models such as \cite{huang2022quasi}), which relies on cross-stream averaged values. Moreover, the flow mixing in the mixing pipe eventually leads to the vanishing of the free streams once the shear layer reaches both the centerline and the wall boundary layers. Therefore, an empirical law for the inter-stream coefficient was proposed by correcting the \eqref{eq:f_ps_correlation} as follows 
\begin{equation}
\label{eq:wx}
    f^*_{ps} = \left(\frac{w_1}{2} \tanh\left(30 (x - w_2)\right) + \frac{w_1}{2} + 1\right) f_{ps}\,,
\end{equation}
where $w_1$ and $w_2$ are calibration parameters that control the amplitude and axial location of the correction. By construction, the factor tends to unity near the pipe inlet, so the original correlation is retained there, while the friction is increased further downstream (for $w_1 > 0$). The constant 30 was chosen to give a smooth step over the length of the constant-area section ($L=0.5$ m); it could also be included in the calibration but behaves similarly to $w_2$ and was therefore fixed.

The calibration parameters influence the total pressure distribution through ~\eqref{eq:pt_i}, making it the clearest indicator of closure accuracy. In on-design operation, inter-stream friction has little effect on entrainment: the sonic point lies near the start of the constant-area section, leaving only a short upstream region where friction can alter the flow. Downstream of the sonic point, the mass flow is fixed, so calibration mainly affects momentum exchange and static-pressure recovery. Accordingly, the model is calibrated by identifying the weights $w_1$ and $w_2$ that minimize the following error measure between model predictions and RANS data:
\begin{equation}\label{eq:cost_calib}
J_{on}(w_1,w_2) = \sqrt{\dfrac{\sum_i \big(p_{t,s}(x_i) - \tilde{p}_{t,s}(x_i)\big)^2}{\sum_i \big(\tilde{p}_{t,s}(x_i)\big)^2}}\,,
\end{equation}
where $p_{t,s}$ depends implicitly on the calibration parameters via \eqref{eq:wx}, and $\tilde{p}_{t,s}$ denotes averaged RANS data. The primary total pressure is excluded, as it is dominated by inlet shock-train losses not captured by the model, leading to overly optimistic predictions. The secondary stream, governed mainly by wall and inter-stream friction, provides the most reliable diagnostic.

The summation in \eqref{eq:cost_calib} is restricted to grid points up to the diffuser inlet, excluding those downstream of the imposed normal shock (section~\ref{sec:practical_shock}), which would otherwise bias the calibration and reduce accuracy in the constant-area section.
\section{Selected test cases}\label{sec:data}
\begin{figure*}%[htb!]
	\centering
	\includegraphics[width=\linewidth]{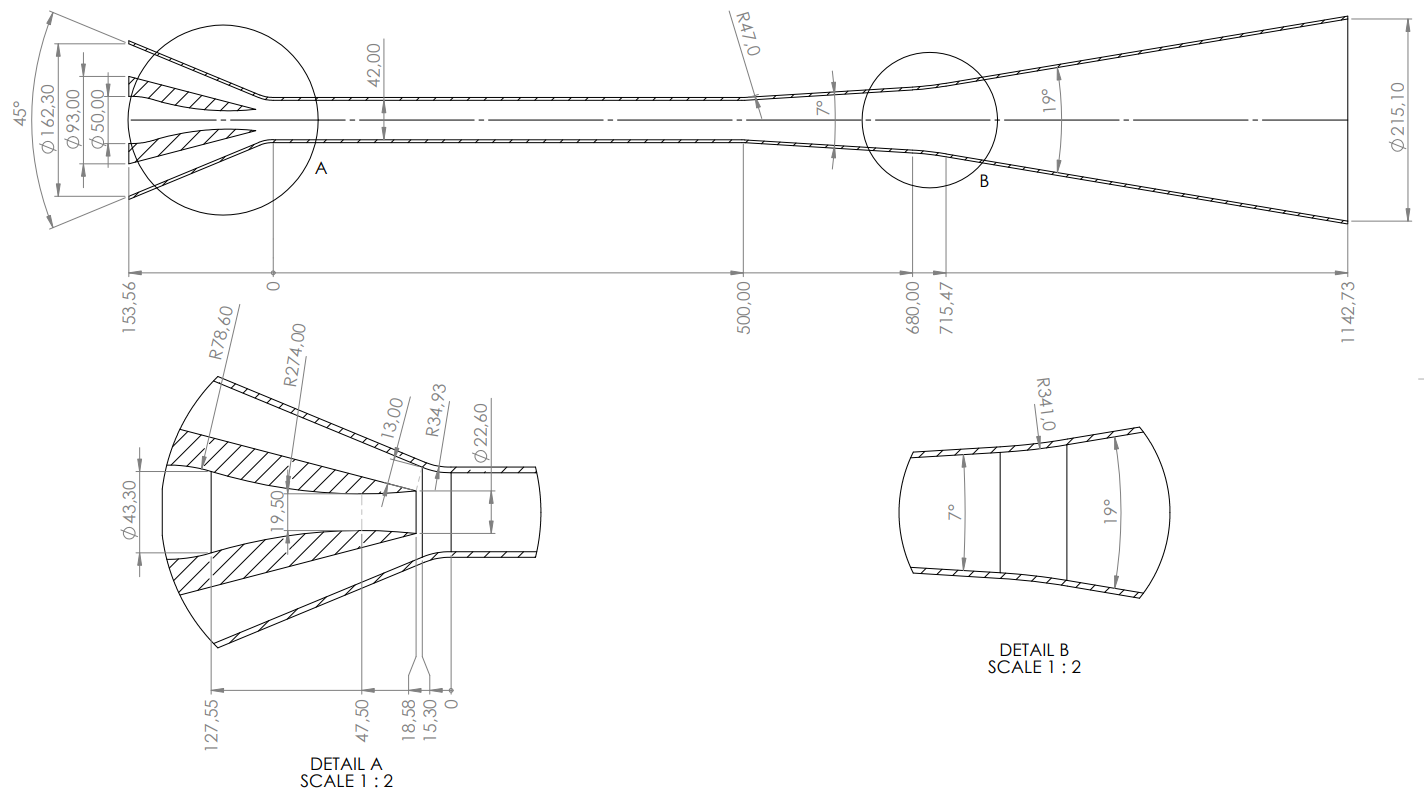}
    \captionsetup{width=\textwidth, justification=justified}
	\caption{The reference data in this work consist of axisymmetric RANS simulations performed by \cite{baguet_querinjean} and \cite{debroeyer2025thesis}. These are post-processed by extracting the dividing streamline and by averaging over the resulting cross-sections using equation \eqref{eq:postproc}. The resulting one-dimensional distributions are compared to the model predictions in section~\ref{sec:results}. The length $L=0.5$ m of the constant-area section is used to normalize the axial coordinate.}
	\label{fig:data_geometry}
\end{figure*}

The reference data and operating conditions have been provided by \cite{baguet_querinjean} and \cite{debroeyer2025thesis}. These authors performed axisymmetric RANS simulations with different inlet pressure ratios and identical total temperatures at the inlets (300 K). The secondary inlet pressure equals 1 bar since the ejector takes air from the atmosphere. We use the results for inlet pressure ratios $p_{t,p}/p_{t,s}$ equal to 4, 5 and 6 in this work. These conditions correspond to a slightly over-expanded and two under-expanded primary jets. The geometry of the ejector is shown in figure~\ref{fig:data_geometry}. The interested reader is referred to the original publications for more details on the mesh and the numerical solver.

The numerical data are post-processed by extracting the dividing streamline and by averaging over the resulting cross-sections following \cite{LAMBERTS201723} and \cite{vandenberghe2024extensioncompoundflowtheory}. The dividing streamline is computed by integrating the mass flux at a given axial position $x$ until the primary mass flow rate as per definition \eqref{eq:def_r_dividing}. This determines the cross-sections $A_i$, over which the density $\rho$, the mass flux $\rho u$ and the total internal energy $\rho e_t=\rho c_V T + 0.5\rho u^2$ are averaged:
\begin{equation}
    \hat{q}_{i} (x) = \frac{1}{A_{i}}\int_{A_{i}}q \,dA_{i} \,,\label{eq:postproc}
\end{equation}
where $q\in[\rho, \rho u, \rho e_t]$. 
Other quantities are computed using the ideal gas law and classic gas dynamic relations. The reported mass flow rates are normalized with the follow reference flow rates:
\begin{align}
    \dot{m}_{p,ref} &= {p_{t,p} A_{th} \sqrt{\frac{\gamma}{R T_{t,p}}}\left(\frac{\gamma + 1}{2}\right)^{-\frac{1}{2}\frac{\gamma+1}{\gamma-1}}}\,,\label{eq:ref_mass_flow_rate_prim}\\
    \dot{m}_{s,ref} &= {p_{t,s} (A_{m} - A_{th}) \sqrt{\frac{\gamma}{R T_{t,s}}}\left(\frac{\gamma + 1}{2}\right)^{-\frac{1}{2}\frac{\gamma+1}{\gamma-1}}}\,.\label{eq:ref_mass_flow_rate_sec}
\end{align}
These correspond to sonic flows with the stagnation conditions $(p_{t,i}, T_{t,i})$ through cross-sections that depend on the primary throat $A_{th}$ and the constant-area section $A_m$ in the mixing pipe. These estimates form an upper limit since they correspond to isentropic flows and the cross-section for the secondary mass flow rate is relatively large. The normalized mass flow rates are thus bounded between zero and one (see \cite{neural2022}).
\section{Results}\label{sec:results}

The results are organized according to the four models introduced in section~\ref{sec:definition_models}. The model 1 is presented first, as it constitutes the only predictive framework. The remaining configurations serve as diagnostic tools to interpret the RANS simulations and to explore potential pathways for improving the predictive model. Models 1 and 2 are based on compound-choking, while models 3 and 4 are based on Fabri-choking. However, their predictions need not be interpreted strictly as such: all cases presented below are compound-sonic ($\beta=0$) but not necessarily Fabri-sonic ($\Ma_s<1$).

\subsection{Model 1: Compound choking with calibrated closures}\label{sec:results_compound_predictive}
\subsubsection{Predictions on the selected operating points}
Table~\ref{tab:results_compound} compares the mass flow rates from the RANS simulations with the predictions of the compound-based model, for the three considered pressure ratios $p_{t,p}/p_{t,s}$. The weights $w_1$ and $w_2$ for the calibration of the inter-stream coefficient \eqref{eq:wx} are also listed for completeness.

The normalized primary mass flow rate is practically constant and slightly lower than unity due to the effect of friction. The model makes an acceptable error of 0.5 \%. The overestimated mass flow rate indicates a slight underestimation of the wall friction. The normalized secondary mass flow rate in on-design operation is further below unity, since the used cross-section for the reference mass flow rate $\dot{m}_{s,ref}$ in Equation~\eqref{eq:ref_mass_flow_rate_sec} is generally larger than the actual cross-section where the secondary stream reaches Mach one.

The compound-based model overestimates the secondary mass flow rate, with the error increasing with the inlet pressure ratio $p_{t,p}/p_{t,s}$ and reaching up to 5.2~\%. This overprediction adds to that of the RANS simulations, which already overestimate the entrainment measured in the experiments of \cite{brosteaux_doucet} by 3--5~\%. The origin of the discrepancy between RANS and experiments remains unclear and is not further investigated here. The cause of the model--simulation mismatch is discussed in section~\ref{sec:issue_ondesign}, based on the local axial distributions.

\begin{table*}%[htb!]
    \centering
    \captionsetup{width=\textwidth, justification=justified}
    \caption{Overview of the mass flow rates predicted by the compound-based model, the used calibration parameters $w_1$ and $w_2$, and the errors with respect to the RANS simulations. The error increases with the inlet pressure ratio $p_{t,p}/p_{t,s}$. An underexpanded primary stream induces a larger model error, as discussed in more detail in section~\ref{sec:issue_ondesign}. Finally, it should be noted that the model error aggravates the overestimation of the RANS simulations with respect to the experiments of \cite{baguet_querinjean}.}
    \begin{tabular}{cccccccccc}
        $p_{t,p}/p_{t,s}$ & $p_b/p_{t,s}$ & $w_1$ & $w_2$ & $\dot{m}_{p} / \dot{m}_{p,ref}$ & $\dot{m}_{p} / \dot{m}_{p,ref}$ & Error & $\dot{m}_{s} / \dot{m}_{s,ref}$ & $\dot{m}_{s} / \dot{m}_{s,ref}$ & Error \\
        {[-]} & [-] & [-] & [-] & (model) [-] & (RANS) [-] & [\%] & (model) [-] & (RANS) [-] & [\%] \\ 
        \hline 
        4 & 1.10 & 1.287 & 0.188 & 0.992 & 0.988 & 0.5 & 0.830 & 0.827 & 0.4 \\
        5 & 1.20 & 1.098 & 0.180 & 0.993 & 0.988 & 0.4 & 0.783 & 0.772 & 1.4 \\
        6 & 1.55 & 0.616 & 0.108 & 0.993 & 0.989 & 0.4 & 0.739 & 0.703 & 5.2
    \end{tabular}
    \label{tab:results_compound}
\end{table*}

\begin{figure*}%[htb!]
  \center
  \begin{subfigure}[t]{\linewidth}
    \center
    \includegraphics[width=0.48\linewidth]{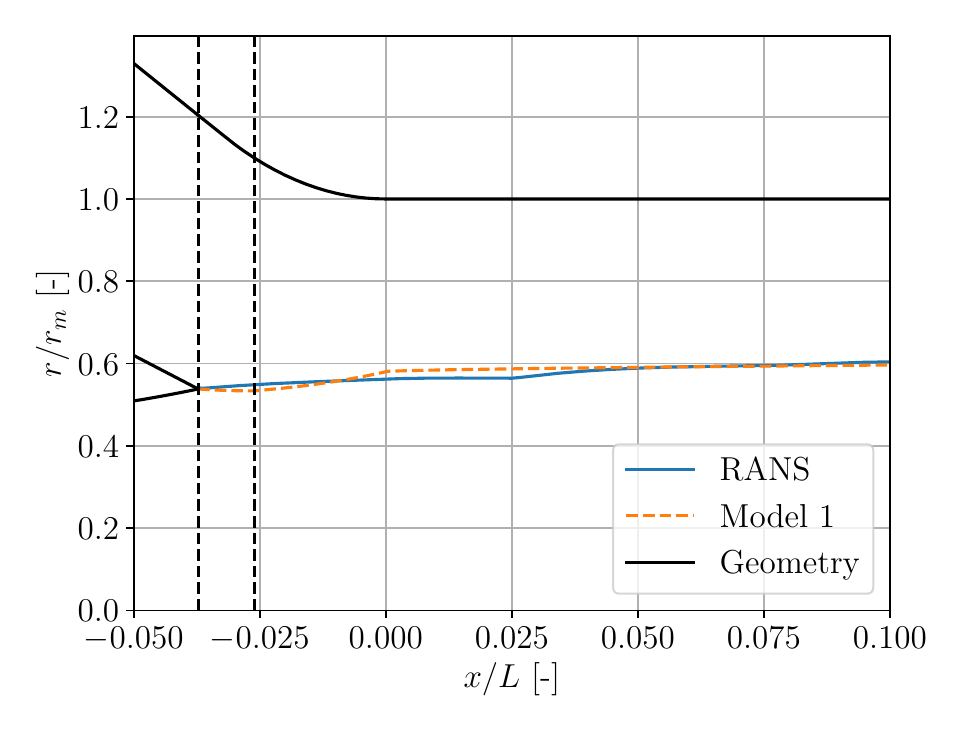}
    \includegraphics[width=0.48\linewidth]{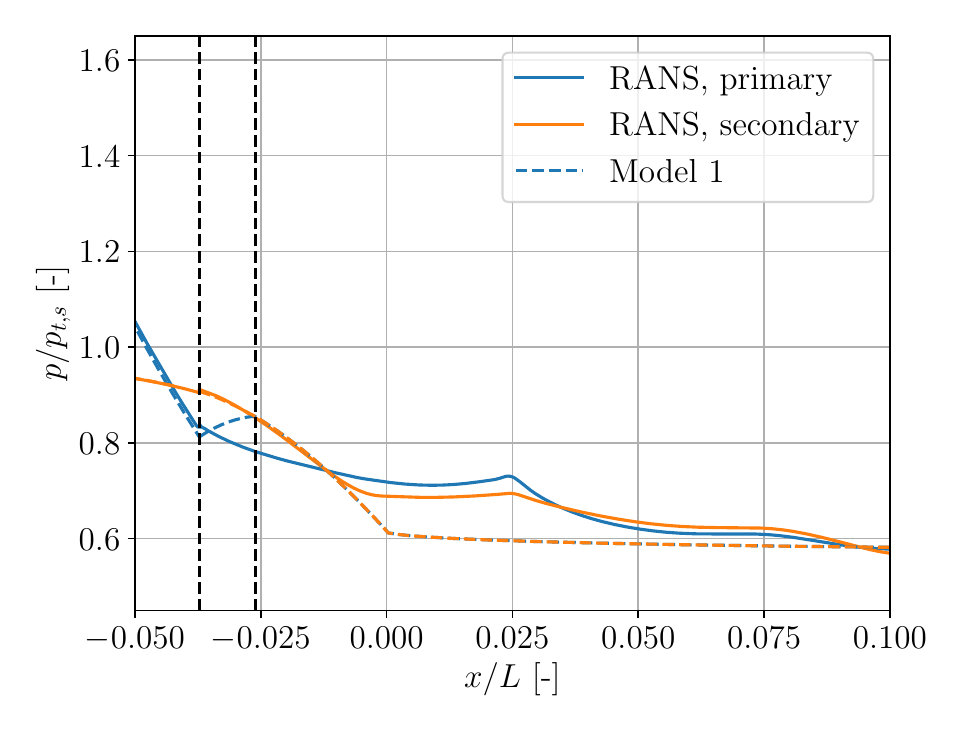}
    \subcaption{$p_{t,p}/p_{t,s}=4$, over-expanded primary stream.}
    \label{fig:results_p_eq_4}
  \end{subfigure}
  \\
  \begin{subfigure}[t]{\linewidth}
    \center
    \includegraphics[width=0.48\linewidth]{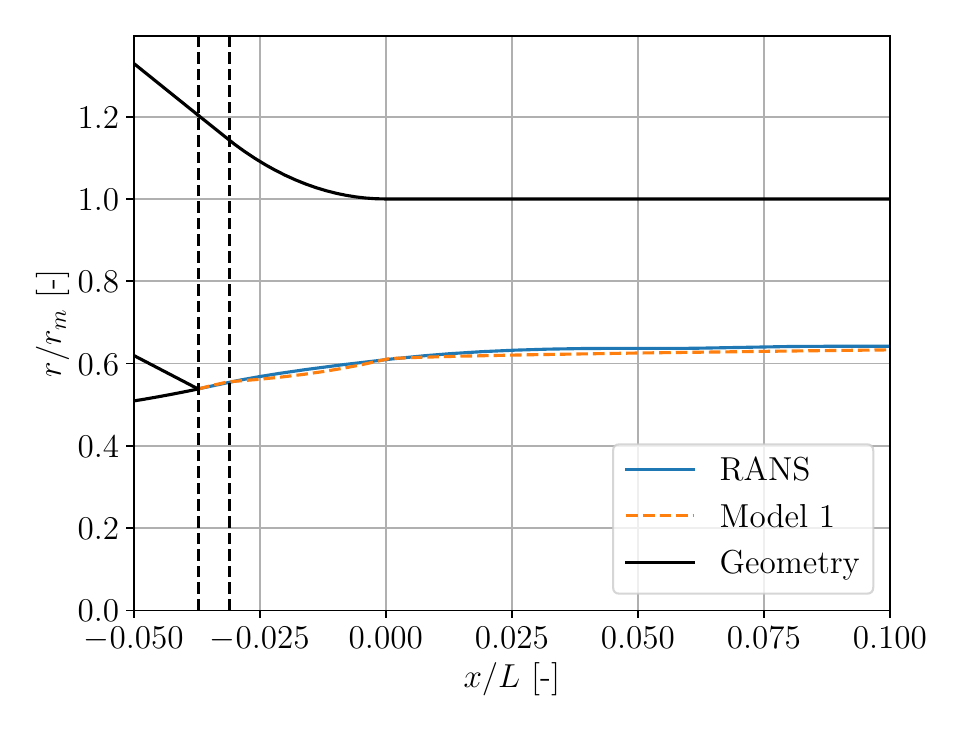}
    \includegraphics[width=0.48\linewidth]{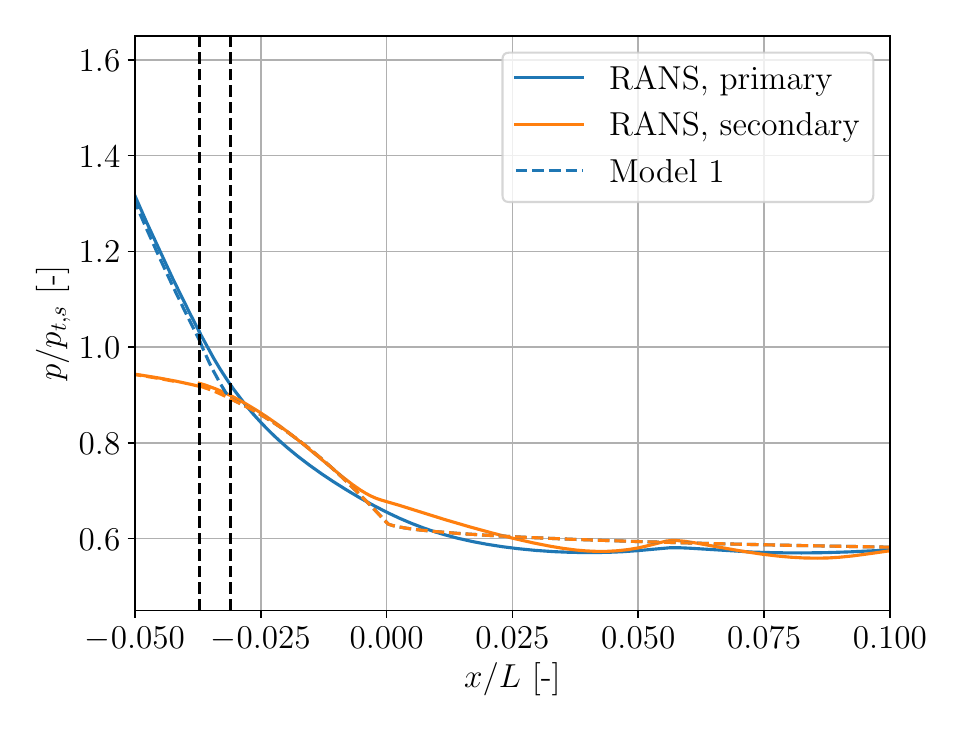}
    \subcaption{$p_{t,p}/p_{t,s}=5$, slightly under-expanded primary stream.}
    \label{fig:results_p_eq_5}
  \end{subfigure}
  \\
  \begin{subfigure}[t]{\linewidth}
    \center
    \includegraphics[width=0.48\linewidth]{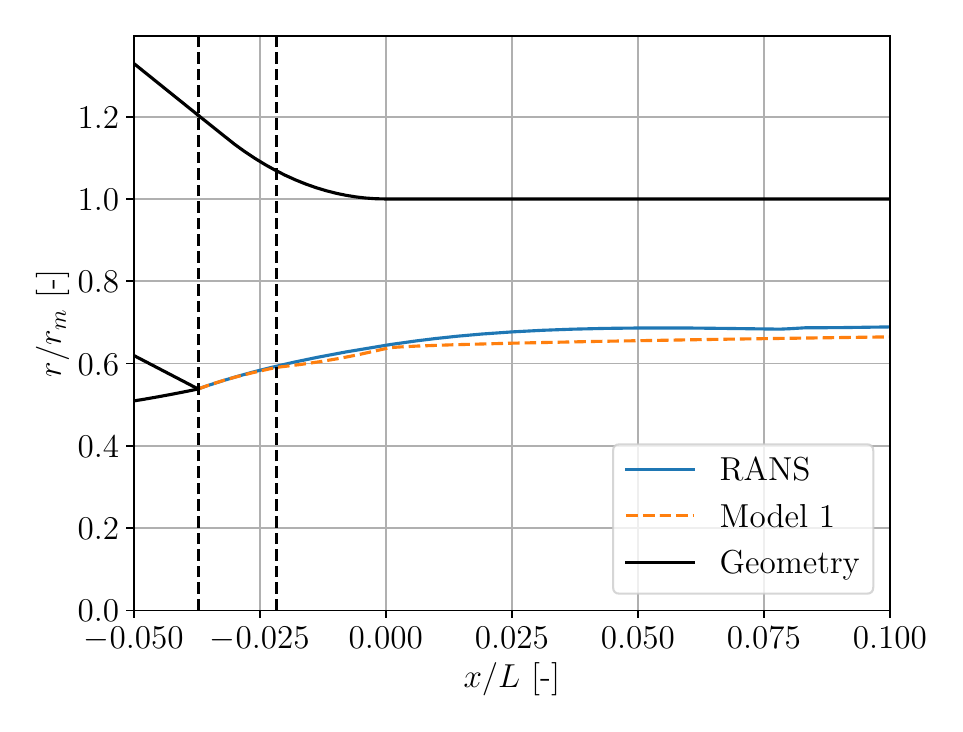}
    \includegraphics[width=0.48\linewidth]{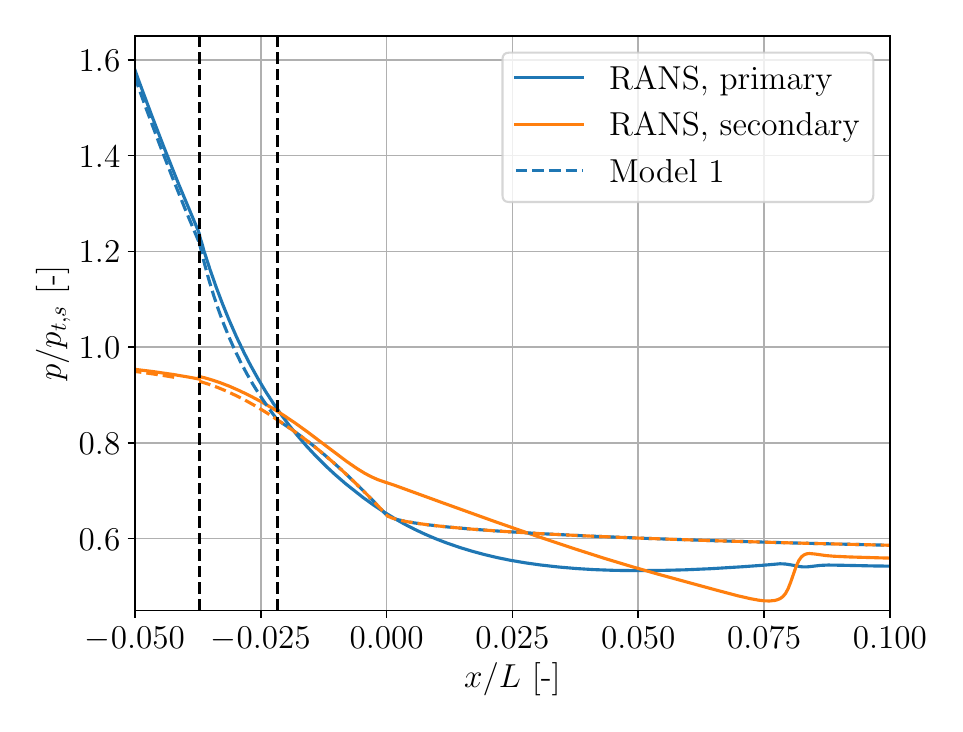}
    \subcaption{$p_{t,p}/p_{t,s}=6$, under-expanded primary stream.}
    \label{fig:results_p_eq_6}
  \end{subfigure}
  \captionsetup{width=\textwidth, justification=justified}
  \caption{The dividing streamline (left) and the cross-stream averaged static pressure (right) from the RANS simulations, and from the compound-based model for three inlet pressure ratios. The pressure equalization acts between the vertical dotted lines and the compound theory predicts the uniform static pressure downstream. The model is most accurate in the near-perfectly expanded case $p_{t,p}/p_{t,s}=5$, where the pressures equalize in the shortest distance. However, it strongly constrains the dividing streamline to keep the pressure uniform, regardless of inertial effects in the primary stream. This leads to sharp and non-physical corners at the point where the pressures equalize and at the inlet of the constant-area section at $x/L=0$. For the colors in this figure, the reader is referred to the online version of this article.}
\label{fig:results_p_eq}
\end{figure*}

Figure~\ref{fig:results_p_eq} shows the distributions of the dividing streamline and the cross-stream averaged static pressures near the inlet of the mixing pipe. The predicted streamline consists of two regions: between the vertical dotted lines, the pressure equalization mechanism from section~\ref{sec:practical_pressure} is active; beyond this region, Equation~\eqref{eq:A_i} is integrated to obtain the cross-sections and hence the dividing streamline.

The angle of the dividing streamline at the outlet of the primary nozzle reflects whether the primary stream is under- or over-expanded. It is deflected inward in the over-expanded case (figure~\ref{fig:results_p_eq_4}), aligns with the nozzle lip in the nearly perfectly expanded case (figure~\ref{fig:results_p_eq_5}), and expands outward in the under-expanded case (figure~\ref{fig:results_p_eq_6}). The predicted streamlines match the reference well for under-expanded primary streams. However, the pressure equalization mechanism becomes less accurate in the over-expanded regime. It enforces a contraction of the primary stream (see Equation~\eqref{eq:p-eq_theta_shock}), whereas the RANS simulation still shows an outward spread. In the RANS solution, this compression occurs through an oblique shock, which cannot be captured by the 1D model. In the model, the oblique shock is only used to compute the inward deflection angle and to compress the supersonic primary stream through a converging cross-section. While a similar inward deflection is present in the RANS solution, it does not overcome the outward inclination imposed by the nozzle lip (figure~\ref{fig:results_p_eq_4}).

In reality, the pressures equalize over a much longer distance, as also visible in figure~\ref{fig:results_p_eq}. The proposed mechanism only mimics the first portion of the shock train. This limitation of the compound-based model is necessary to address pressure equalization and choking separately. A generalized compound choking theory with non-uniform static pressure would render this mechanism redundant, but this has not been presented in literature. The error on the dividing streamline has a direct consequence on the static pressure distributions on the right of figure~\ref{fig:results_p_eq}, which show a superior agreement in the region of pressure equalization (between the vertical lines) for under-expanded primary streams. Note that the distance between the nozzle exit and the predicted point of equal pressure is minimal in the near-perfectly expanded case in figure~\ref{fig:results_p_eq_5}.

A sharp corner appears in the predicted dividing streamlines at the point of equal static pressure, deviating from the reference in the under-expanded cases. In reality, inertia carries the primary stream further outward, causing it to overexpand and triggering a shock wave that sustains the shock train. A different gradient of the dividing streamline is required to enforce identical pressures. This local constraint makes the compound theory oblivious to inertial effects from upstream, and thus permits sharp corners. This limitation is an inherent consequence of capturing choking within the compound framework, as no generalized compound theory for non-uniform static pressure is currently available.

The predicted uniform static pressure is a reasonable `average' of the primary and secondary pressures in the near-perfectly expanded case $p_{t,p}/p_{t,s}=5$, as the identical static pressure is reached far upstream due to the small initial difference. The flow in the RANS simulations expands further than in the compound prediction in the stronger under-expanded case $p_{t,p}/p_{t,s}=6$, as evidenced by the stronger divergence of the dividing streamline and the lower static pressures in figure~\ref{fig:results_p_eq_6}. A second sharp corner appears in the compound predictions at the inlet of the constant-area section ($x/L=0$). This arises because the gradient $dA/dx$ of the total cross-section disappears from the numerator in Equation~\eqref{eq:p_compound} and only the force terms remain. Upstream, the convergent geometry of the mixing pipe (considered to start at the outlet of the primary nozzle at $x/L=-0.0372$) thus dominates.

\begin{figure*}%[htb!]
  \center
  \begin{subfigure}[t]{\linewidth}
    \center
    \includegraphics[width=\linewidth]{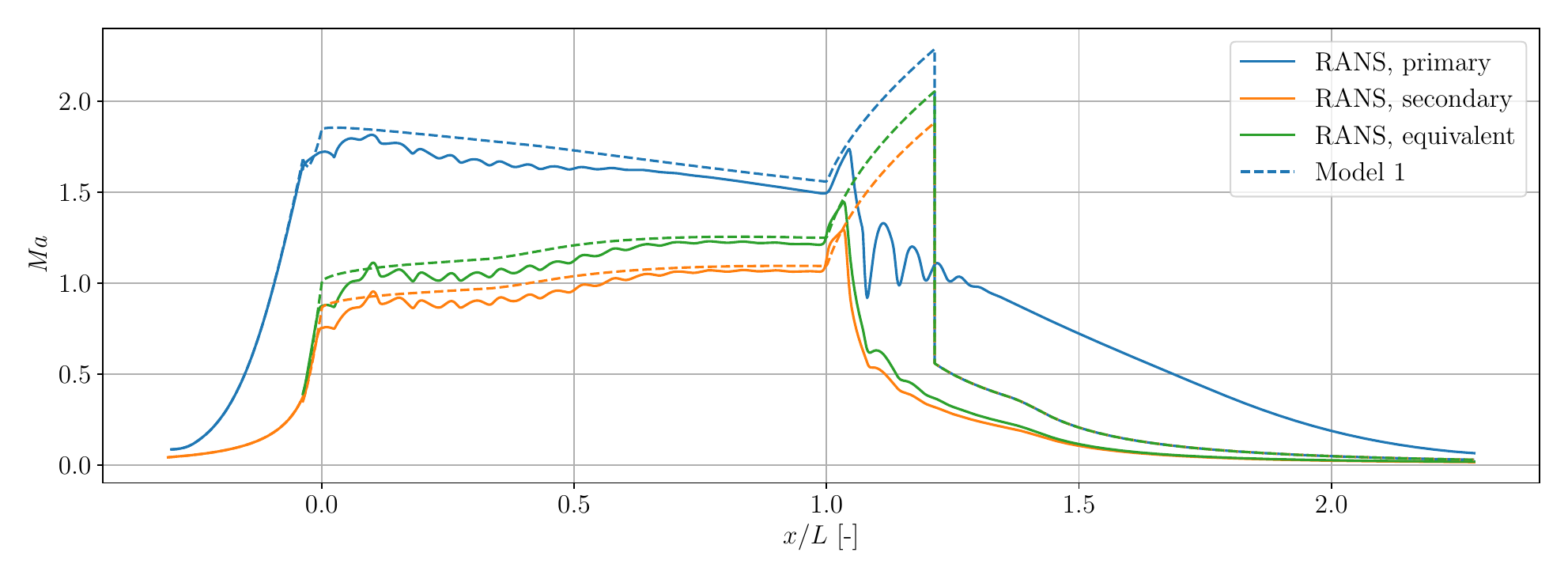}
    \subcaption{$p_{t,p}/p_{t,s}=4$}
    \label{fig:results_compound_M_4}
  \end{subfigure}
  \\
  \begin{subfigure}[t]{\linewidth}
    \center
    \includegraphics[width=\linewidth]{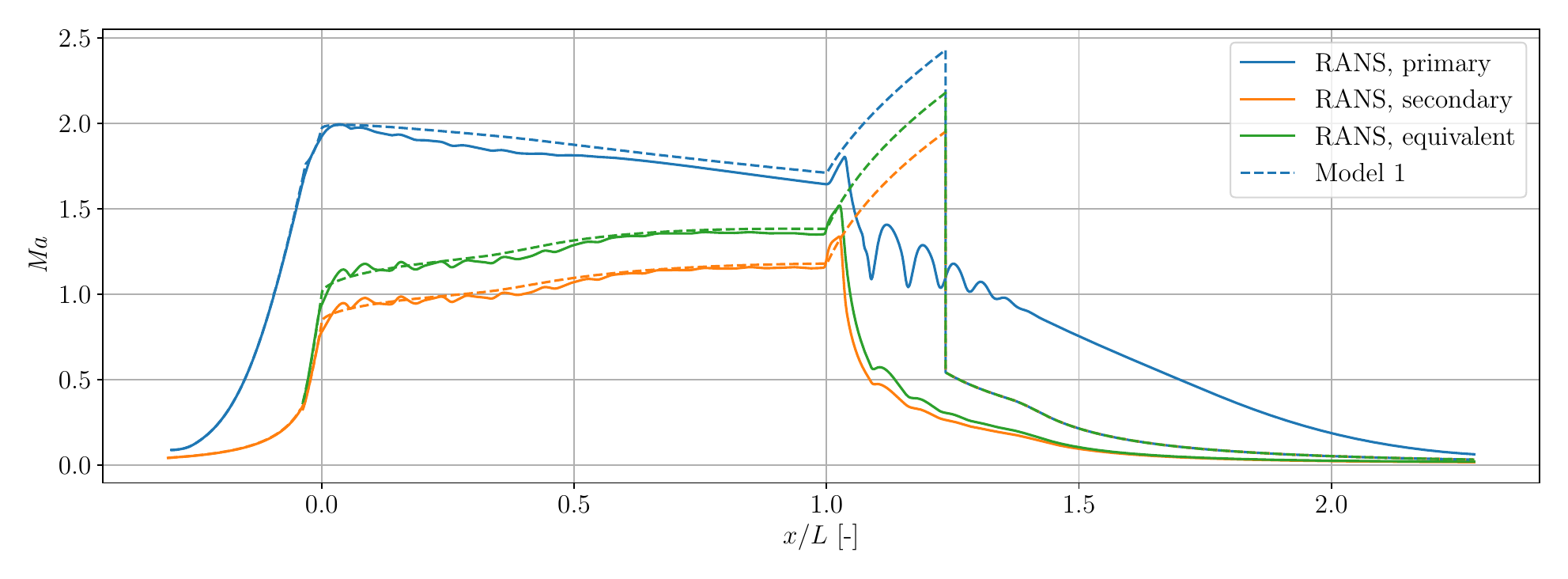}
    \subcaption{$p_{t,p}/p_{t,s}=5$}
    \label{fig:results_compound_M_5}
  \end{subfigure}
  \\
  \begin{subfigure}[t]{\linewidth}
    \center
    \includegraphics[width=\linewidth]{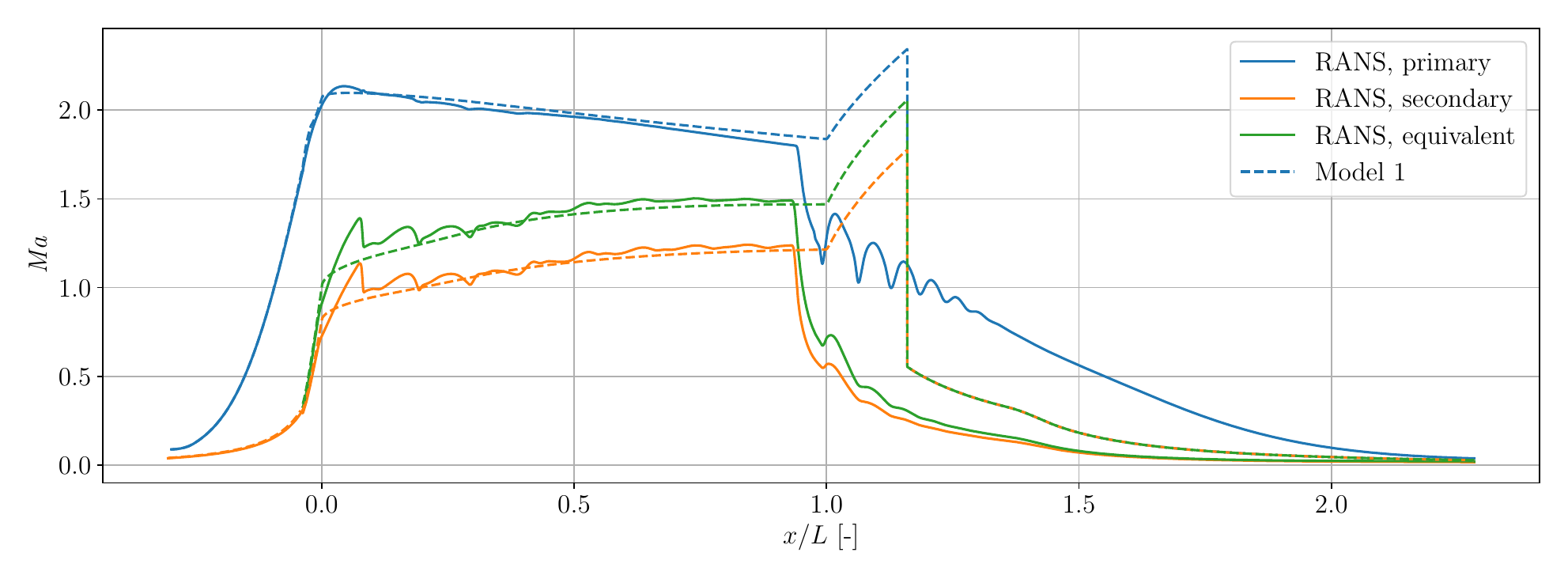}
    \subcaption{$p_{t,p}/p_{t,s}=6$}
    \label{fig:results_compound_M_6}
  \end{subfigure}
  \captionsetup{width=\textwidth, justification=justified}
  \caption{Axial distributions of the Mach numbers from the RANS simulations, and from the compound-based model for the three inlet pressure ratios. The primary stream is over-expanded in the case $p_{t,p}/p_{t,s}=4$ and under-expanded otherwise as evidenced by the higher primary Mach number in the latter cases. The compound-sonic point is located near the inlet of the constant-area section at $x/L=0$. For the colors in this figure, the reader is referred to the online version of this article.}
\label{fig:results_compound_M}
\end{figure*}

\begin{figure*}%[htb!]
  \center
  \begin{subfigure}[t]{\linewidth}
    \center
    \includegraphics[width=\linewidth]{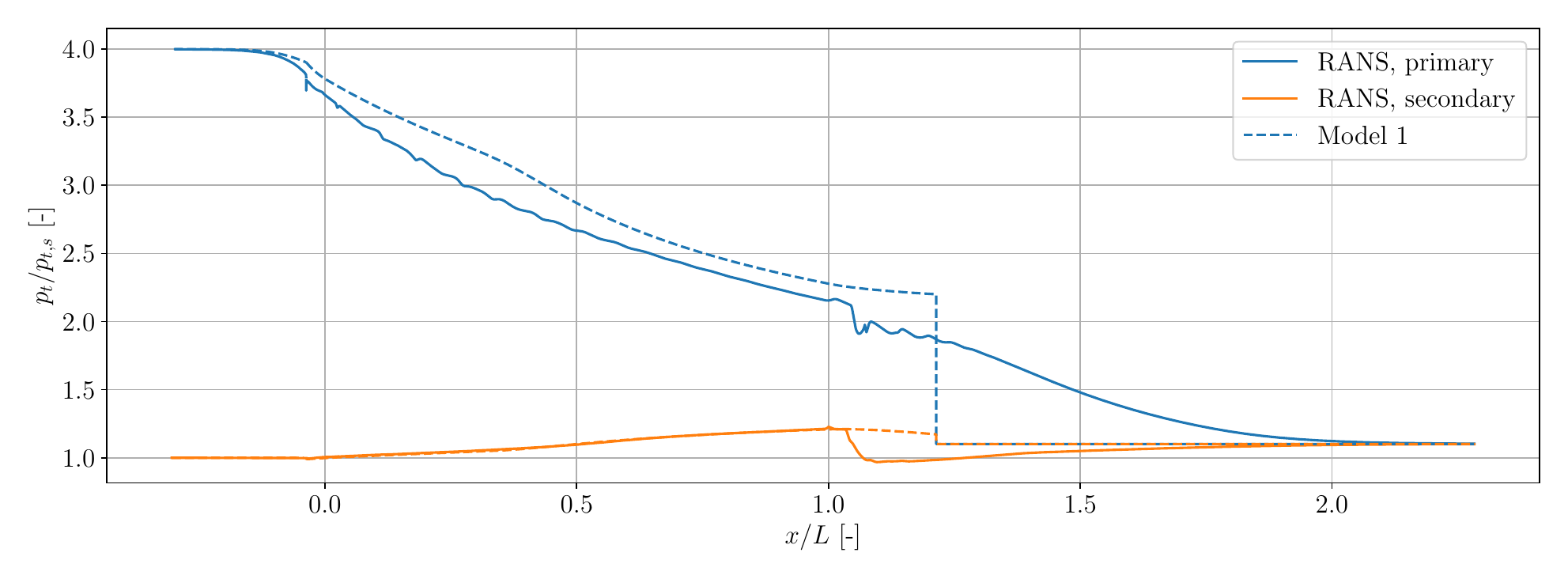}
    \subcaption{$p_{t,p}/p_{t,s}=4$}
    \label{fig:results_compound_pt_4}
  \end{subfigure}
  \\
  \begin{subfigure}[t]{\linewidth}
    \center
    \includegraphics[width=\linewidth]{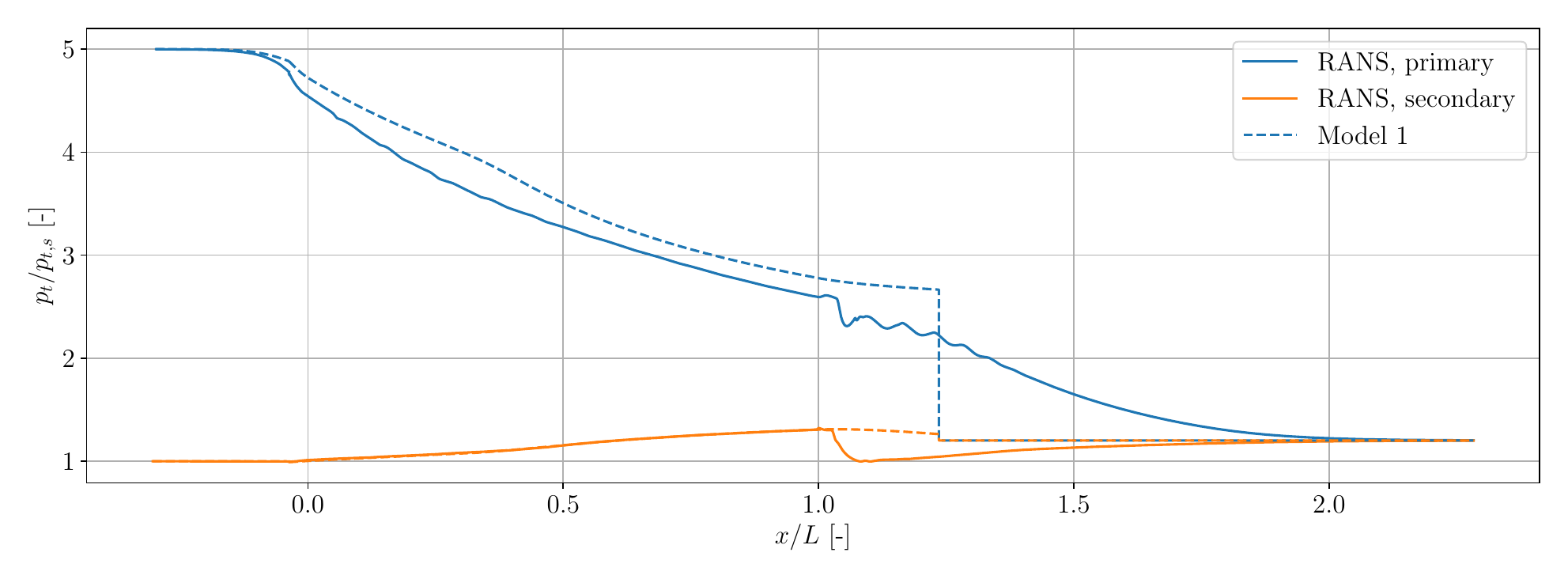}
    \subcaption{$p_{t,p}/p_{t,s}=5$}
    \label{fig:results_compound_pt_5}
  \end{subfigure}
  \\
  \begin{subfigure}[t]{\linewidth}
    \center
    \includegraphics[width=\linewidth]{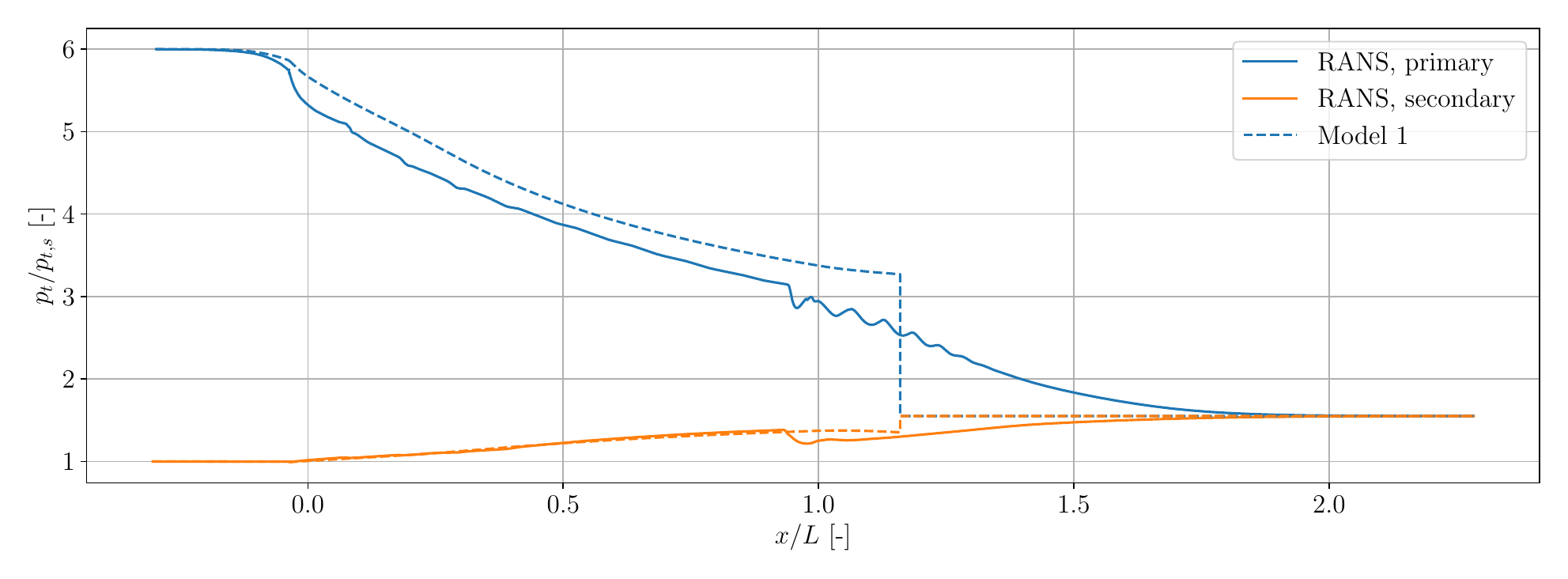}
    \subcaption{$p_{t,p}/p_{t,s}=6$}
    \label{fig:results_compound_pt_6}
  \end{subfigure}
  \captionsetup{width=\textwidth, justification=justified}
  \caption{Axial distributions of the total pressure from the RANS simulations, and from the compound-based model for the three inlet pressure ratios. The primary and secondary total pressures tend to converge through momentum exchange and reach an identical value at complete mixing. The inter-stream friction has been calibrated to minimize the error on the secondary total pressure. The primary total pressure is overestimated due to underestimated friction in the primary nozzle (with the outlet at $x/L=-0.0372$), but mostly due to the losses in the shock train that are not included in the model. For the colors in this figure, the reader is referred to the online version of this article.}
\label{fig:results_compound_pt}
\end{figure*}

Figures~\ref{fig:results_compound_M} and \ref{fig:results_compound_pt} show the distributions of the Mach number and the total pressure for the three investigated inlet pressure ratios (see table~\ref{tab:results_compound}). The total temperature remains uniform at $T_{t,p} = T_{t,s} = 300$ K with deviations below 1 \% relative to the RANS simulations. All cases are characterized by a choked flow, as evidenced by a unitary equivalent Mach number.

The model shows generally good agreement with the RANS simulations. The main differences arise from substituting the complex two-dimensional shock trains in the diffuser with normal shocks. These serve only to match the outlet pressure; the instantly mixed state downstream does not reflect the actual flow in the diffuser.

The compound-sonic point is predicted slightly upstream of the constant-area inlet ($x/L=0$). However, the sonic points in the RANS simulations are consistently located downstream in the constant-area section. The origin of this discrepancy is unclear. It may arise from inaccuracies in the closure modeling, altering the balance of the force terms in Equation~\eqref{eq:p_compound}, or from the strong assumption of uniform static pressure in the compound flow theory. Indeed, the static pressures at the compound-sonic points generally differ, as seen near $x/L=0$ in figure~\ref{fig:results_p_eq}.

The primary and the secondary total pressures converge from their boundary conditions toward a unique value, as shown in figure~\ref{fig:results_compound_pt}. This indicates mixing, since a uniform flow cannot be distinguished as originating from a single reservoir or from two reservoirs as in ejectors. This is also apparent from the modeled normal shock which involves instant mixing. The inter-stream friction is strongest at the inlet of the mixing pipe, where the velocity difference is largest. Consequently, the primary total pressure decreases while the secondary total pressure increases, consistent with the signs of the forces in Equation~\eqref{eq:forces_mix}. As the secondary stream accelerates, the wall friction gains importance, ultimately leading to a reduction in the secondary total pressure.

A small mismatch in primary total pressure appears between the RANS and the model at the outlet of the primary nozzle ($x/L=-0.0372$) due to underestimated friction. The overestimated primary total pressure is likely due to losses in the shock train, which are not included in the momentum balance and mainly affect the primary stream. The secondary total pressure shows excellent agreement since its error has been minimized by calibrating the inter-stream friction (see Equation~\eqref{eq:cost_calib}).

\subsubsection{Overestimated entrainment using the compound theory}\label{sec:issue_ondesign}
The compound model overestimates the entrainment in the under-expanded case $p_{t,p}/p_{t,s}=6$ (see table~\ref{tab:results_compound}). To reach the same secondary mass flow rate as in the RANS simulation, the model therefore predicts off-design operation. This prediction is obtained here by imposing the averaged RANS flow state as a boundary condition at the inlet of the mixing pipe and integrating the governing equations downstream.

\begin{figure}%[htb!]
  \center
  \begin{subfigure}[t]{0.48\linewidth}
    \includegraphics[width=\linewidth]{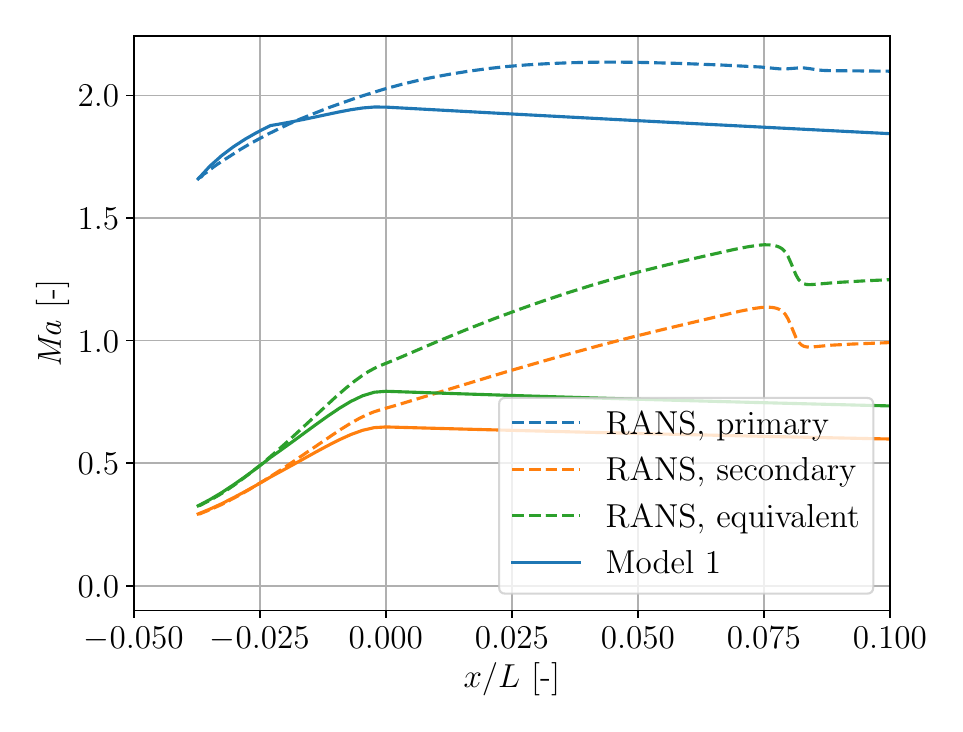}
    \subcaption{Mach number}
    \label{fig:issue2_ondesign_Mach}
  \end{subfigure}
  \begin{subfigure}[t]{0.48\linewidth}
    \includegraphics[width=\linewidth]{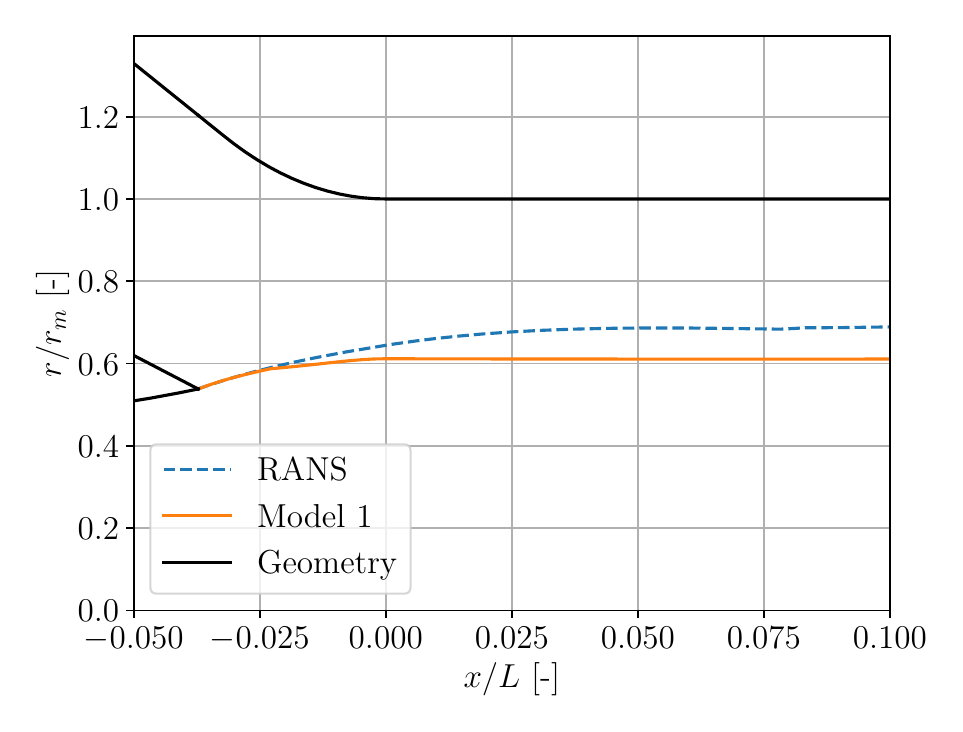}
    \subcaption{Dividing streamline}
    \label{fig:issue2_ondesign_rdiv}
  \end{subfigure}
  \captionsetup{width=\textwidth, justification=justified}
  \caption{The model predicts a compound-subsonic solution when using the averaged states from the RANS simulations as boundary conditions at the inlet of the mixing pipe ($p_{t,p}/p_{t,s}=6$). The expansion of the primary stream in the constant area section is not captured by the compound theory since the gradients of the primary and the total cross-sections are directly related through Equations~\eqref{eq:A_i} and \eqref{eq:p_compound}. The missed expansion is a consequence of the assumption of uniform static pressure and causes the overestimated entrainment. For the colors in this figure, the reader is referred to the online version of this article.}
\label{fig:issue2_ondesign}
\end{figure}

Figure~\ref{fig:issue2_ondesign} shows the resulting distributions of the Mach number and the dividing streamline. The flow remains compound-subsonic and stops expanding at the inlet of the constant area section ($x/L=0$). The predicted dividing streamline departs from the reference curve with a relatively sharp corner as in figure~\ref{fig:results_p_eq_6} due to the constraint of uniform static pressure. Consequently, both streams miss the stronger expansion in the constant area section, and the iterative procedure from section~\ref{sec:solution} incorrectly detects off-design operation.  The secondary flow rate is then increased until the sonic condition is reached, as shown in figure~\ref{fig:results_compound_M_6}.

The overestimated flow rate in on-design operation is thus inherent to the assumption of uniform static pressure in the mixing pipe, challenging the foundations of the compound flow theory. This limitation is particularly noteworthy, since uniform static pressure in the mixing pipe is a common assumption in 0D (\cite{huang1999, METSUE2021121856}, among others) and 1D ejector models (see \cite{clark1995application, BANASIAK20112235, grazzini2015constructal}).

\subsection{Model 2: Compound choking with closure from CFD}\label{sec:results_compound_analysis}
The closure model incorporating friction forces performs reasonably well but tends to overestimate the primary total pressure due to unmodeled pressure losses in the shock train. To assess potential improvements in accuracy, the total pressure gradient extracted from the RANS simulations is imposed within the mixing pipe (see section~\ref{sec:definition_compound_analysis}).

Globally, the error on the flow rate decreased only slightly (from 5.2 to 4.8 \% in the case $p_{t,p}/p_{t,s}=6$), indicating that further refinement of the closure model yields negligible improvement in entrainment prediction. This limited effect arises from the short distance between the outlet of the primary nozzle and the sonic section, across which friction forces can influence the mass flow rate. Variations downstream affect the mixing, the associated pressure recovery and ultimately the critical back pressure, but not the flow rate itself.

\begin{figure*}%[htb!]
  \center
  \begin{subfigure}[t]{\linewidth}
    \center
    \includegraphics[width=\linewidth]{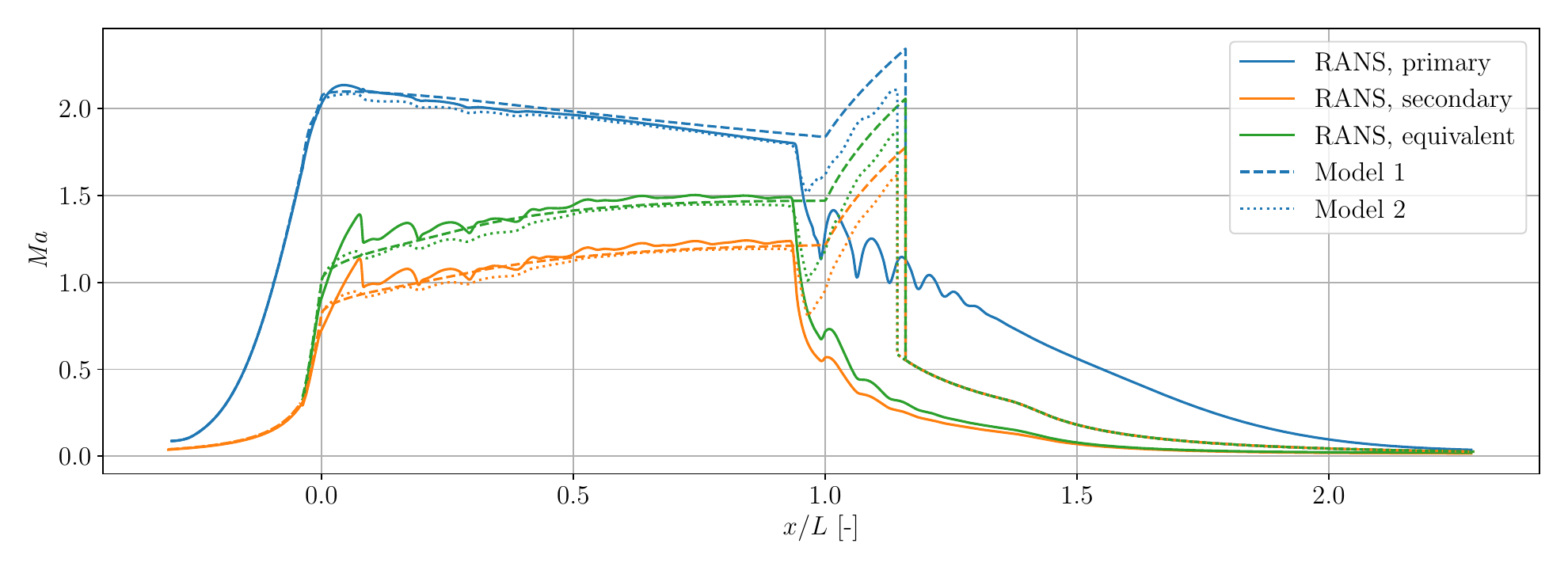}
    \subcaption{Mach number}
    \label{fig:results_compound_dptdx_Mach}
  \end{subfigure}
  \\
  \begin{subfigure}[t]{\linewidth}
    \center
    \includegraphics[width=\linewidth]{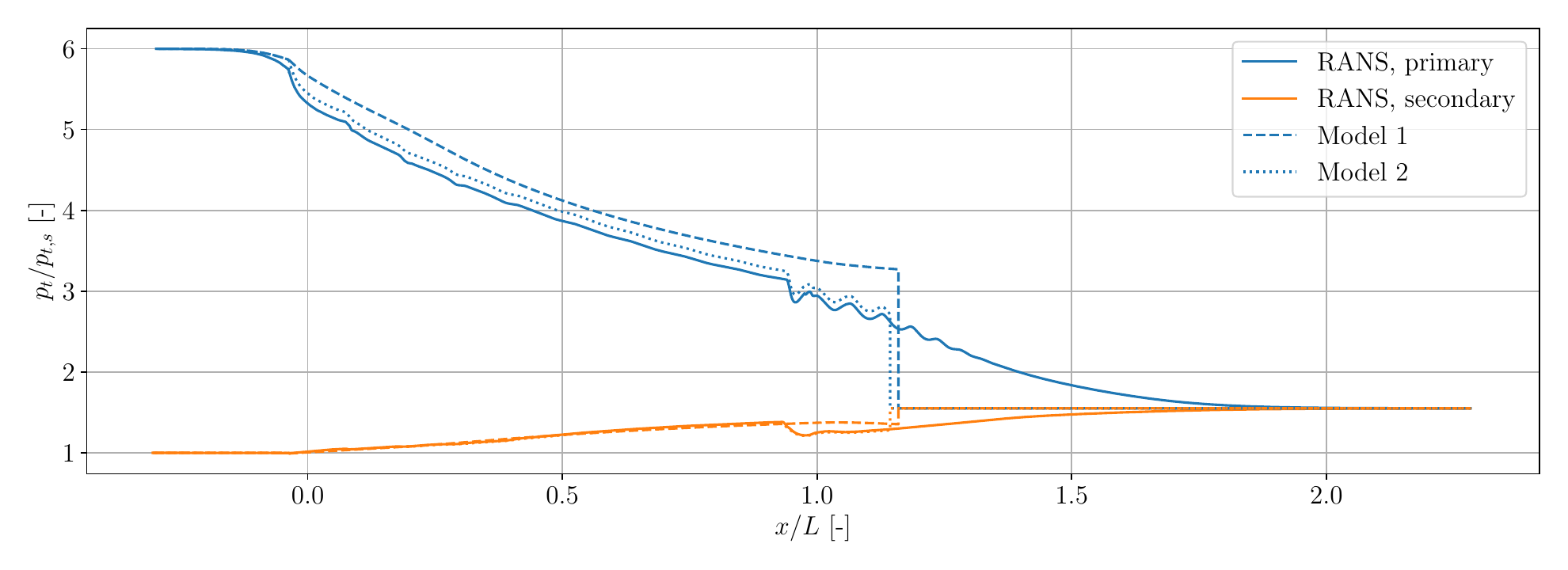}
    \subcaption{Total pressure}
    \label{fig:results_compound_dptdx_pt}
  \end{subfigure}
  \captionsetup{width=\textwidth, justification=justified}
  \caption{Axial distributions of the Mach number and the total pressure from the RANS simulations, and from the compound-based model using the correlations (model 1) or the imposed total pressure gradient (model 2) for the case $p_{t,p}/p_{t,s}=6$. Imposing the total pressure gradient of the RANS simulation in the mixing pipe yields higher accuracy on the local distributions, but marginally reduces the error on the secondary flow rate, which is therefore linked to the dividing streamline and not to the closure modeling of the exchanges. For the colors in this figure, the reader is referred to the online version of this article.}
\label{fig:results_compound_dptdx}
\end{figure*}

Figure~\ref{fig:results_compound_dptdx} compares the distributions of the Mach number and the total pressure with their RANS references and to the original predictions using the friction forces. The improvement in total pressure is evident, though unsurprising, since the model is no longer predictive but rather used to interpret and diagnose the RANS results. The constant offset in primary total pressure originates from the primary nozzle, where the wall friction remained modeled through the correlation of \cite{vandriest1951turbulent}. An error persists in the secondary Mach number due to the overestimated flow rate, as the missed expansion mechanism discussed in section~\ref{sec:issue_ondesign} remains active. The static pressure continues to be equalized and constrained to uniformity, as in the original compound model.

This analysis shows that the prediction error on the entrainment is fundamentally governed by the dividing streamline, and ultimately linked to the constraint of uniform pressure. It can therefore not be substantially reduced through improvements of closures modeling. In the following section, the dividing streamlines from the RANS simulations are imposed in the frame of the  Fabri approach.

\subsection{Model 3: Fabri choking with closures from CFD}\label{sec:results_fabri_analysis}

The dividing streamline and total pressure gradients obtained from the RANS simulations are now imposed within the model. This gives rise to the Fabri-sonic condition, as the constraint of uniform static pressure is relaxed, leading to a singularity in the one-dimensional equations if $\Ma_s~=~1$. The Van Driest correlation continues to be used for the wall friction in the inlets (section~\ref{sec:definition_fabri_analysis}). We recall that the model is highly sensitive to the gradients of the dividing streamline and the total pressures near the sonic point, where the sum of the corresponding terms in Equation~\eqref{eq:p_i} should tend to zero. This sensitivity prevents the use of predictive correlations in conjunction with the dividing streamline extracted from the RANS simulations.

\subsubsection{Predictions on the selected operating points}\label{sec:results_fabri_analysis_choked}
\begin{figure*}%[htb!]
  \center
  \begin{subfigure}[t]{\linewidth}
    \center
    \includegraphics[width=\linewidth]{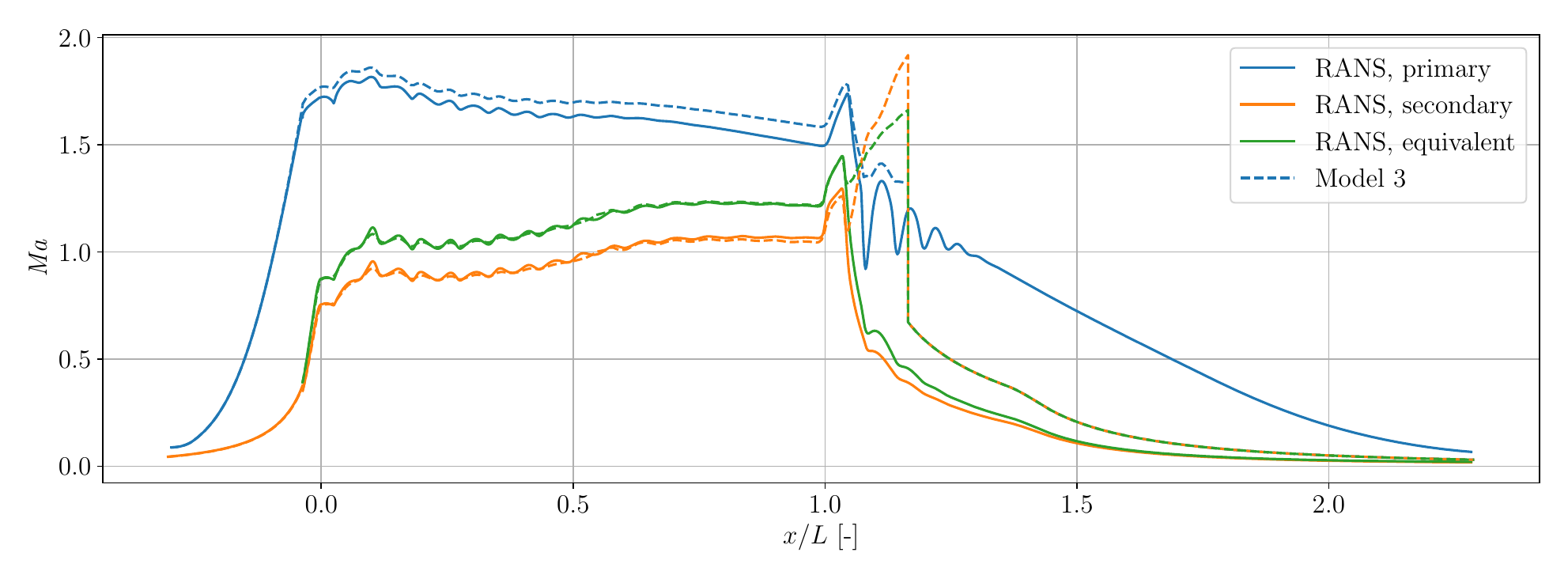}
    \subcaption{Mach number}
    \label{fig:results_fabri_4_Mach}
  \end{subfigure}
  \\
  % \begin{subfigure}[t]{\linewidth}
  %   \center
  %   \includegraphics[width=\linewidth]{figures/results_UCL_fabri/elong_ptp_pts_400_pb_pts_110_total_pressure.pdf}
  %   \subcaption{Total pressure}
  %   \label{fig:results_fabri_4_pt}
  % \end{subfigure}
  % \\
  \begin{subfigure}[t]{\linewidth}
    \center
    \includegraphics[width=\linewidth]{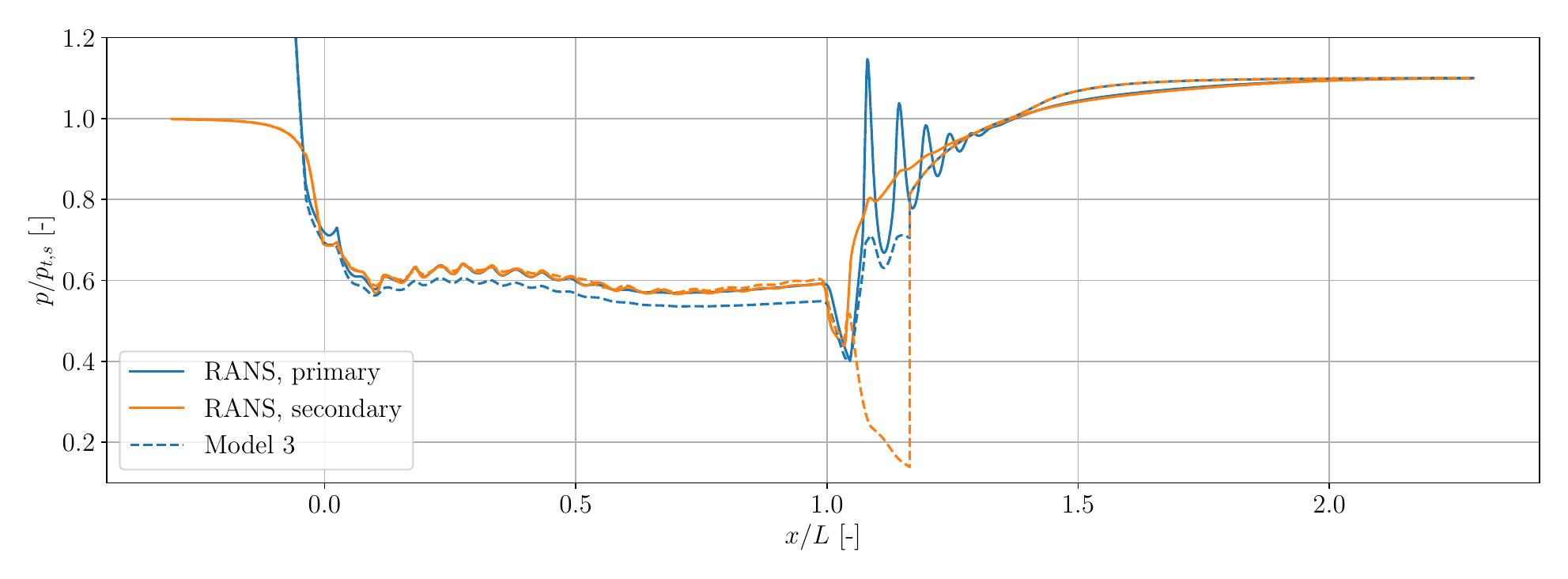}
    \subcaption{Static pressure}
    \label{fig:results_fabri_4_p}
  \end{subfigure}
  \captionsetup{width=\textwidth, justification=justified}
  \caption{Axial distributions of the Mach number, the total pressure and the static pressure from the RANS simulations, and from the Fabri-based model using the imposed total pressure gradient ($p_{t,p}/p_{t,s}=4$). The distributions are in excellent agreement, but a small pressure difference persists at the inlet of the diffuser ($x/L=1$). For the colors in this figure, the reader is referred to the online version of this article.}
\label{fig:results_fabri_4}
\end{figure*}

\begin{figure*}%[htb!]
  \center
  \begin{subfigure}[t]{\linewidth}
    \center
    \includegraphics[width=\linewidth]{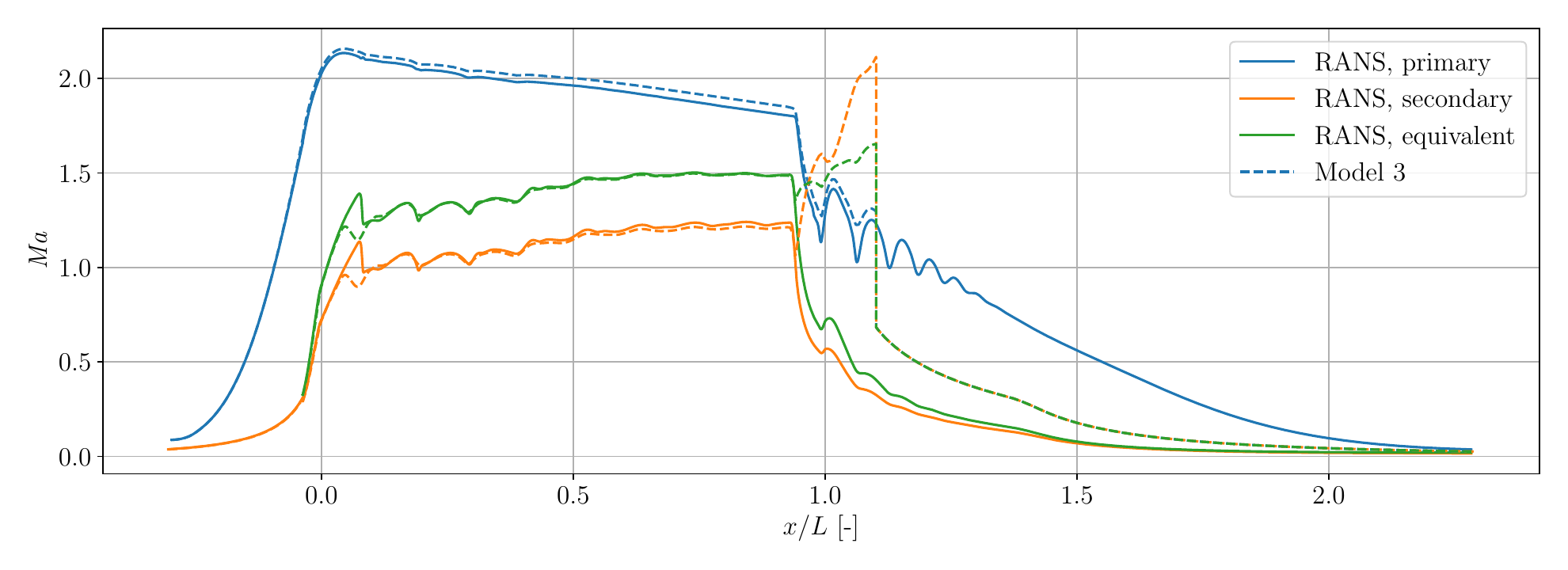}
    \subcaption{Mach number}
    \label{fig:results_fabri_6_Mach}
  \end{subfigure}
  \\
  % \begin{subfigure}[t]{\linewidth}
  %   \center
  %   \includegraphics[width=\linewidth]{figures/results_UCL_fabri/elong_ptp_pts_600_pb_pts_155_total_pressure.pdf}
  %   \subcaption{Total pressure}
  %   \label{fig:results_fabri_6_pt}
  % \end{subfigure}
  % \\
  \begin{subfigure}[t]{\linewidth}
    \center
    \includegraphics[width=\linewidth]{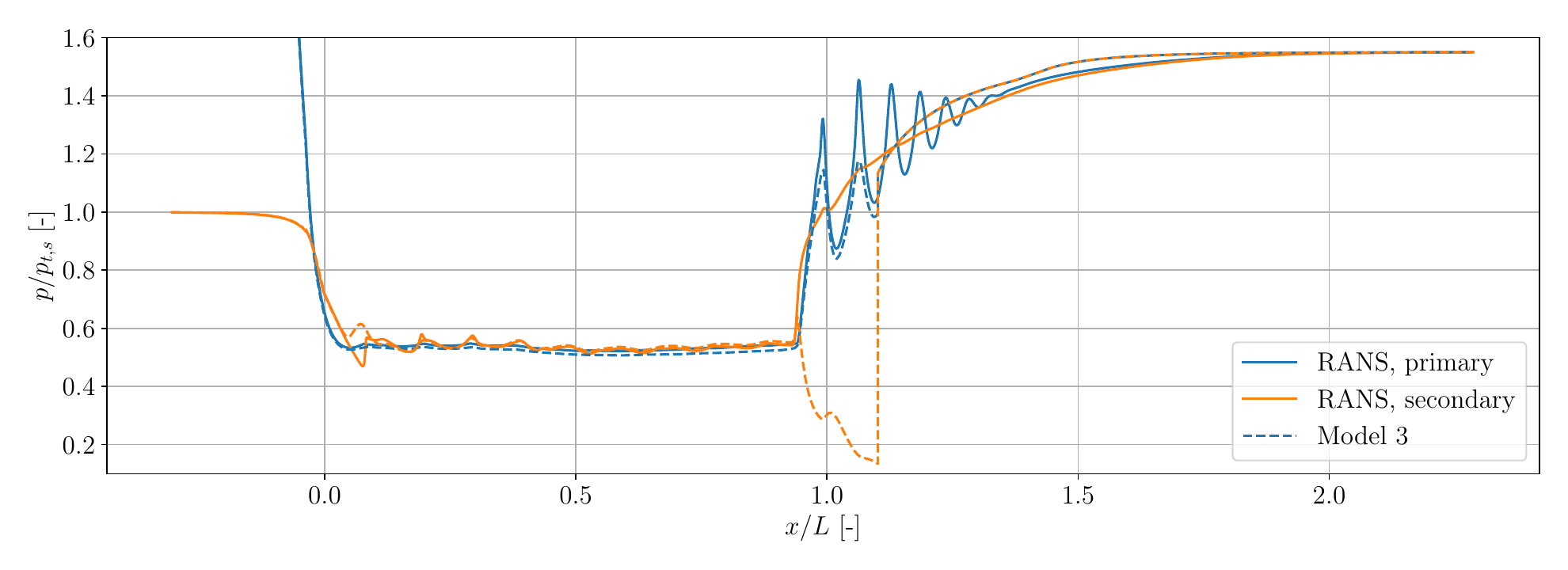}
    \subcaption{Static pressure}
    \label{fig:results_fabri_6_p}
  \end{subfigure}
  \captionsetup{width=\textwidth, justification=justified}
  \caption{Axial distributions of the Mach number, the total pressure and the static pressure from the RANS simulations, and from the Fabri-based model using the imposed total pressure gradient ($p_{t,p}/p_{t,s}=6$). The Fabri-sonic condition is reached in the second shock cell, but overall, the Fabri-based model predictions show excellent agreement with the reference distributions. For the colors in this figure, the reader is referred to the online version of this article.}
\label{fig:results_fabri_6}
\end{figure*}

Table~\ref{tab:results_fabri} summarizes the predicted mass flow rates in the three previously studied operating points (see also table~\ref{tab:results_compound}). The primary mass flow rates are identical to those predicted by the compound model, since the primary stream is assumed to be choked and the same wall friction is applied. The error on the secondary flow rate remains below 1 \% in all cases. This low discrepancy is also obvious from the local distributions, shown in figures~\ref{fig:results_fabri_4} and \ref{fig:results_fabri_6} for the inlet pressure ratios $p_{t,p}/p_{t,s}=4$ and $6$ respectively.

The predicted distributions are nearly coincident with those from the RANS simulations, except near the sonic point in figure~\ref{fig:results_fabri_6_Mach}, which lies in the second shock cell of the RANS simulation. The secondary Mach number reaches 0.96 near the reference sonic point at $x/L=0.05$. This slight discrepancy is likely sufficient to explain the underestimation of 0.9 \% on the secondary flow rate, given the strong sensitivity of Equation~\eqref{eq:p_i} close to unity Mach number. The total pressure distributions match those from section~\ref{sec:results_compound_analysis} (compound model with extracted closures), where the same total pressure gradients were imposed, they are therefore not shown. The corresponding 4.8 \% error in the secondary mass flow rate in the under-expanded case ($p_{t,p}/p_{t,s}=6$) can therefore be attributed to the dividing streamline predicted by the compound model, since all imposed exchange terms are otherwise identical in both approaches.

Note that the predicted static pressures do not fully equalize. For instance, a difference of about 0.05 between the normalized static pressures remains in figure~\ref{fig:results_fabri_4_p} near the diffuser inlet ($x/L = 1$), whereas the reference distributions intersect at several points and eventually equalize. Unlike in the compound model, no pressure equalization mechanism is active here. The discrepancy remains small because the dividing streamline and total pressures are imposed from the RANS simulations. However, in a fully predictive setting, small errors in the dividing streamline or in the momentum exchange could lead to significant, non-physical pressure gaps. This underlines the need for strong coupling between the closure models for momentum exchange and dividing streamline location, ensuring that the primary and secondary static pressures converge toward a common value.
 
The accurate performance of the Fabri-based model demonstrates that the cross-stream averaged framework (cf. figure~\ref{fig:schematics}) can yield reliable predictions when the correct dividing streamline and corresponding closure relations are provided. Currently, the compound-based model remains the only formulation capable of predicting the dividing streamline. As discussed in sections~\ref{sec:results_compound_predictive} and~\ref{sec:results_compound_analysis}, its assumption of uniform static pressure proves overly restrictive, leading to inaccuracies in the dividing streamline and, consequently, in the predicted mass flow rates. The compound model provided the most accurate axial distributions for the case nearest to a perfectly expanded primary stream (see figure~\ref{fig:results_compound_M_5}).

\begin{table*}%[htb!]
    \centering
    \captionsetup{width=\textwidth, justification=justified}
    \caption{Overview of the mass flow rates predicted by the Fabri-based model and the errors with respect to the RANS simulations. The error on both mass flow rates remains below 1 \% in all conditions.}
    \begin{tabular}{cccccccc}
        $p_{t,p}/p_{t,s}$ & $p_b/p_{t,s}$ & $\dot{m}_{p} / \dot{m}_{p,ref}$ & $\dot{m}_{p} / \dot{m}_{p,ref}$ & Error & $\dot{m}_{s} / \dot{m}_{s,ref}$ & $\dot{m}_{s} / \dot{m}_{s,ref}$ & Error \\
        {[-]} & [-] & (model) [-] & (RANS) [-] & [\%] & (model) [-] & (RANS) [-] & [\%] \\ 
        \hline 
        4 & 1.10 & 0.992 & 0.988 & 0.5 & 0.821 & 0.827 & -0.7 \\
        5 & 1.20 & 0.993 & 0.988 & 0.4 & 0.767 & 0.772 & -0.7 \\
        6 & 1.55 & 0.993 & 0.989 & 0.4 & 0.696 & 0.703 & -0.9
    \end{tabular}
    \label{tab:results_fabri}
\end{table*}

\subsubsection{A choked case with a subsonic secondary stream}\label{sec:results_fabri_analysis_critpoint}
\cite{LAMBERTS2018_compound, baguet_querinjean, vandenberghe2024extensioncompoundflowtheory} have reported compound-choked ejector flows with a subsonic secondary stream. The Fabri-sonic condition is therefore not universal, despite the accurate results obtained in the previous section. An intuitive reason for the broader validity of the compound theory lies in the fact that Fabri-sonic parallel streams are compound-supersonic by definition ($\Ma_i\geq1$ in Equation~\eqref{eq:beta}), and therefore compound-sonic farther upstream. If the shock train in the diffuser migrates upstream between these two sonic locations due to increasing back pressure, the flow remains choked in the compound sense, even though the secondary stream does not reach a unitary Mach number.

\begin{figure*}%[htb!]
  \center
  \begin{subfigure}[t]{\textwidth}
    \centering
    \includegraphics[height=0.37\textwidth]{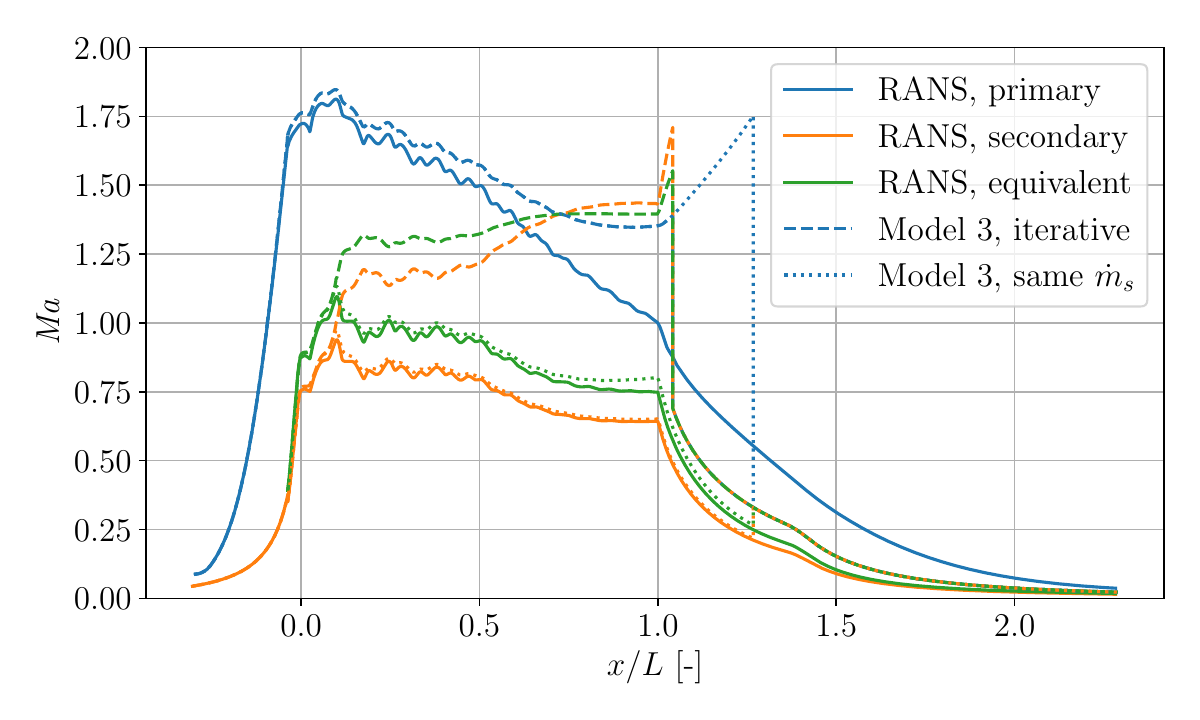}
    \hfill
    \includegraphics[height=0.37\textwidth]{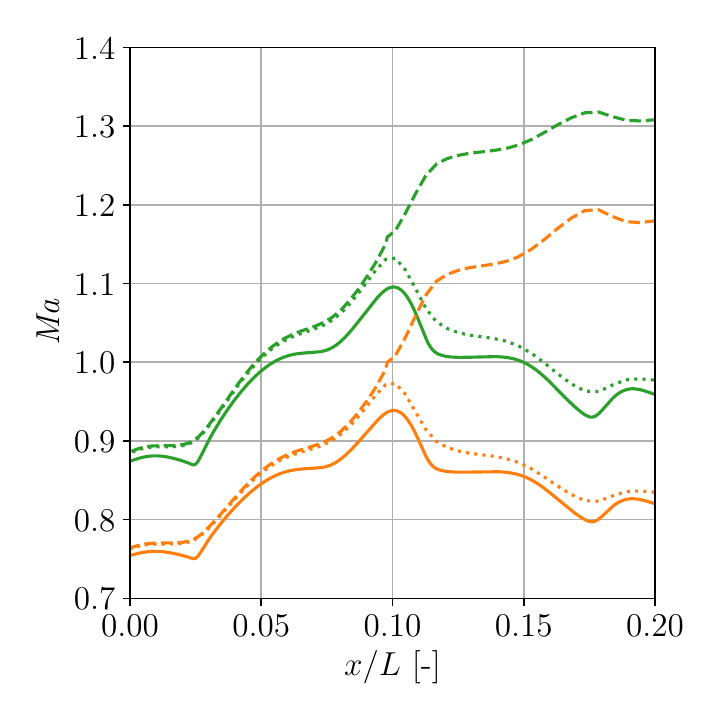}
    \subcaption{Axial distributions of the averaged Mach number}
  \end{subfigure}
  \\
  \begin{subfigure}[t]{0.8\textwidth}
    \centering
    \includegraphics[width=\textwidth]{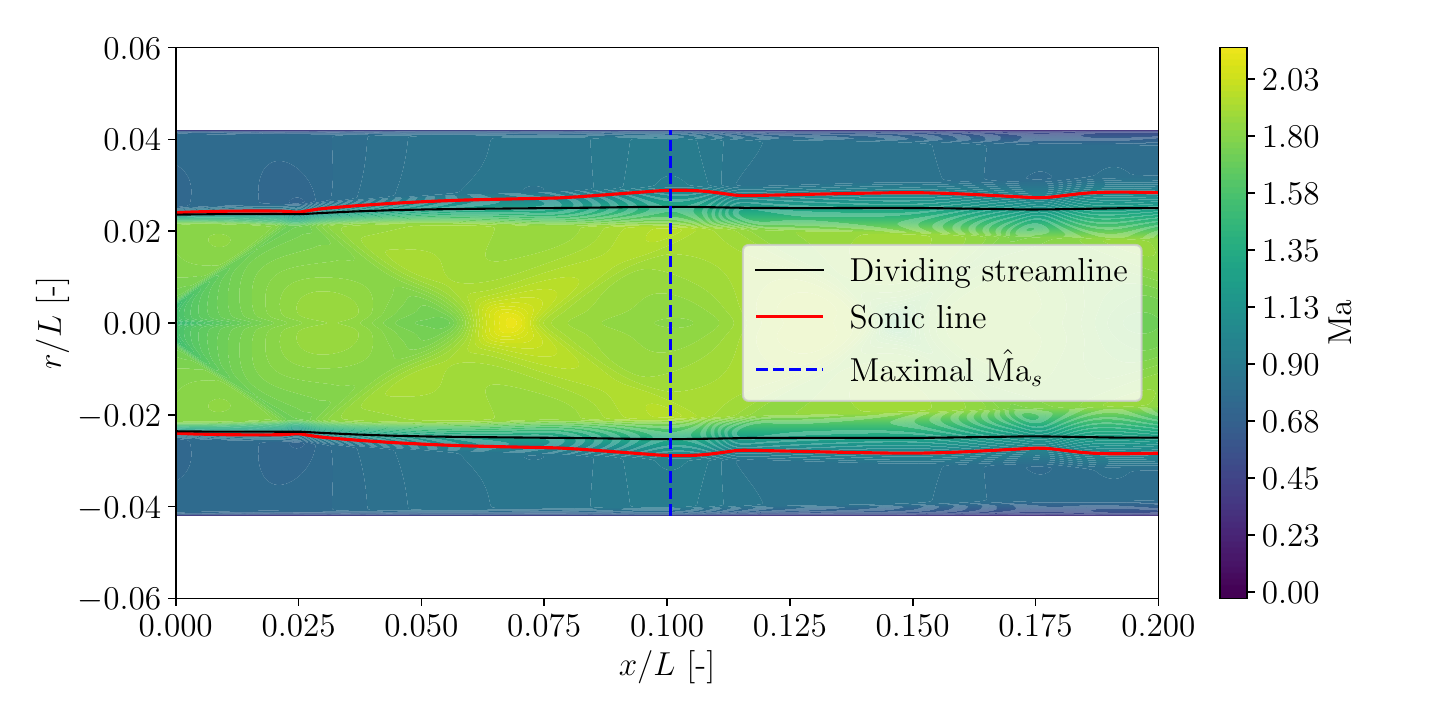}
    \captionsetup{width=\textwidth, justification=justified}
    \subcaption{Mach number field from the RANS simulation (mirrored about $r=0$).}
  \end{subfigure}
  \captionsetup{width=\textwidth, justification=justified}
  \caption{The Mach number from the RANS simulations at the critical operating point ($p_{t,p}/p_{t,s}=4$, $p_b/p_{t,s}=1.48$), from model 3 iterating to the choked solution, and from model 3 imposing the same mass flow rate. The reference flow becomes compound-sonic, but not Fabri-sonic, as as evidenced by the averaged distributions and the sonic line, which penetrates only a small portion of the secondary stream. This indicates the generality of the compound theory, since the flow rate is unchanged compared to figure~\ref{fig:results_fabri_4}.}
\label{fig:results_fabri_crit}
\end{figure*}

Such a case is presented in figure~\ref{fig:results_fabri_crit} for the case $p_{t,p}/p_{t,s}=4$, with an increased back pressure from $p_b/p_{t,s}=1.10$ in figure~\ref{fig:results_fabri_4} to $p_b/p_{t,s}=1.48$. The shock train in the simulation with the lower back pressure has moved from $x/L=1.1$ (in the diffuser) upstream to $x/L=0.1$ in figure~\ref{fig:results_fabri_crit}, which lies beyond the Fabri-sonic from figure~\ref{fig:results_fabri_4} point near $x/L=0.5$. Consequently, the secondary stream no longer reaches a unitary Mach number in figure~\ref{fig:results_fabri_crit}. This also becomes apparent from the sonic line, which does not penetrate the bulk of the secondary stream in the two-dimensional flow field. Nevertheless, the mass flow rates in the RANS simulations with both back pressures are identical, confirming that compound choking is the more general condition governing flow blockage (see also~\cite{vandenberghe2024extensioncompoundflowtheory}).

The choked case with a subsonic secondary stream gives rise to the question how model 3, based on the Fabri-sonic condition, behaves in this case. Two predictions are shown in figure~\ref{fig:results_fabri_crit}; one following the iterative procedure from section~\ref{sec:solution} to find the choked solution, and one where the secondary mass flow rate is imposed to match the RANS simulations.

The Fabri-sonic point in the iterative case is reached at $x/L=0.1$, which coincides with the position where the secondary Mach number is maximal in the RANS simulations due to a local minimum of the cross-section. The sonic point has thus moved upstream with respect to the case with the lower back pressure and the same inlet conditions in figure~\ref{fig:results_fabri_4}. This shift is accompanied by an increase of the predicted secondary flow rate by 1 \%. The Fabri-based model 3 thus predicts a higher flow rate near the critical point than at lower back pressures when imposing the corresponding dividing streamlines and exchanges from RANS simulations. This non-physical behavior has been described earlier in 0D modeling by \cite{METSUE2021121856}, who observed a bump in the operating curves near the critical point. The cause of the different choked solutions in this work lies in the different dividing streamlines (and consequently, the flow variables) in the RANS simulations for $p_b/p_{t,s}=1.10$ and $p_b/p_{t,s}=1.48$, which start differing from the compound-sonic point at $x/L=0.1$. The prediction with matching flow rate remains Fabri-subsonic, and is hence classified as off-design by the iterative procedure in section~\ref{sec:solution}. These observations further indicate the close link between the dividing streamline and the accuracy on the secondary flow rate.

The results presented above indicate that developing a predictive model based on the Fabri-sonic condition---without assuming uniform static pressure---is inherently challenging. This difficulty arises from the strong coupling required between the dividing streamline and the closure relations, as well as from the sensitivity of Equation~\eqref{eq:p_i} near the sonic point. An additional complication occurs when the primary stream decelerates to subsonic speeds. This situation occurs in the RANS simulation in figure~\ref{fig:results_fabri_crit} at $x/L=1$, but not in the model, which instead strays away from the reference and stays supersonic. This discrepancy gradually grows from $x/L=0$ and substantially increases near the second sonic point of the primary stream at $x/L=1$, introducing a similar sensitivity in Equation~\eqref{eq:p_i} as in the Fabri-sonic point of the secondary stream. This situation inevitably manifests itself also under off-design conditions, where the static pressure increases monotonically along the mixing duct (see for example \cite{debroeyer2024analysis}). In such cases, the primary stream passes through its sonic point between the outlet of the primary nozzle (where it is supersonic) and the diffuser exit (where it becomes subsonic). This transition poses a fundamental problem within the proposed framework, since the numerator of the pressure Equation~\eqref{eq:p_i} may not vanish at that point, leading to an infinite pressure gradient.

In contrast, the compound framework does not suffer from this limitation: the corresponding Equation~\eqref{eq:p_compound} remains well-behaved when $\Ma_p = 1$ and $\Ma_s < 1$ ($\beta > 0$). The secondary stream does not exhibit this issue either; encountering a singularity instead signifies an incorrect estimate of the secondary mass flow rate in the iterative procedure described in section~\ref{sec:solution}. To the authors’ knowledge, no existing one-dimensional model in the literature can operate in off-design regimes without assuming uniform static pressure. For instance, the Fabri-based model of \cite{delvalle2012} is explicitly limited to on-design operation.

\subsection{Model 4: Fabri choking with closure from compound choking theory}\label{sec:results_fabri_rdiv_compound}
The final model seeks to reconcile both choking conditions by enforcing the dividing streamline and momentum exchange predicted by the compound model within the Fabri-based framework. The governing equations and closures are identical, but the stopping criteria differ: the compound model converges to $(N\to0,\beta\to0)$ in Equation~\eqref{eq:p_compound}, while the Fabri-based model seeks $(N\to0,\Ma_s\to1)$ in Equation~\eqref{eq:p_i}. Consequently, the choked solutions are not necessarily identical.

Two situations were examined: the fully supersonic and the critical solutions predicted by the compound model for $p_{t,p}/p_{t,s}=6$ (see table~\ref{tab:results_compound}). The critical solution mimics a case with higher back pressure compared to the case shown in figure~\ref{fig:results_compound_M_6}. These were obtained using the positive and negative roots of the quadratic equation for the static pressure gradient at the sonic point, analogous to the two isentropic solutions in a choked nozzle flow. The supersonic solution corresponds to the compound prediction in figure~\ref{fig:results_compound_M_6} prior to applying the normal shock procedure described in section~\ref{sec:practical_shock}.

The Fabri-based model was then initialized using the gradients of the cross-sections from Equation~\eqref{eq:A_i}, evaluated by interpolating the distributions predicted by the compound model. The exchange terms were obtained from the closure relations in section~\ref{sec:definition_compound_predictive}. This interpolation ensures that the geometric gradients remain independent of the flow field computed in the Fabri-based framework.

\begin{figure}%[htb!]
  \center
  \begin{subfigure}[t]{0.48\linewidth}
    \includegraphics[width=\linewidth]{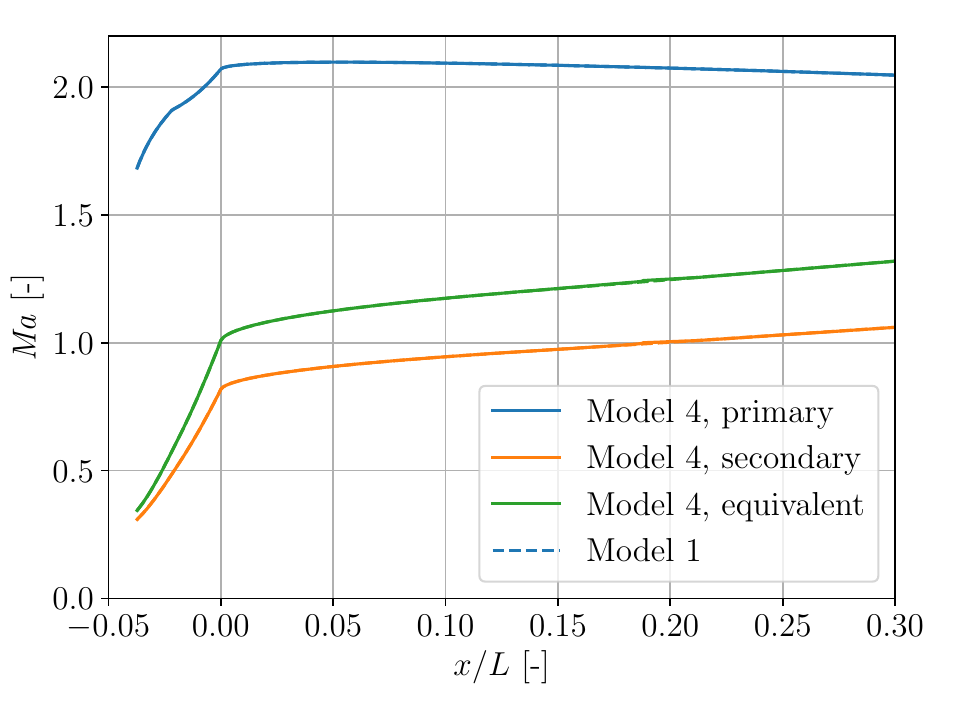}
    \subcaption{Supersonic solution}
    \label{fig:common_supersonic}
  \end{subfigure}
  \begin{subfigure}[t]{0.48\linewidth}
    \includegraphics[width=\linewidth]{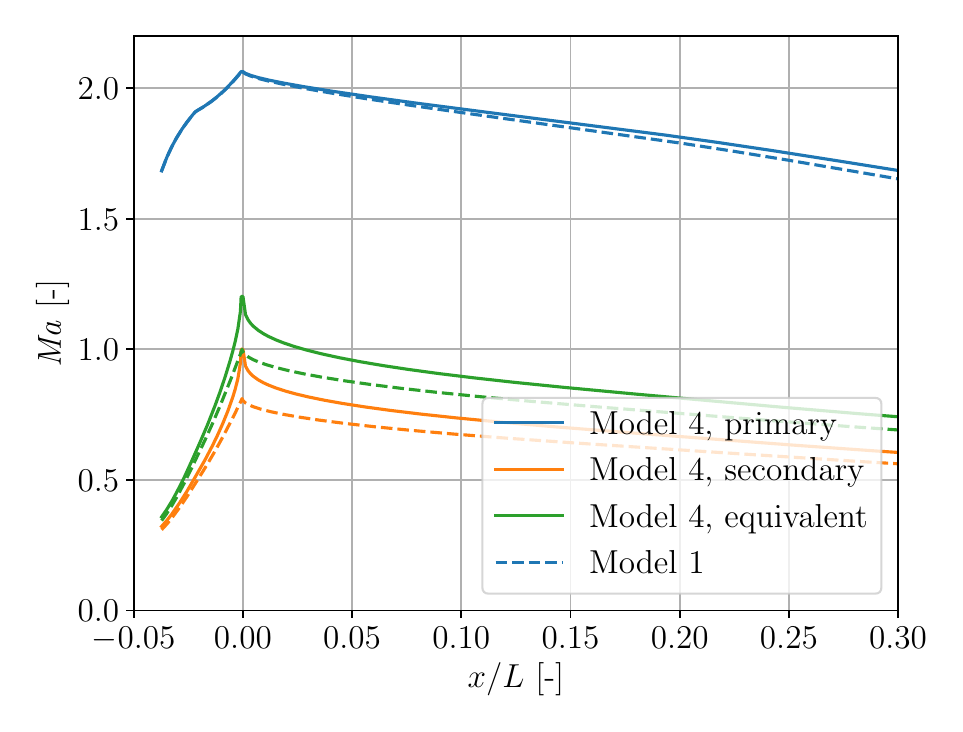}
    \subcaption{Subsonic solution}
    \label{fig:common_subsonic}
  \end{subfigure}
  \captionsetup{width=\textwidth, justification=justified}
  \caption{Predicted distributions of the Mach number based on the compound model, and the Fabri model using the same dividing streamline and the same closure relations. Both theories agree if the Fabri-sonic point is reached. The Fabri-based model overestimates the flow rate with respect to the critical operating point predicted by the compound model.}
\label{fig:common}
\end{figure}

Figure~\ref{fig:common} presents the resulting Mach number distributions. In the supersonic case, both model predictions coincide, reaching the Fabri-sonic condition near $x/L=0.2$, consistent with figure~\ref{fig:results_compound_M_6}. The compound solution thus also satisfies the vanishing-numerator condition in the Fabri-sonic point. 
This equivalence becomes evident when substituting the compound relation for the cross-section, Equation~\eqref{eq:A_i}, in the static pressure Equation~\eqref{eq:p_i}: The $(1-\Ma_s^2)$ term in the numerator of Equation~\eqref{eq:A_i} effectively regularizes the singularity that would otherwise occur at the Fabri-sonic point. As a result, both sonic conditions yield an identical solution, if the same dividing streamline and closure relations are imposed.

The agreement confirms that the 5.2~\% flow-rate overprediction of the compound model (cf. table~\ref{tab:results_compound}) stems from inaccuracies in the dividing streamline rather than from the choking condition itself. The matching distributions in figure~\ref{fig:common_supersonic} confirm that the Fabri-based model predicts the same overestimated flow rate as the compound model. This discrepancy is therefore unrelated to the choking condition, which is the only formal distinction between the two predictions shown in figure~\ref{fig:common_supersonic}.

In contrast, for the critical case, the Fabri-sonic point is not reached in the prediction of model 1, consistent with the RANS results in figure~\ref{fig:results_fabri_crit}. The iterative process then drives $\Ma_s\to1$, leading to a 9~\% overestimation of the secondary flow rate relative to the compound model.
The likely cause of this discrepancy is the artificial nature of the dividing streamline predicted by the compound theory at the critical condition. The abrupt reversal of the static pressure gradient required to restore subsonic conditions in the compound model does not accurately represent the complex un-choking process in reality involving a two-dimensional shock train (cf. figure~\ref{fig:results_fabri_crit}). These findings underline the limitations of the one-dimensional framework defined by Equations~\eqref{eq:p_i}–\eqref{eq:Tt_i}, especially near the critical operating point.

\section{Conclusion} \label{sec:conclusion}

This work provides a detailed one-dimensional analysis of the choking mechanisms in supersonic ejectors, focusing on the interplay between the compound and Fabri formulations. By implementing both mechanisms within the same two-stream framework, the study disentangles their respective assumptions and effects on the predicted flow evolution. The analysis shows that compound and Fabri choking can coexist within the same device and operating map. However, all Fabri-choked cases are also compound-choked, while the reverse is not true, indicating that the compound mechanism represents the more general condition governing flow-rate limitation in ejectors.

The compound model predicts the secondary mass flow rate with an accuracy of approximately 5~\% on the secondary flow rate compared to reference CFD results across all investigated cases. The observed overestimation of entrainment under on-design conditions stems from the assumption of equal static pressure, which neglects the inertial influence of the primary stream and its associated overshooting expansion. The iterative scheme compensates for this simplification by converging to a higher secondary flow rate. Imposing the momentum exchange from the RANS simulations did not improve this error, showing that it can not be reduced further through calibration of the forces. Although the accuracy of the local flow distributions can be slightly improved, the correlation-based predictions remain remarkably accurate---particularly considering that the model executes within seconds on a standard laptop, compared to several hours required for equivalent RANS simulations

In contrast, the Fabri formulation allows static pressures to evolve freely but requires the dividing streamline to be prescribed by an external model or data source. Using the filtered dividing streamline extracted from RANS simulations reduced the error in secondary mass flow rate to below 1~\%. However, the integration procedure proved extremely sensitive to both the dividing streamline and the closure relations small perturbations in the cross-sections induced large fluctuations in the Mach number and in some cases to infinite pressure gradients. This sensitivity is expected to persist if the dividing streamline is obtained from a predictive model. Notably, even when RANS-based streamlines were applied, complete static pressure equalization was not achieved within the Fabri framework. This limitation is absent from the compound model, whose cross-sectional formulation inherently accounts for the effects of forces, regardless of their precise definition.

Both choking formulations yield identical results when employing the same dividing streamline and closure relations, provided the Fabri-sonic condition is reached. In that case, the Fabri-based model reproduces the same prediction error as the compound model, confirming that discrepancies in entrainment arise from inaccuracies in the dividing streamline rather than from the choking condition itself. Nonetheless, distinct solutions appear when using the dividing streamline from the critical point as predicted by the compound model. Such cases—featuring a subsonic secondary stream—have been presented here and in literature. Although it should be kept in mind that these flow fields near the critical point remain strongly two-dimensional, thereby testing the limits of the present one-dimensional framework, the compound sonic condition emerges as the more general criterion governing flow rate blockage.

However, the investigated test cases exhibit nonuniform static pressure profiles, which conflict with the fundamental assumption of uniform static pressure in compound models. This suggests the existence of a \emph{generalized} compound condition, where the sonic condition depends on both streams without requiring uniform static pressure. Such an extension could significantly enhance one-dimensional ejector modeling, though at present, no straightforward research path exists for developing a compound formulation without this simplifying assumption.

\vspace{3mm}
\textbf{Acknowledgements} J. Van den Berghe is supported by F.R.S.-FNRS FRIA grant number 47455. Declaration of Interests: The authors report no conflict of interest.
\appendix
\section{First and second order approximations of the pressure gradient in a single stream}\label{sec:appendix_single_stream}
In this section, the first and second order approximations are computed for the pressure gradient of a single stream, using Taylor expansions of the numerator and denominator of Equation~\eqref{eq:p_i_denom}. The notation and the governing equations are provided in section~\ref{sec:app_single_def}, which are required for computing the first and second derivatives in the Taylor expansions in sections~\ref{sec:app_single_first} and \ref{sec:app_single_second}.

\subsection{Definitions}\label{sec:app_single_def}
We reformulate the definition \eqref{eq:p_i_denom} of the static pressure gradient as follows:
\begin{equation}
    \frac{d\ln(p)}{dx} = \frac{N_A + N_F}{D}\,,
\end{equation}
where
\begin{equation}
    N_A = \gamma \Ma^2 \frac{d\ln(A)}{dx} \,,
    \quad
    N_F = \left(1 + \left(\gamma  - 1\right)\Ma^2\right)\frac{F}{A p}\,,
    \quad \text{and} \quad
    D = 1-\Ma^2\,.
\end{equation}
The indices $p$ have been omitted since the analysis that follows is valid for any single stream in a channel $A(x)$. The following equation for the Mach number can be derived from the governing Equations~\eqref{eq:p_i}-\eqref{eq:Tt_i} (see \cite{shapiro1953dynamics}):
\begin{equation}\label{eq:dMa2_dx}
    \frac{d\Ma^2}{dx} = c_{M0} \frac{d\ln(p)}{dx} + c_{M^21} F\,,
\end{equation}
where
\begin{equation}
    c_{M^20} = -\frac{2}{\gamma}\left(1+\frac{\gamma - 1}{2}\Ma^2\right)\,,
    \quad \text{and} \quad
    c_{M^21} = \frac{2}{\gamma}\left[1 + \frac{\gamma - 1}{2}\Ma^2\right] \frac{1}{A p}\,.
\end{equation}
For conciseness, we also define:
\begin{equation}
    c_1 = 1 + \left(\gamma  - 1\right)\Ma^2\,.
\end{equation}

\subsection{First order approximation}\label{sec:app_single_first}
The first order approximation consists of using the pressure gradient in the sonic section as a constant in the neighbourhood. It is computed with de l'Hôpital's rule as in \eqref{eq:hopital_main}:
\begin{equation}\label{eq:single_approx_1}
        \left(\frac{d\ln(p)}{dx}\right)^* = \frac{N^*}{D^*} = \frac{0}{0} = \frac{(dN/dx)^*}{(dD / dx)^*}\,.
\end{equation}
We therefore need the derivatives of the numerator and the denominator, as computed below.
The derivative of $N_A$ reads:
\begin{equation}
    \frac{dN_A}{dx} = c_{NA0} \frac{d\ln(p)}{dx} + c_{NA1} \,,
\end{equation}
where:
\begin{equation}
    c_{NA0} = \gamma \frac{d\ln(A)}{dx} c_{M^20}\,,
    \quad \text{and} \quad
    c_{NA1} = \gamma \frac{d\ln(A)}{dx} c_{M^21} F + \gamma \Ma^2 \frac{d^2\ln(A)}{dx^2}\,.
\end{equation}
The derivative of $N_F$ reads:
\begin{equation}\label{eq:dNF_dx_1}
    \frac{dN_F}{dx} = c_{NFp} \frac{d\ln(p)}{dx} + c_{NFF} \frac{dF}{dx} + c_{NF2}\,,
\end{equation}
where
\begin{gather}
    c_{NFp} = \left[\left(\gamma  - 1\right)c_{M^20} -c_1\right]\frac{F}{A p}\,,
    \quad
    c_{NFF} = \left(1 + \left(\gamma  - 1\right)\Ma^2\right)\frac{1}{A p}\,,\nonumber\\
    c_{NF2} = \left(\gamma  - 1\right)c_{M^21} \frac{F^2}{A p}-c_1\frac{F}{A p}\left(\frac{d\ln(A)}{dx}\right)\,.
\end{gather}
We introduce the following decomposition for the derivative of the force $F$:
\begin{equation}
    \frac{dF}{dx} = c_{F0} \frac{d\ln(p)}{dx} + c_{F1}\,.\label{eq:app_decompF_single}
\end{equation}
This operation is general: any definition of the force $F$ can be reduced to this form using the governing equations. The definition of these coefficients depends on the closure of the conservation equations; the definitions used in this work are given in section~\ref{sec:app_single_forces}. Introducing this decomposition in \eqref{eq:dNF_dx_1} leads to:
\begin{equation}\label{eq:dN_dx}
    \frac{dN}{dx} = c_{N0} \frac{d\ln(p)}{dx} + c_{N1}\,,
\end{equation}
where
\begin{equation}
    c_{N0} = c_{NA0} + c_{NFp} + c_{NFF} c_{F0}\,,
    \quad \text{and} \quad
    c_{N1} = c_{NA1} + c_{NFF} c_{F1} + c_{NF2}\,.
\end{equation}
The derivative of the denominator equals:
\begin{equation}\label{eq:dD_dx}
    \frac{dD}{dx} = -\frac{d\Ma^2}{dx} = c_{D0} \frac{d\ln(p)}{dx} + c_{D1}\,,
\end{equation}
where:
\begin{equation}
    c_{D0} = -c_{M^20}\,,
    \quad \text{and} \quad
    c_{D1} = -c_{M^21} F\,.
\end{equation}
Introducing the computed derivatives of the numerator and the denominator in Equation~\eqref{eq:single_approx_1} leads to a quadratic equation for the pressure gradient with a positive root (compressing to subsonic speed) and a negative root (expanding to supersonic speed):
\begin{equation}\label{eq:quadratic_dpdx}
     c_{D0}^*\left(\left(\frac{d\ln(p)}{dx}\right)^*\right)^2 +  \left(c_{D1}^* - c_{N0}^*\right)\left(\frac{d\ln(p)}{dx}\right)^*
     -c_{N1}^* = 0\,.
\end{equation}

\subsection{Second order approximation}\label{sec:app_single_second}
The full Taylor expansions of the numerator and denominator provide the following exact equation (recall that $N^*=D^*=0$):
\begin{equation}
        \frac{d\ln(p)}{dx} = \dfrac{N^* + (x-x^*)\left(\dfrac{dN}{dx}\right)^* + \dfrac{(x-x^*)^2}{2}\left(\dfrac{d^2N}{dx^2}\right)^* + ...}{D^* + (x-x^*)\left(\dfrac{dD}{dx}\right)^* + \dfrac{(x-x^*)^2}{2}\left(\dfrac{d^2D}{dx^2}\right)^* + ...}\,.
\end{equation}
Truncating after second derivatives allows to build a second order approximation near the sonic point:
\begin{equation}\label{eq:single_second_order}
        \dfrac{d\ln(p)}{dx} \approx \dfrac{\left(\dfrac{dN}{dx}\right)^* + \dfrac{(x-x^*)}{2}\left(\dfrac{d^2N}{dx^2}\right)^*}{\left(\dfrac{dD}{dx}\right)^* + \dfrac{(x-x^*)}{2}\left(\dfrac{d^2D}{dx^2}\right)^*}\,.
\end{equation}
The second derivatives of the numerator and the denominator are computed below.
The derivatives of the coefficients $c_{M^20}$ and $c_{M^21}$ are given by:
\begin{align}
    \frac{dc_{M^20}}{dx} =  &-\frac{\gamma - 1}{\gamma}\frac{d\Ma^2}{dx} \,,\\
    \frac{dc_{M^21}}{dx} = &\frac{\gamma-1}{\gamma}\frac{d\Ma^2}{dx}\frac{1}{A p} - c_{M^21}\left(\frac{d\ln(A)}{dx} + \frac{d\ln(p)}{dx}\right)\,.
\end{align}
These first derivatives are directly available from the equations in the section above, so are not expanded further for conciseness. We differentiate the coefficients related to $N_A$:
\begin{align}
    \frac{dc_{NA0}}{dx} = &\gamma \frac{d^2\ln(A)}{dx^2} c_{M^20} + \gamma \frac{d\ln(A)}{dx} \frac{dc_{M^20}}{dx}\,,\\
    \frac{dc_{NA1}}{dx} = & \gamma \Ma^2 \frac{d^3\ln(A)}{dx^3} + \gamma \frac{d^2\ln(A)}{dx^2} \left[ \frac{d\Ma^2}{dx} + c_{M^21} F\right]\nonumber\\
    & + \gamma  \frac{d\ln(A)}{dx}\left[ F \frac{dc_{M^21}}{dx} + c_{M^21} \frac{dF}{dx}\right]\,.
\end{align}
The second derivative of $N_A$ becomes:
\begin{equation}
    \frac{d^2N_A}{dx^2} = c_{NA0} \frac{d^2\ln(p)}{dx^2} + c_{NA21} \,,
    \quad \text{where} \quad
    c_{NA21} = \frac{dc_{NA0}}{dx} \frac{d\ln(p)}{dx} + \frac{dc_{NA1}}{dx}\,.
\end{equation}
We differentiate the coefficients related to $N_F$:
\begin{align}
    \frac{dc_{NFp}}{dx} = &\left(\gamma  - 1\right)\frac{F}{A p}\left(\frac{dc_{M^20}}{dx} -\frac{d\Ma^2}{dx}\right) + \left[\left(\gamma  - 1\right)c_{M^20} -c_1\right] \frac{d}{dx}\left(\frac{F}{Ap}\right)\,,
    \\
    \frac{dc_{NFF}}{dx} = &\left(\gamma  - 1\right)\frac{d\Ma^2}{dx}\frac{1}{A p} + c_1\frac{d}{dx}\left(\frac{1}{A p}\right)\,,
    \\
    \frac{dc_{NF2}}{dx} = &\left(\gamma  - 1\right)\left[ \frac{F^2}{A p} \frac{dc_{M^21}}{dx}
    + c_{M^21} \frac{d}{dx}\left(\frac{F^2}{A p}\right)\right] -\left(\gamma  - 1\right)\frac{d\Ma^2}{dx}\frac{F}{A p}\frac{d\ln(A)}{dx}\nonumber \\
    & - c_1\frac{F}{A p}\frac{d^2\ln(A)}{dx^2}-c_1\left(\frac{d\ln(A)}{dx}\right)\frac{d}{dx}\left(\frac{F}{A p}\right)\,.
\end{align}
Differentiating $N_F$ twice leads to:
\begin{equation}
    \frac{d^2N_F}{dx^2} = c_{NFp} \frac{d^2\ln(p)}{dx^2} + c_{NFF} \frac{d^2F}{dx^2} + c_{NF22}\,,
\end{equation}
where
\begin{equation}
    c_{NF22} = \frac{dc_{NFp}}{dx} \frac{d\ln(p)}{dx} + \frac{dc_{NFF}}{dx} \frac{dF}{dx} + \frac{dc_{NF2}}{dx}\,.
\end{equation}
Differentiating the decomposition \eqref{eq:app_decompF_single} gives:
\begin{equation}\label{eq:app_decompdFdx_single}
    \frac{d^2F}{dx^2} = c_{F0} \frac{d^2\ln(p)}{dx^2} + c_{F21}\,,
    \quad \text{where} \quad
    c_{F21} = \frac{dc_{F0}}{dx} \frac{d\ln(p)}{dx} + \frac{dc_{F1}}{dx}
\end{equation}
The second derivative of $N_F$ becomes:
\begin{equation}
    \frac{d^2N_F}{dx^2} = c_{NF20} \frac{d^2\ln(p)}{dx^2} + c_{NF21}\,,
\end{equation}
where
\begin{equation}
    c_{NF20} = c_{NFp} + c_{NFF} c_{F0}\,,
    \quad \text{and} \quad
    c_{NF21} = c_{NFF}c_{F21} + c_{NF22} \,.
\end{equation}
The second derivative of the numerator equals:
\begin{equation}\label{eq:d2N_dx2}
    \frac{d^2N}{dx^2} = c_{N20} \frac{d^2\ln(p)}{dx^2} + c_{N21}\,,
\end{equation}
where
\begin{equation}
    c_{N20} = c_{NA0} + c_{NF20}\,,
    \quad \text{and} \quad
    c_{N21} = c_{NA21} + c_{NF21}\,.
\end{equation}
The second derivative of the denominator equals:
\begin{equation}\label{eq:d2D_dx2}
    \frac{d^2D}{dx^2} = c_{D0} \frac{d^2\ln(p)}{dx^2} + c_{D21}\,,
    \quad \text{where} \quad
    c_{D21} = \frac{dc_{D0}}{dx} \frac{d\ln(p)}{dx} + \frac{dc_{D1}}{dx}\,,
\end{equation}
and the derivatives of the coefficients are given by:
\begin{align}
    \frac{dc_{D0}}{dx} &= -\frac{dc_{M^20}}{dx}\,,\\
    \frac{dc_{D1}}{dx} &= -\frac{dc_{M^21}}{dx} F - c_{M^21} \frac{dF}{dx}\,.
\end{align}
The second derivative of the static pressure is given by:
\begin{equation}
    \frac{d^2\ln(p)}{dx^2} = \frac{d}{dx}\left(\frac{N}{D}\right) = \frac{D (dN/dx) - N (dD/dx)}{D^2}\,.
\end{equation}
If $(N \to 0, D \to 0)$, de l'Hôpital's rule yields the following equation:
\begin{equation}
    \left(\frac{d^2\ln(p)}{dx^2}\right)^* = \frac{0}{0} = \frac{(d^2N/dx^2)^*}{2 (dD/dx)^*} - \frac{N^*}{D^*}\frac{(d^2D/dx^2)^*}{2(dD/dx)^*}\,,
\end{equation}
where $N^*/D^*$ equals the static pressure gradient in the sonic section. Substituting the second derivatives $d^2N/dx^2$ and $d^2D/dx^2$ eventually leads to:
\begin{equation}\label{eq:d2p_dx2}
    \left(\dfrac{d^2\ln(p)}{dx^2}\right)^* = \dfrac{- c_{N21}^* + \left(\dfrac{d \ln(p)}{dx}\right)^* c_{D21}^*}{c_{N20}^*  - \left(\dfrac{d \ln(p)}{dx}\right)^* c_{D0}^* - 2 \left(\dfrac{dD}{dx}\right)^*} \,.
\end{equation}
The terms in \eqref{eq:single_second_order} can be computed with Equations~\eqref{eq:dN_dx}, \eqref{eq:dD_dx}, \eqref{eq:d2N_dx2}, \eqref{eq:d2D_dx2}, which can be evaluated in the sonic point using the derivatives of the static pressure given by \eqref{eq:quadratic_dpdx} and \eqref{eq:d2p_dx2}. The result is a the second order approximation of the static pressure gradient.

\subsection{Derivatives of the friction force}\label{sec:app_single_forces}
The decompositions \eqref{eq:app_decompF_single} and \eqref{eq:app_decompdFdx_single} introduce the coefficients $c_{F0}$ and $c_{F1}$ and their derivatives, which are given by the equations below for the definition in Equation~\eqref{eq:F_i_inlets}:
\begin{equation}
    c_{F0} =  F\left(1 + \frac{c_{M0}}{\Ma^2}\right) \,,
    \quad \text{and} \quad
    c_{F1} = c_{M1} \frac{F^2}{\Ma^2} + \frac{F}{l_w}\frac{dl_w}{dx}\,.
\end{equation}
\begin{align}
    \frac{dc_{F0}}{dx} = &\left(1 + \frac{c_{M0}}{\Ma^2}\right) \frac{dF}{dx} + \frac{F}{\Ma^2}\frac{dc_{M0}}{dx} - \frac{Fc_{M0}}{\Ma^4} \frac{d\Ma^2}{dx}\,,\\
    \frac{dc_{F1}}{dx} = &c_{F1}\left(\frac{1}{c_{M1}}\frac{dc_{M1}}{dx} + 2 \frac{1}{F}\frac{dF}{dx} - \frac{1}{\Ma^2}\frac{d\Ma^2}{dx}\right)+\frac{1}{l_w}\frac{dl_w}{dx}\frac{dF}{dx} - \frac{F}{l_w^2}\left(\frac{dl_w}{dx}\right)^2 + \frac{F}{l_w}\frac{d^2l_w}{dx^2}
\end{align}

\bibliographystyle{jfm}
\bibliography{bibliography}
%Use of the above commands will create a bibliography using the .bib file. Shown below is a bibliography built from individual items.

% \bibliographystyle{jfm}
%\bibliography{jfm2esam}

% \begin{thebibliography}{99}

% \expandafter\ifx\csname natexlab\endcsname\relax
% \def\natexlab#1{#1}\fi
% \expandafter\ifx\csname selectlanguage\endcsname\relax
% \def\selectlanguage#1{\relax}\fi

% \end{thebibliography}

%% End of file `jfm2esam.bib'.

\end{document}